\documentclass[preprint,3p]{elsarticle}
\usepackage[T1]{fontenc}
\usepackage{syntax, amsmath, amssymb, amsthm, stmaryrd, mathtools, semantic, enumitem, booktabs, listings, textcomp, placeins, algorithm, algorithmicx, algpseudocode, subcaption, xr, multicol, float, array, xspace, graphicx, hyperref}
\usepackage[dvipsnames]{xcolor}
\usepackage{todonotes}
\usepackage[export]{adjustbox}
\usepackage{arydshln}
\usetikzlibrary{arrows,calc}
\newfloat{Syntax}{htp}{syn}
\lstdefinelanguage{ADL}{
	keywords = [1]{struct, int, bool, nat, String, if, then, Fix},
	morekeywords=[2]{null, this, true, false},
	sensitive=false, 
	morecomment=[l]{//},
	morecomment=[s]{/*}{*/},
	morestring=[b]"
}

\lstdefinelanguage{gmcommands}{
	morekeywords = {push, rd, wr, cons, if, not, op, this, Fix},
	sensitive=true, 
	morecomment=[l]{//},
	morecomment=[s]{/*}{*/},
	morestring=[b]"
}

\definecolor{codegray}{rgb}{0.5,0.5,0.5}

\lstdefinestyle{gmcommands}{
	language={gmcommands},
	showstringspaces=false,
	basicstyle=\small,
	keywordstyle=\textbf,
	numberstyle=\tiny\color{codegray},
	stringstyle=\slshape,
	commentstyle=\color{codegray},
	emph={
		val
	},
	emphstyle = \rmfamily\itshape,
	breaklines=true
}

\lstdefinestyle{CUDA}{
	language=[ANSI]C++,
	showstringspaces=false,
	basicstyle=\small,
	keywordstyle=\color{MidnightBlue},
	numberstyle=\tiny\color{codegray},
	stringstyle=\slshape,
	commentstyle=\color{codegray},
	emph={
		cudaMalloc, cudaFree, cudaMemcpy, cudaMemcpyHostToDevice, cudaMemcpyDeviceToHost, cudaDeviceSynchronize,
		__global__, __shared__, __device__, __host__,
		__syncthreads,
	},
	emphstyle = \color{OliveGreen},
	breaklines=true
}

\lstdefinestyle{OpenCL}{
	language=[ANSI]C++,
	showstringspaces=false,
	basicstyle=\small,
	keywordstyle=\color{MidnightBlue},
	numberstyle=\tiny\color{codegray},
	stringstyle=\slshape,
	commentstyle=\color{codegray},
	emph={
		kernel, local, barrier
	},
	emphstyle = \color{OliveGreen},
	breaklines=true
	escapeinside={(*}{*)},
}

\newcommand{\etal}{\emph{et al.}}
\newcommand{\Lname}{AuDaLa\xspace}

\newcommand{\vempty}[0]{\varepsilon}
\DeclareMathSymbol{\sm}{\mathbin}{AMSa}{"39}
\newcommand{\nil}[0]{\mathit{null}}
\newcolumntype{?}{!{\vrule width 1.5pt}}

\newtheorem{theorem}{Theorem}[section]
\newtheorem{lemma}[theorem]{Lemma}
\newtheorem{corollary}[theorem]{Corollary}

\newdefinition{definition}[theorem]{Definition}
\newdefinition{problem}{Problem}[section]

\newcommand{\type}[1]{\langle \mathit{#1}\rangle}
\newcommand{\word}[1]{\text{`\texttt{#1}'}}
\newcommand{\SynTypes}[0]{\mathcal{T}}
\newcommand{\Id}{\mathit{ID}}
\newcommand{\tr}{\vdash}
\newcommand{\true}[0]{\mathit{true}}
\newcommand{\false}[0]{\mathit{false}}

\newcommand{\Sched}[0]{\mathit{Sc}}
\newcommand{\Structs}[0]{\sigma}

\newcommand{\Env}[0]{\xi}
\newcommand{\Labels}[0]{\mathcal{L}}
\newcommand{\StructsSet}[0]{\mathcal{S}}
\newcommand{\Values}[0]{\mathcal{V}}
\newcommand{\getValue}{\mathit{val}}
\newcommand{\defaultVal}{\mathit{defaultVal}}

\newcommand{\rarr}[0]{\rightarrow}
\newcommand{\Rarr}[0]{\Rightarrow}
\newcommand{\StateSpace}[0]{S_\mathcal{G}}
\newcommand{\Literals}[0]{\mathit{LT}}
\newcommand{\Schedules}[0]{\mathit{SC}}
\newcommand{\Program}[0]{\mathcal{P}}
\newcommand{\Programs}[0]{\Pi}

\newcommand{\String}[0]{\mathit{String}}
\newcommand{\SynString}[0]{\texttt{String}}
\newcommand{\Expr}[0]{\mathit{EX}}

\newcommand{\Nat}[0]{\mathbb{N}}
\newcommand{\SynNat}[0]{\texttt{Nat}}
\newcommand{\Int}[0]{\mathbb{Z}}
\newcommand{\SynInt}[0]{\texttt{Int}}
\newcommand{\Bool}[0]{\mathbb{B}}
\newcommand{\SynBool}[0]{\texttt{Bool}}

\newcommand{\StructTypes}[0]{\Theta}
\newcommand{\StructType}[0]{\theta}

\newcommand{\SynType}[0]{\mathit{T}}
\newcommand{\SemTypes}[0]{\mathbb{T}}
\newcommand{\interp}[1]{\llbracket #1 \rrbracket}
\newcommand{\sr}[0]{\Rarr_\Program}

\newcommand{\Stab}[0]{\mathit{s\chi}}
\newcommand{\Stack}[0]{\chi}
\newcommand{\readg}[1]{\mathbf{rd}(#1)}
\newcommand{\writev}[1]{\mathbf{wr}(#1)}
\newcommand{\cons}[1]{\mathbf{cons}(#1)}
\newcommand{\push}[1]{\mathbf{push}(#1)}

\newcommand{\Stat}[0]{\mathit{ST}}
\newcommand{\Commands}[0]{\mathcal{C}}
\newcommand{\Operator}[0]{\mathbf{op}}
\newcommand{\Ifc}[1]{\mathbf{if}(#1)}
\newcommand{\Notc}[0]{\mathbf{not}}
\newcommand{\this}[0]{\mathbf{this}}
\newcommand{\CList}[0]{\gamma}
\newcommand{\StatList}[0]{\mathcal{S}}
\newcommand{\ComList}[0]{\gamma}
\newcommand{\StateP}[0]{\mathit{P}}
\newcommand{\Par}[1]{\mathit{Pars}(#1)}
\newcommand{\nilL}[1]{\ell^0_{#1}}
\newcommand{\SynOp}[0]{O}

\newcommand{\NilLabels}[0]{\Labels^{0}}

\newcommand{\val}[1]{\mathit{val}(#1)}

\begin{document}
	\title{The Autonomous Data Language -- Concepts, Design and Formal Verification}
	%
	%
	\author{Tom T.P. Franken}
	\ead{t.t.p.franken@tue.nl}
	\author{Thomas Neele}
	\ead{t.s.neele@tue.nl}
	\author{Jan Friso Groote}
	\ead{j.f.groote@tue.nl}
	\affiliation{Eindhoven University of Technology, The Netherlands}
	\begin{abstract}
		Nowadays, the main advances in computational power are due to parallelism.
		However, most parallel languages have been designed with a focus on processors and threads. 
		This makes dealing with data and memory in programs hard, which distances the implementation from its original algorithm.
		We propose a new paradigm for parallel programming, the \emph{data-autonomous} paradigm, where computation is performed by autonomous data elements. 
		Programs in this paradigm are focused on making the data collaborate in a highly parallel fashion.
		We furthermore present \Lname, the first data autonomous programming language, and provide a full formalisation that includes a type system and operational semantics.
		Programming in \Lname is very natural, as illustrated by examples, albeit in a style very different from sequential and contemporary parallel programming. 
		Additionally, it lends itself for the formal verification of parallel programs, which we demonstrate.
		
	\end{abstract}
	
	
	\maketitle
	%
	%
	%
\section{Introduction}\label{section:introduction}
As increasing the speed of sequential processing becomes more difficult~\cite{leiserson-theres-2020}, exploiting parallelism has become one of the main means of obtaining further performance improvements in computing. 
Thus, languages and frameworks aimed at parallel programming play an increasingly important role in computation.
Many existing parallel languages use a \emph{task-parallel} or a \emph{data-parallel} paradigm~\cite{ciccozzi-comprehensive-2022}.

Task-parallelism mostly focuses on the computation carried out by individual threads, scheduling tasks to threads depending on which threads are idle.
In data-parallelism, threads execute the same function but are distributed over the data, thus performing a parallel computation on the collection of all data.

In a shared memory setting, programs in both paradigms require careful design of memory layout, memory access and movement of data to facilitate the threads used by the program. Examples of this are the use of barriers and data access based on thread id's, as well as access protocols.
Not only is extensive data movement costly and hinders some performance optimizations~\cite{giles-trends-2014,geist-survey-2017}, the memory handling necessary throughout the entire program due to the focus on threads only widens the gap between algorithms and implementation as noted by, for instance, Leiserson \etal~\cite{leiserson-theres-2020}. 
Therefore, to promote memory locality and more algorithmic code, a new data-focused paradigm is needed.

\begin{figure}[t]
	\centering
	\setlength{\tabcolsep}{1pt}
	\begin{tikzpicture}[>=stealth',semithick,auto]
		\draw
		(0,0.2) -- (0,-0.2)
		(12,0.2) -- (12,-0.2)
		(0,0) -- (12,0)
		;
		\node[circle,inner sep=1.3pt,fill=black] (tp) at (0.5,0) {};
		\node[circle,inner sep=1.3pt,fill=black] (dp) at (6,0)   {};
		\node[circle,inner sep=1.3pt,fill=black] (da) at (11.5,0) {};
		\def\ph{1.5}
		\node at ($(tp.center) + (0.4,\ph)$) {task-parallel};
		\node at ($(dp.center) + (0.0,\ph)$) {data-parallel};
		\node at ($(da.center) + (-0.7,\ph)$) {data-autonomous};
		\draw[<-] (tp.north) -- ++(0,1.2);
		\draw[<-] (dp.north) -- ++(0,1.2);
		\draw[<-] (da.north) -- ++(0,1.2);
		
		
		\def\chm{-0.5}
		\def\chp{0.5}
		\draw[<-] (11.5,0) -- ++(0,\chm) node[xshift=-0.4mm,yshift=-3mm] {\Lname~\cite{franken-autonomous-2023}\hspace{0.1cm}};
		\draw[<-] (9,0) -- ++(0,\chp) node[xshift=-5mm,yshift=6mm] {\begin{tabular}{rl}
				{\tiny POOL}&\cite{america-parallel-1990}\\
				{\tiny Gamma}&\cite{gannouni-gamma-calculus-2015}\\	
				{\tiny Act.Peb.}&\cite{willcock-active-2011}
		\end{tabular}};
		\draw[<-] (9,0) -- ++(0,\chm) node[xshift=-5mm,yshift=-6mm] {\begin{tabular}{rl}
				{\tiny AL-1}&\cite{marcoux-1-1988}\\
				{\tiny Ly}&\cite{ungar-harnessing-2010}\\
				{\tiny Parcel-1}&\cite{vialle-parcel-1-1996}
		\end{tabular}};
		\draw[<-] (7.4,0) -- ++(0,\chp) node[xshift=-5mm,yshift=3.5mm] {\begin{tabular}{rl}
				{\tiny Chestnut}&\cite{stromme-chestnut-2012}\\
				{\tiny RELACS}&\cite{raimbault-relacs-1993}
		\end{tabular}};
		\draw[<-] (7.4,0) -- ++(0,\chm) node[xshift=-3mm,yshift=-6mm] {\begin{tabular}{rl}
				{\tiny PPC}&\cite{maresca-programming-1993}\\
				{\tiny Halide}&\cite{ragan-kelley-halide-2017}\\
				{\tiny OP2}&\cite{mudalige-op2-2012}\\	
		\end{tabular}};
		\draw[<-] (6,0) -- ++(0,\chm) node[xshift=-5mm,yshift=-4mm] {\begin{tabular}{rl}
				{\tiny CUDA}&\cite{garland-parallel-2008}\\
				{\tiny OpenCL}&\cite{chong-sound-2014}
		\end{tabular}};
		\draw[<-] (5.25,0) -- ++(0,\chp) node[xshift=-5mm,yshift=3.5mm] {\begin{tabular}{rl}
				
				{\tiny Parcel-2}&\cite{cagnard-parcel-2-2000}
		\end{tabular}};
		\draw[<-] (4.5,0) -- ++(0,\chm) node[xshift=-3mm,yshift=-2.2mm] {\begin{tabular}{rl}
				{\tiny MPI}&\cite{clarke-mpi-1994}\\
		\end{tabular}};
		\draw[<-] (3.5,0) -- ++(0,\chp) node[xshift=-7mm,yshift=6mm] {\begin{tabular}{rl}
				{\tiny Legion}&\cite{bauer-legion-2012}\\
				{\tiny Graphgrind}&\cite{sun-graphgrind-2017}\\
				{\tiny DDG}&\cite{tran-parallel-2000}
		\end{tabular}};
		%
	\end{tikzpicture}
	\caption{
		Approximate placement of related work on an axis from process-focused (left) to data-focused (right) paradigms.
	}
	\label{fig:paradigms}
\end{figure}
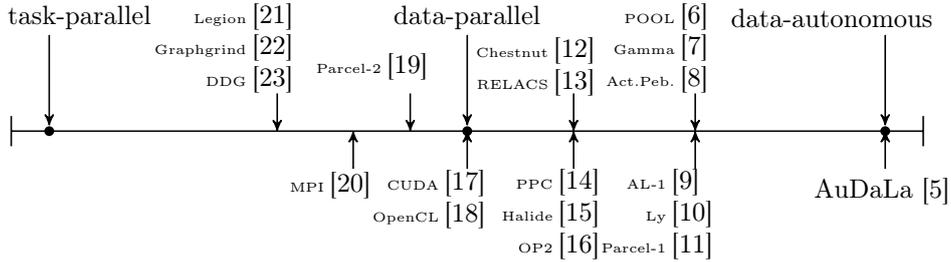

In the conference version of this paper~\cite{franken-autonomous-2023}, we proposed the new \emph{data-autonomous} paradigm, where \emph{data elements} not only locally store data and references, but also execute their own computations.
Computations are always carried out in parallel by all data elements; this is governed by a \emph{schedule}.
Data elements can cooperate through stored references.
The paradigm completely abstracts away from processors and memory and is fully focused on data, compared to task- and data-parallelism (see Figure~\ref{fig:paradigms}).

The paradigm provides several benefits.
First, it results in a \emph{separation of concerns}: code concerning data structures, algorithms and orchestration is properly separated.
Furthermore, parallelism is encouraged by always running computations concurrently on groups of data elements. 
Finally, the paradigm promotes a bottom-up design process, from data structure to computations to schedule.
Through these benefits, the paradigm and languages based on the paradigm facilitate a more intuitive understanding of how the data interacts in a parallel program, and therefore what that program actually does.
We foresee that the data-autonomous paradigm is particularly well suited for graph algorithms and other procedures that operate on unstructured data.

We also introduced \Lname~\cite{franken-autonomous-2023}, the \emph{Autonomous Data Language}, the first data-autonomous programming language, which explores what programming in the data-autonomous paradigm would be like.
As opposed common programming languages like C++/CUDA~\cite{garland-parallel-2008, harris-parallel-2007} or theoretical frameworks like the PRAM~\cite{fortune-parallelism-1978} or PPM~\cite{cook-parallel-1993}, the focus of \Lname is neither the performance of parallel programming, nor is (mainly) the expression of parallel algorithms. 
Instead, \Lname's focus is to be a simple yet implementable and usable parallel programming language which follows the data autonomous paradigm and promotes the creation of accessible and understandable parallel programs.
\Lname facilitates the understanding of the effects of a parallel program by its enforced structure, and its decoupling of loops from parallelism. 
By keeping \Lname compact and formally defining \Lname's semantics, \Lname is small and independent of specific hardware for both execution and formal verification.

In \Lname, \emph{structs}, \emph{steps} and a \emph{schedule} are responsible for data, computation and orchestration, respectively. 
We explained \Lname's design principles using an example, and gave further examples of \Lname programs implementing basic parallel methods.
\Lname's behaviour is completely formalised in an operational semantics, enabled by its compact syntax.
The semantics enabled us to show~\cite{franken-audala-2024} that \Lname is Turing complete.

The conference version of the paper focused mostly on the requirements of keeping \Lname simple, accessible and understandable. While the operational semantics allowed for an implementation, we provide an alternative semantics suitable for execution on systems with a weak memory model in~\cite{leemrijse-formalisation-2024}, thereby making \Lname more feasible for implementation on most modern processors.
The two semantics are equivalent under the absence of read-write race conditions.
A prototype compiler from \Lname to CUDA is presented in~\cite{leemrijse-2023}, along with other implemented algorithms.

This work extends~\cite{franken-autonomous-2023} by providing a type system, additional example programs and correctness proofs.
In particular, we make the following contributions:
\begin{itemize}
	\item We provide a formal type system for \Lname and prove it \emph{safe} (Corollary~\ref{cor: safety}).
	\item We extend the set of examples with three additional \Lname programs. The first is a program for sorting with a worst case time complexity of $O(\log{m})$, where $m$ is the length of the to be sorted list. This program is followed by two programs for the \emph{3SUM} problem, one of which uses a nested fixpoint.
	\item We provide a theorem which specifies the exact effect of the execution of commands in well-formed programs, as well as some other properties (Theorem~\ref{thm: wellformed}).
	This significantly simplifies the notation in later theorems and proofs.
	Notably, the type safety also follows from Theorem~\ref{thm: wellformed}.
	\item We prove that both our \Lname programs for \emph{prefix sum} and the new $O(\log{m})$ sorting algorithm are correct (Theorems~\ref{thm:correctness-prefix-sum} and~\ref{thm:correctness-sorting}, respectively).
	These proofs demonstrate that \Lname's separation of concerns also helps to separate the proofs into two independent parts.
\end{itemize}
Our basic lemmas and correctness proofs serve as inspiration for the techniques required for reasoning about \Lname programs.
They can thus form the basis for a future proof system.

\paragraph{Overview}
We first discuss the concepts of \Lname in Section~\ref{section:motivation}.
We then give the syntax of \Lname in Section~\ref{section:syntax}, the type system in Section~\ref{section:types} and a semantics in Section~\ref{section:semantics}.
We discuss more examples in Section~\ref{section:standardalgos}. We then prove some properties of \Lname in Section~\ref{section:props} before proving two example programs correct in Section~\ref{section:proving}.
Lastly, we review related work in Section~\ref{section:relatedwork} and conclude in Section~\ref{section:conclusion}.

All sections benefit from having read Section~\ref{section:motivation}. Sections~\ref{section:types}, \ref{section:semantics} and \ref{section:standardalgos} require knowledge of Section~\ref{section:syntax}, and Sections~\ref{section:props} and \ref{section:proving} require knowledge of Sections~\ref{section:syntax}, \ref{section:types} and \ref{section:semantics}. Section~\ref{section:proving} also requires familiarity with the code for prefix sum presented in Section~\ref{section:motivation} and the code for the $O(\log{m})$ sorting algorithm presented in Section~\ref{section:sortex2}. Sections~\ref{section:relatedwork} and \ref{section:conclusion} benefit from any section, but only require the introduction to have been read.

\section{Concepts And Motivating Example}\label{section:motivation}
\begin{figure}[t]
	\centering
	\includegraphics[width=0.6\textwidth]{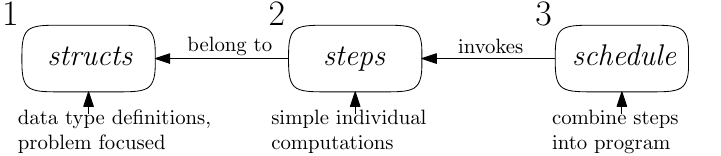}
	\caption{The three main components of an \Lname program.}
	\label{fig:design}
\end{figure}
In this section, we first discuss the concepts of \Lname, and subsequently we design a program for the prefix sum problem in \Lname as a motivating example.

\Lname has three main components: \emph{structs}, \emph{steps} and a \emph{schedule}.
The relation between these components is shown in Figure~\ref{fig:design}.
Structs are data type definitions from which data elements are instantiated during runtime. 
They contain the name of the data type and the parameters available to data elements of that type.
See Listing~\ref{ex: PS1} for an example of a struct definition.

When starting an \Lname program, a special $\nil$-instance is created for every struct, which is a data element representing non-existing instances of that struct. 
It can be considered a constant \emph{default} instance of a data type, which is inherently stable in the values of its parameters. 
This is distinct from the use of null-pointers in other common languages, which do not point to an existing address.
Among other things null-instances can be used for initialisation, since it already exists when launching the program. 
Note that this means that reading a null-instance is a valid operation in \Lname and does not warrant an interruption to halt the program, like reading null-pointers does for other programming languages. 

Each struct contains zero or more \emph{steps}, which represent operations a data element instantiated from that struct can do.
A step contains simple, algorithmic code, consisting of conditions and assignments, without loops. 
This makes steps finite and easy to reason about.
Within a step, it is possible to access the parameters of the surrounding struct and also to follow references stored in those parameters.
Since these access patterns are known at compile-time, we can increase memory locality by grouping struct instances in a suitable manner.

The \emph{schedule} prescribes an execution order on the steps.
It contains step references and fixpoint operators ($\mathit{Fix}$).
The occurrences of step references and fixpoint operators are separated by synchronization barriers (`$<$'). 
Execution only proceeds past a barrier when all computations that precede the barrier have concluded.
Whenever a step occurs in the schedule, it is executed in parallel by all data elements which contain that step, although it is also possible to invoke a step for data elements of a specific type.
\Lname programs are thus inherently parallel.

We do not make assumptions about a global execution order of statements executed in parallel.
In particular, code is not necessarily executed by multiple struct instances in \emph{lock-step}.
Furthermore, we allow the occurrence of race conditions within one step, see also  Section~\ref{section:standardalgos}.
Thus, barriers (and implicit barriers, see below) are the main method of synchronisation.

Iterative behaviour is achieved through a fixpoint operator, which executes its body repeatedly until an iteration occurs in which no data is changed.
At this point, a fixpoint is reached and the schedule continues past the fixpoint operator.
Between the iterations of a fixpoint, there is an implicit synchronisation barrier.
For an example schedule, see Listing~\ref{ex: PS3}.

\begin{figure}[t]
	\centering
	\includegraphics[width=0.7\textwidth]{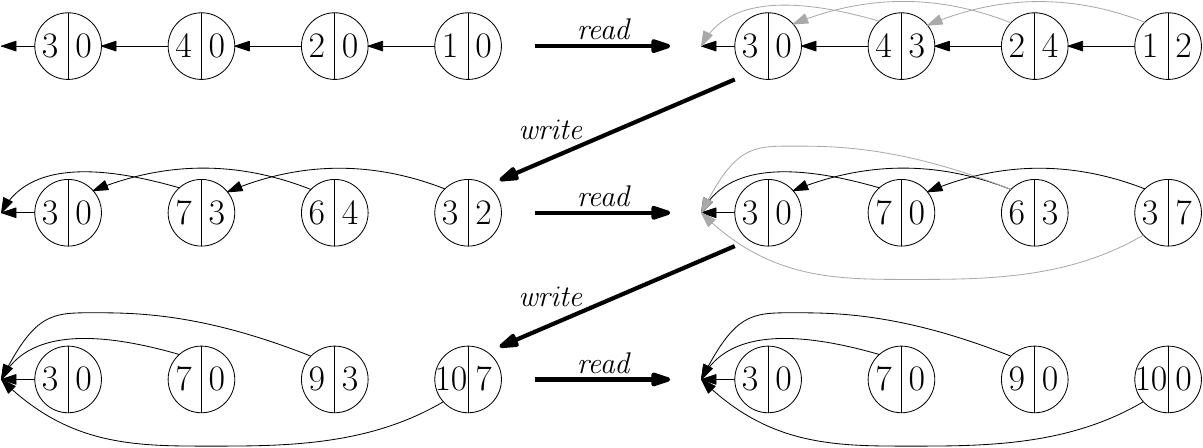}
	\caption{Execution of Prefix Sum on a small list. The left side of a list element holds the parameter \textit{val}, while the right side holds the parameter \textit{auxval}. The parameter \textit{prev} is shown as unmarked black arrows, while the parameter \textit{auxprev} is shown as unmarked grey arrows.}
	\label{gr: PrefixSum}
\end{figure}

To give an example of these components in action, we consider the \emph{prefix sum} problem: given a sequence of integers $x_1,\dots,x_n$, we compute for each position $1 \leq k \leq n$ the sum $\Sigma_{i = 1}^k x_i$.
We have included OpenCL and CUDA implementations of the problem that previously occurred in the literature~\cite{chong-sound-2014,harris-parallel-2007}, see Listings~\ref{ex: KSOCL} and~\ref{ex: PSCUDA}.
Here, we omit the initialization to focus on the kernels. 
In Listing~\ref{ex: KSOCL}, the kernel first copies the input array to the output array, after which it uses an offset variable to check whether an element of the array needs to keep updating itself with other elements. It uses an auxiliary element ``temp'' to facilitate this update.
In Listing~\ref{ex: PSCUDA}, the kernel uses a single array which is twice as big as the original array to alternate the updates between the front half and the back half of the array. The method is similar to the method of the the OpenCL example, but it does need to synchronise less due to the alternation in the array.
Both kernels require synchronization barriers in their code, as well as an offset variable to check which data needs to be operated on, against which the thread ids need to be checked multiple times per execution. 
This makes the intended use of the program opaque to those not initially familiar with CUDA or OpenCL.
\begin{lstlisting}[float = t,caption={OpenCL kernel for Prefix Sum (from~\cite{chong-sound-2014})}, style=OpenCL, label={ex: KSOCL}]
	kernel void koggeStone(const local T *in, local T *out) {
		out[tid] = in[tid];
		barrier();
		for (unsigned offset = 1; offset < (*$n$*); offset *= 2){
			T temp;
			if (tid (*$\geq$*) offset) temp = out[tid - offset];
			barrier();
			if (tid (*$\geq$*) offset) out[tid] = temp (*$\oplus$*) out[tid];
			barrier();	
	}}
\end{lstlisting}
\begin{lstlisting}[float=t,caption={CUDA kernel for Prefix Sum (or Scan) (from~\cite{harris-parallel-2007})}, style=CUDA, label={ex: PSCUDA}]
	__global__ void scan(float *g_odata, float *g_idata, int n){
		extern __shared__ float temp[];
		int thid = threadIdx.x;
		int pout = 0, pin = 1;
		temp[pout*n + thid] = (thid > 0) ? g_idata[thid-1] : 0;
		__syncthreads();
		for (int offset = 1; offset < n; offset *= 2){
			pout = 1 - pout; // swap double buffer indices
			pin = 1 - pout;
			if (thid >= offset)
				temp[pout*n+thid] += temp[pin*n+thid - offset];
			else
				temp[pout*n+thid] = temp[pin*n+thid]
			__syncthreads();
		}
		g_odata[thid] = temp[pout*n+thid];
	}
\end{lstlisting}
\FloatBarrier
To design a corresponding \Lname program, we follow the design structure suggested in Figure~\ref{fig:design}.
As before, we omit the initialization.
In the prefix sum problem, the input is a sequence of integers.
We model an element of this sequence with a struct \textit{Position} containing a value $\mathit{val}$.
We also give every \textit{Position} a reference to the preceding \textit{Position}, contained in parameter \textit{prev}, as seen in Listing~\ref{ex: PS1}.
This is needed to compute the prefix sum.
The value of $\mathit{prev}$ for the first position in the list is set to $\nil$, referencing the null-\textit{Position}. 
This null-instance has the values $0$ for $\mathit{val}$ and $\nil$ for $\mathit{prev}$. 
\begin{lstlisting}[float = h, caption={Partial \Lname code for the structs for Prefix Sum}, label={ex: PS1}]
	struct (*\textit{Position}*)((*\textit{val}*): Int, (*\textit{prev}*): (*\textit{Position}*)){ (*\textcolor{gray}{[...]}*) }
\end{lstlisting}
In Listing~\ref{ex: PS3} the steps \textit{read} and \textit{write} of the \textit{Position} struct are shown.
These steps are based on the method for computing prefix sum in parallel, as shown in Figure~\ref{gr: PrefixSum}, which was introduced by Hillis and Steele~\cite{hillis-data-1986}.
Every \textit{Position} first reads \textit{prev.prev} and \textit{prev.val} from their predecessor in the step \textit{read}, and after synchronisation, every position updates their \textit{prev} to \textit{prev.prev} and their \textit{val} to \textit{val $+$ prev.val} in the step \textit{write}.
As the scope of local variables in \Lname does not exceed a step, the use of additional parameters \textit{auxprev} and \textit{auxval} in the \textit{read} step is required to recover the value in the \textit{write} step. 
The steps do not need an \textit{offset} variable like the CUDA and OpenCL kernels, as \textit{Positions} which reached the beginning of the list have a null-instance as predecessor and can still execute the steps.

For our program schedule, we want to repeat \textit{read} and then \textit{write} until all \textit{Positions} have reached the beginning of the list, which results in the schedule as shown in Listing~\ref{ex: PS3}. 
Eventually, all \textit{Positions} will have $\nil$ as their predecessor and no parameters will change further (as $\nil.\mathit{val} = 0$ and $\nil.\mathit{prev} = \nil$), causing the fixpoint to terminate.

\begin{lstlisting}[float=h,caption={\Lname code for Prefix Sum with steps and a schedule}, label={ex: PS3}]
	struct (*\textit{Position}*)((*\textit{val}*): Int, (*\textit{prev}*): (*\textit{Position}*), (*\textit{auxval}*): Int,  (*\textit{auxprev}*): (*\textit{Position}*)){
		(*\textit{read}*) { (*\hfill*)/*step definition*/
			(*\textit{auxval}*) := (*\textit{prev.val}*);
			(*\textit{auxprev}*) := (*\textit{prev.prev}*);
		}
		(*\textit{write}*) { (*\hfill*)/*step definition*/
			(*\textit{val}*) := (*\textit{val}*) + (*\textit{auxval}*);
			(*\textit{prev}*) := (*\textit{auxprev}*);
	}}
	
	Fix((*\textit{read}*) < (*\textit{write}*)) (*\hfill*)/*schedule*/
\end{lstlisting}

As illustrated by Listing~\ref{ex: PS3} and by Figure~\ref{fig:design}, \Lname has a high \emph{separation of concerns}: structs model data and their attributes, steps contain the algorithmic code and the schedule contains the execution.
This approach requires no synchronization barriers in the user code for the steps, no variables to find the right indices for memory access and no offset variables to avoid going out of bounds. While it may still be opaque to people not familiar with \Lname, we are confident that the threshold for attaining this familiarity is generally lower than hardware-focused languages like CUDA and OpenCL.
	\section{Syntax}\label{section:syntax}
In this section, we highlight the most important parts of the concrete syntax of \Lname.
In the definitions below, non-terminals are indicated with $\langle \text{--} \rangle$ and symbols with quotes; the empty word is $\vempty$.
The non-terminal \emph{Id} describes identifiers, and the non-terminal \emph{Type} describes type names, which are either \texttt{Int}, \texttt{Nat} (natural number), \texttt{Bool}, \texttt{String} or an identifier (the name of a struct).

An \Lname \emph{Program} consists of a list of definitions of \emph{structs} and a schedule:
\begin{equation*}
	\begin{aligned}
		&\type{Program} ::= \type{Defs}\ \type{Sched}\\
		&\type{Defs} ::= \type{Struct}\mid\type{Struct}\ \type{Defs}
	\end{aligned}
\end{equation*}
A \emph{struct} definition gives the struct a type name (\textit{Id}), a list of parameters (\textit{Pars}) and a number of steps (\textit{Steps}):
\begin{equation*}
	\begin{aligned}
		&\type{Struct} ::= \word{struct}\ \type{Id}\  \word{(}\ \type{Pars}\ \word{)}\ \word{\{}\ \type{Steps}\ \word{\}},\\
		&\type{Pars} ::= \type{Par} \type{ParList} \mid \vempty\\
		&\type{ParList} ::= \word{,}\ \type{Par} \type{ParList} \mid \vempty\\
		&\type{Par} ::= \type{Id}\ \word{:}\ \type{Type}
	\end{aligned}
\end{equation*}
\emph{Steps} are defined with a step name (\textit{Id}) and a list of statements:
\begin{equation*}
	\begin{aligned}
		&\type{Steps} ::= \type{Id}\ \word{\{}\ \type{Stats}\ \word{\}}\ \type{Steps} \mid \vempty\\
		&\type{Stats} ::= \type{Stat}\ \type{Stats}\mid \vempty
	\end{aligned}
\end{equation*}
A \emph{statement} adheres to the following syntax:
\begin{center}
	\begin{tabular}{cl@{\hspace{1cm}}l}
		$\type{Stat} ::=$ &$\word{if}\ \type{Exp}\ \word{then}\ \word{\{}\ \type{Stats}\ \word{\}}$ & if-then statement\\
		&$\mid \type{Type}\ \type{Id}\ \word{:=}\ \type{Exp}\ \word{;}$ & variable assignment\\
		&$\mid \type{Var}\ \word{:=}\ \type{Exp}\ \word{;}$ & variable update\\
		&$\mid \type{Id}\ \word{(}\ \type{Exps}\ \word{)}\ \word{;}$ & constructor statement 
	\end{tabular}
\end{center}
The \textit{Id} in the variable assignment is a variable name.
The constructor statement spawns a new data element of the type determined by \textit{Id}, with parameter values determined by the expressions \textit{Exps}.
The syntax of \textit{Exps} is similar to that of \textit{Pars}, using \textit{ExpList} and \textit{Exp}. 
The syntax for a single \emph{expression} \textit{Exp} is as follows: 
\begin{center}
	\begin{tabular}{cl@{\hspace{1cm}}l}
		$\type{Exp} ::=$ &$\type{Exp}\ \type{BOp}\ \type{Exp}$ & binary operator expression\\
		&$\mid \word{(}\ \type{Exp}\ \word{)}$ & brackets\\
		&$\mid \word{!}\ \type{Exp}$ & negation\\
		&$\mid \type{Id}\ \word{(}\ \type{Exps}\ \word{)}$ & constructor expression\\
		&$\mid \type{Var}$& variable expression\\
		&$\mid \type{Literal}$& literal expression\\
		&$\mid \word{null}$& null expression\\
		&$\mid \word{this}$& this expression
	\end{tabular}
\end{center}
We consider \texttt{this} to refer to the data element from which the code which includes \texttt{this} is executed.

\emph{Binary Operators} follow the syntax $$\type{BOp} ::= \word{=} \mid \word{!=}\mid \word{<=}\mid\word{>=}\mid\word{<}\mid\word{>}\mid\word{*}\mid\word{/}\mid\word{\%}\mid\word{+}\mid\word{-}\mid\word{\^}\mid\word{\&\&}\mid\word{||},$$ which represent the binary operations for which the notation is commonly used.

A \emph{variable} reference follows the syntax:
\begin{equation*}
	\type{Var} ::= \type{Id}\ \word{.} \type{Var} \mid \type{Id},
\end{equation*}
where in both cases \textit{Id} is the name of a variable.
Through the first case, one can access the parameters of parameters.
For example, \textit{prev}.\textit{prev}.\textit{val} would have been valid \Lname in Listing~\ref{ex: PS3}, and would access the value of the \textit{Position} before the previous \textit{Position} of the current \textit{Position}.

Lastly, the schedule consists of the variants as given in the following syntax: 
\begin{center}
	\begin{tabular}{cl@{\hspace{1cm}}l}
		$\type{Sched} ::=$ &$\type{Id}$ & step execution\\
		&$\mid \type{Id}\ \word{.}\ \type{Id}$& typed step execution\\
		&$\mid \type{Sched}\ \word{<}\ \type{Sched}$ & barrier composition\\
		&$\mid \word{Fix}\ \word{(}\ \type{Sched}\ \word{)}$ & fixpoint calculation
	\end{tabular}
\end{center}
The \textit{Id} in the step execution is a step name. In the typed step execution, the first \textit{Id} is a type name, while the second is a step name.
The typed step execution is used to schedule a step executed by only one specific struct type.


	\section{Well-Formedness of \Lname programs}\label{section:types}
Like many other programming languages, we put type constraints and other syntactical constraints on the syntax of an acceptable \Lname program. To formalise what is and what is not considered well-formed for an \Lname program, we give a \emph{type system}~\cite{pierce2002types, cardelli-type-1996, cardelli-type-2004, mitchell-chapter-1990} for \Lname in this section.
With the type system of \Lname, we root out standard \emph{type errors}, where there is a mismatch between the expected type of an expression or a variable and the actual type presented, as well as any breach of the following requirements:
\begin{enumerate}[noitemsep]
	\item Identifiers may not be keywords.
	\item A step name is declared at most once within each struct definition.
	\item A parameter name is used at most once within each struct definition.
	\item Names of local variables do not overlap with parameter names of the surrounding struct definition.
	\item Variable assignment statements (as defined in the syntax) are only used to declare new local variables.
	\item A local variable is only used after its declaration in a variable assignment statement.
\end{enumerate}

\begin{table}[p]
	\centering
	\fbox{
		\resizebox{\linewidth}{!}{
{\renewcommand{\arraystretch}{1} 
	\begin{tabular}{c}
		\fbox{
			Environment
		} \\
		\begin{minipage}{90px}
			\begin{equation*}
				\inference[(\textbf{Empty})]{
				}{
					\emptyset, \emptyset\tr \diamond
				}
			\end{equation*}
		\end{minipage}\hfill
		\begin{minipage}{225px}
			\begin{equation*}
				\inference[(\textbf{AddType})]{
					\Gamma, \Omega\tr \diamond & \StructType\notin\mathit{dom}(\Omega) & \StructType\notin K \\ \forall p_i, p_j\in (p_1, \ldots, p_n).(p_i\neq p_j)
				}{
					\Gamma, \Omega;\StructType(p_1:T_1, \ldots, p_n:T_n) \tr \diamond
				}
			\end{equation*}
		\end{minipage} \hfill
		\begin{minipage}{200px}
			\begin{equation*}
				\inference[(\textbf{AddVar})]{
					\Gamma, \Omega\tr T & v\notin \mathit{dom}(\Gamma) & v\notin K
				}{
					\Gamma;v:T, \Omega \tr \diamond
				}
			\end{equation*}
		\end{minipage}
		\\ \\\hdashline
		\fbox{
			Types
		} \\
		\begin{minipage}{275px}
			\begin{equation*}
				\inference[(\textbf{TBasic})]{
					\Gamma, \Omega\tr\diamond & T\in\{\texttt{Nat}, \texttt{Int}, \texttt{Bool}, \texttt{String}, \texttt{Step}\}
				}{
					\Gamma, \Omega \tr T
				}
			\end{equation*}
		\end{minipage}\\ \\
		\begin{minipage}{125px}
			\begin{equation*}
			\inference[(\textbf{TStruct})]{
				\Gamma, \Omega;\StructType(\mathit{Pars});\Omega'\tr\diamond
			}{
				\Gamma, \Omega;\StructType(\mathit{Pars});\Omega' \tr \StructType
			}
			\end{equation*}
		\end{minipage} \hspace{100px}
		\begin{minipage}{200px}
			\begin{equation*}
				\inference[(\textbf{DefStruct})]{
					\Gamma, \Omega\tr\StructType &
					\forall_{1\leq i\leq n} T_i.\Gamma, \Omega \tr T_i
				}{
					\Gamma, \Omega \tr \StructType(p_1: T_1, \ldots, p_n: T_n)
				}
			\end{equation*}
		\end{minipage}
		\\ \\\hdashline
		\fbox{
			Standard Values
		} \\
		\begin{minipage}{200px}
			\begin{equation*}
				\inference[(\textbf{Bool})]{
					\Gamma, \Omega\tr\diamond & x\in\{\true, \false\}
				}{
					\Gamma, \Omega \tr x: \texttt{Bool}
				}
			\end{equation*}
		\end{minipage}\hfill
		\begin{minipage}{175px}
			\begin{equation*}
				\inference[(\textbf{String})]{
					\Gamma, \Omega \tr\diamond
				}{
					\Gamma, \Omega \tr ``\mathit{Str}": \texttt{String}
				}
			\end{equation*}
		\end{minipage} \hfill
		\begin{minipage}{150px}
			\begin{equation*}
				\inference[(\textbf{NInt})]{
					\Gamma, \Omega \tr \diamond & c < 0
				}{
					\Gamma, \Omega \tr c:\texttt{Int}
				}
			\end{equation*}
		\end{minipage}
		\\\\
		\begin{minipage}{175px}
			\begin{equation*}
				\inference[(\textbf{PNat})]{
					\Gamma, \Omega \tr \diamond & c \geq 0
				}{
					\Gamma, \Omega\tr c:\texttt{Nat}
				}
			\end{equation*}
		\end{minipage}\hfill
		\begin{minipage}{150px}
			\begin{equation*}
				\inference[(\textbf{Null})]{
					\Gamma, \Omega \tr T
				}{
					\Gamma, \Omega \tr \texttt{null}:T
				}
			\end{equation*}
		\end{minipage} \hfill
		\begin{minipage}{150px}
			\begin{equation*}
				\inference[(\textbf{This})]{
					\Gamma, \Omega \tr \StructType 
				}{
					\Gamma, \Omega \tr \texttt{this}_\StructType: \StructType
				}
			\end{equation*}
		\end{minipage}\\ \\ \hdashline
		\fbox{
			Expressions
		} \\
		\begin{minipage}{170px}
			\begin{equation*}
				\inference[(\textbf{Eq})]{
					\Gamma, \Omega \tr E_1:T \land \Gamma, \Omega \tr E_2:T\\
					\circ\in\{=, \text{!=}\}
				}{
					\Gamma, \Omega\tr E_1=E_2:\texttt{Bool}
				}
			\end{equation*}
		\end{minipage}\hfill
		\begin{minipage}{190px}
			\begin{equation*}
				\inference[(\textbf{Comp})]{
					\Gamma, \Omega \tr E_1:N_1& \Gamma, \Omega \tr E_2:N_2\\
					\circ\in\{< =, >=, <, >, =, \text{!=}\}
				}{
					\Gamma, \Omega\tr E_1\circ E_2:\texttt{Bool}
				}
			\end{equation*}
		\end{minipage} \hfill
		\begin{minipage}{190px}
			\begin{equation*}
				\inference[(\textbf{NArith})]{
					\Gamma, \Omega \tr E_1:\texttt{Nat}& \Gamma, \Omega \tr E_2:\texttt{Nat}\\
					\circ\in\{*, /, \%, +, \text{\textasciicircum}\}
				}{
					\Gamma, \Omega\tr E_1\circ E_2:\texttt{Nat}
				}
			\end{equation*}
		\end{minipage}
		\\ \\
		\begin{minipage}{200px}
			\begin{equation*}
				\inference[(\textbf{IArith})]{
					\Gamma, \Omega \tr E_1:N_1& \Gamma, \Omega \tr E_1:N_2\\
					\circ\in\{*, /, \%, +, \text{\textasciicircum}, -\}
				}{
					\Gamma, \Omega\tr E_1\circ E_2:\texttt{Int}
				}
			\end{equation*}
		\end{minipage}\hfill
		\begin{minipage}{200px}
			\begin{equation*}
				\inference[(\textbf{BinLog})]{
					\Gamma, \Omega \tr E_1:\texttt{Bool}& \Gamma, \Omega \tr E_2:\texttt{Bool}\\
					\circ\in\{\&\&, ||\}
				}{
					\Gamma, \Omega\tr E_1\circ E_2:\texttt{Bool}
				}
			\end{equation*}
		\end{minipage} \hfill
		\begin{minipage}{150px}
			\begin{equation*}
				\inference[(\textbf{Brackets})]{
					\Gamma, \Omega \tr E:T
				}{
					\Gamma, \Omega\tr (E):T
				}
			\end{equation*}
		\end{minipage}
		\\ \\
		\begin{minipage}{150px}
			\begin{equation*}
				\inference[(\textbf{Neg})]{
					\Gamma, \Omega \tr E:\texttt{Bool}
				}{
					\Gamma, \Omega\tr !E:\texttt{Bool}
				}
			\end{equation*}
		\end{minipage}\hfill
		\begin{minipage}{250px}
			\begin{equation*}
					\inference[(\textbf{ConsE})]{
					\forall_{0\leq i\leq n}. \Gamma, \Omega \tr E_i: T_i
				}{
					\Gamma, \Omega;\StructType(p_1:T_1, \ldots p_n:T_n);\Omega' \tr \StructType(E_1, \ldots, E_n): \StructType
				}
			\end{equation*}
		\end{minipage} \hfill
		\begin{minipage}{200px}
			\begin{equation*}
				\inference[(\textbf{VarS})]{
					\Gamma_1;(x:T);\Gamma_2, \Omega\tr \diamond
				}{
					\Gamma_1;(x:T);\Gamma_2, \Omega\tr x: T
				}
			\end{equation*}
		\end{minipage}
		\\ \\
		\begin{minipage}{250px}
			\begin{equation*}
				\inference[(\textbf{VarR})]{
					\Gamma, \Omega;\StructType(p_1:T_1, \ldots p_n:T_n);\Omega' \tr X: \StructType\\
					\Gamma, \Omega;\StructType(p_1:T_1, \ldots p_n:T_n);\Omega' \tr T_i
				}{
					\Gamma, \Omega;\StructType(p_1:T_1, \ldots p_n:T_n);\Omega' \tr X.p_i: T_i
				}
			\end{equation*}
		\end{minipage}
		\\ \\ \hdashline
		\fbox{
			Statements
		} \\
		\begin{minipage}{175px}
			\begin{equation*}
				\inference[(\textbf{IfThen})]{
					\Gamma, \Omega \tr E: \texttt{Bool}& \Gamma, \Omega \tr \mathcal{S}
				}{
					\Gamma, \Omega \tr\text{if}\ E\ \text{then}\{\mathcal{S}\}
				}
			\end{equation*}
		\end{minipage}\hfill
		\begin{minipage}{200px}
			\begin{equation*}
				\inference[(\textbf{LVar})]{
					\Gamma, \Omega \tr E: T & \Gamma_1;(x:T);\Gamma_2, \Omega\tr\diamond 
				}{
					\Gamma, \Omega \tr T x := E
				}
			\end{equation*}
		\end{minipage} \hfill
		\begin{minipage}{175px}
			\begin{equation*}
				\inference[(\textbf{Update})]{
					\Gamma, \Omega \tr X: T & \Gamma, \Omega \tr E: T
				}{
					\Gamma, \Omega \tr X := E
				}
			\end{equation*}
		\end{minipage}
		\\\\
		\begin{minipage}{240px}
			\begin{equation*}
				\inference[(\textbf{ConsS})]{
					\forall_{0\leq i\leq n}. \Gamma, \Omega \tr E_i: T_i
				}{
					\Gamma, \Omega;\StructType(p_1:T_1, \ldots p_n:T_n);\Omega' \tr \StructType(E_1, \ldots, E_n)
				}
			\end{equation*}
		\end{minipage}\hfill
		\begin{minipage}{140px}
			\begin{equation*}
				\inference[(\textbf{Seq})]{
					\Gamma, \Omega\tr S& \Gamma,\Omega\tr\mathcal{S} 
				}{
					\Gamma, \Omega\tr S;\mathcal{S}
				}
			\end{equation*}
		\end{minipage} \hfill
		\begin{minipage}{210px}
			\begin{equation*}
				\inference[(\textbf{SeqLVar})]{
					\Gamma, \Omega\tr T x := E & \Gamma;(x: T), \Omega\tr\mathcal{S} 
				}{
					\Gamma, \Omega\tr T x := E;\mathcal{S}
				}
			\end{equation*}
		\end{minipage}\\ \\ 
		\begin{minipage}{210px}
			\begin{equation*}
				\inference[(\textbf{SeqIT})]{
					\Gamma, \Omega\tr \text{if}\ E\ \text{then}\{\mathcal{S}_1\}& \Gamma,\Omega\tr\mathcal{S}_2 
				}{
					\Gamma, \Omega\tr \text{if}\ E\ \text{then}\{\mathcal{S}_1\}\ \mathcal{S}_2
				}
			\end{equation*}
		\end{minipage}\\ \\ \hdashline
		\fbox{
			Schedule
		} \\
		\begin{minipage}{150px}
			\begin{equation*}
				\inference[(\textbf{SBar})]{
					\Gamma, \Omega\tr \mathit{sc}_1 & \Gamma, \Omega\tr\mathit{sc}_2 
				}{
					\Gamma, \Omega\tr \mathit{sc}_1<\mathit{sc}_2
				}
			\end{equation*}
		\end{minipage}\hfill
		\begin{minipage}{115px}
			\begin{equation*}
				\inference[(\textbf{SFix})]{
					\Gamma, \Omega\tr \mathit{sc}
				}{
					\Gamma, \Omega\tr \texttt{Fix}(\mathit{sc})
				}
			\end{equation*}
		\end{minipage} \hfill
		\begin{minipage}{140px}
			\begin{equation*}
				\inference[(\textbf{STypCall})]{
					\Gamma, \Omega\tr F_\StructType: \texttt{Step}
				}{
					\Gamma, \Omega\tr \StructType.F
				}
			\end{equation*}
		\end{minipage}\hfill
		\begin{minipage}{130px}
			\begin{equation*}
				\inference[(\textbf{SNormCall})]{
					\Gamma, \Omega\tr F_\StructType: \texttt{Step}
				}{
					\Gamma, \Omega\tr F
				}
			\end{equation*}
		\end{minipage}\\ \\ 
		\hdashline
		\fbox{
			Program
		} \\
		\begin{minipage}{400px}
			\begin{equation*}
				\inference[(\textbf{Prog})]{
					\forall \StructType(\mathit{Pars})\in\mathit{Structs}(D).\Gamma, \Omega;\mathit{Structs}(D)\tr \StructType(\mathit{Pars})\\
					\forall (F_\StructType,\mathcal{S})\in\mathit{StepDef}(D). \Gamma;\mathit{Pars}(D, \StructType), \Omega;\mathit{Structs}(D) \tr \mathcal{S} &
					\Gamma;\mathit{Steps}(D), \Omega \tr \Sched
				}{
					\Gamma, \Omega\ \tr\ D\ \Sched
				}
			\end{equation*}
		\end{minipage}
	\end{tabular}
}
		}
	}
	\caption{The type system of \Lname, ordered by category.}
	\label{fig: typeSystem}
\end{table}

In Table~\ref{fig: typeSystem}, we give the type rules of the type system. These rules use a \emph{syntax variable environment} $\Gamma$ and a \emph{struct type environment} $\Omega$, both of which are ordered lists separated by $;$. The environment $\Gamma$ consists of pairs of variable names and their types; the variable names present in $\Gamma$ can be accessed using $\mathit{dom}(\Gamma)$. The environment $\Omega$ saves struct type definitions and their parameters, of which the struct types can be accessed using $\mathit{dom}(\Omega)$.

We denote struct types using variations of the letter $\StructType$. We assume that for every occurrence of \texttt{this}, the struct in which it occurs is recorded in such a way that every occurrence of \texttt{this} is annotated with the name of its struct, like $\texttt{this}_\StructType$. We also assume that any list, like a list of parameters, is separated by semicolons even if the syntax specifies commas. We use the shorthand $\mathit{Pars}$ to denote a list of parameters $p_1: T_1; \ldots$, as per the non-terminal $\type{Pars}$. To make working with the schedule easier, we treat occurrences of $s_\StructType$ for some step $s$ as a variable name for a variable of a special newly introduced type \texttt{Step}.

We use $\mathit{Str}$ to denote any sequence of characters admissible by the syntax and let $N_1$ and $N_2$ s.t. $N_1, N_2\in\{\texttt{Nat}, \texttt{Int}\}$. Additionally, let $S$ be a statement and let $\mathcal{S}$ be a sequence of statements separated with $;$. Let $E, E_1, E_2$ be expressions, let $x$ be a single variable name and let $X$ be a variable following the non-terminal $\type{Var}$. Lastly, let $\Sched$, $\Sched_1$ and $\Sched_2$ be schedules and let $K = \{\texttt{this}, \texttt{if}, \texttt{then}, \texttt{null}, \texttt{Fix}, \texttt{struct}\}$ be the set of keywords. Let $\vempty_\Sched$ be the empty schedule.

There are 7 categories of type rules. The first category, \textit{Environment}, deals with the well-formedness of the environment and the types and struct types contained in it. The category \textit{Types} deals with checking whether types are well-formed and are admitted by the environment. Note that the rule \textbf{DefStruct} uses the `for all' notation, but that the predicate can be decomposed into a predicate of the form $\Gamma, \Omega\tr T_1\land \ldots \land \Gamma, \Omega \tr T_n$. Alternatively, the rule can be decomposed in a set of rules which peel of the parameter list one by one.

The third category, \textit{Expressions}, deals with checking whether expressions are well-typed, while the fourth category, \textit{Statements}, deals with checking whether statements are admitted according to the type system. Note that expressions have types, while statements do not, which is exemplified by the two distinct rules for constructor expressions and constructor statements. Also note that variable assignment statements have their own sequence rule which adds the newly assigned variable to the environment, and that the if-then statements have their own sequence rule because they do not end with an `;' in the syntax. The sixth category, \textit{Schedule}, deals with the schedule by checking whether all step names in the schedule have been assigned the type \texttt{Step}. We consider $\vempty$ to be the empty schedule, which we allow.

The last category, \textit{Program}, consists of only one rule, \textbf{Prog}. This is the rule which decomposes checking a program into multiple subtasks to check. In this rule, we define that a program consisting of definitions and a schedule is well-formed iff all struct types are defined well (top line), all steps are well defined (second line left) and the schedule is well defined (second line right). We denote the topmost syntactic $\mathit{Defs}$ element as $D$ and the schedule as $\Sched$. We use the function $\mathit{Structs}(D)$ to extract struct type definition information from $D$, which consists of all struct types and their parameters formatted as $\StructType(\mathit{Pars})$. We overload $\mathit{Pars}$ with a function s.t. $\mathit{Pars}(D, \StructType)$ returns the parameters of $\StructType$ in $D$. We use the function $\mathit{Steps}(D)$ to extract step definition information from $D$, consisting of the step and its struct type. We consider these steps to have the type $\texttt{Step}$, so the output is formatted as a list consisting of elements with format $F_\StructType: \texttt{Step}$, where $F$ is a step. We use the function $\mathit{StepDef}(D)$ to extract step definition information coupled with their statements, which outputs a list of elements of the format $(F_\StructType, \mathcal{S})$.

\begin{definition}[\Lname well-formedness]
	An \Lname program $\Program$ is well-formed iff $\Program$ is derivable using the rules in Table~\ref{fig: typeSystem}.
\end{definition}

	\section{Semantics}\label{section:semantics}
In this section, we present the semantics of \Lname. These semantics take the form of a stack machine, as stacks are naturally suited for expressing the evaluation of expressions, and the stack machine also exposes the interleaving in the semantics very well, which we think suitable for a concurrent programming language.

To define the semantics, we start by defining some sets used in the semantics and which syntactical sets they mirror, if any. We then define \emph{commands}, used as primitives for the behaviour encoded in the syntax, and a function from the syntax to commands. We define the states of an \Lname program, and then use the commands and the schedule to define operational behaviour of an \Lname program. We conclude with Definition~\ref{def: os}.

In the semantics, we regularly use lists. 
List concatenation is denoted with a semicolon, and we identify a singleton list with its only element.
The empty list is denoted~$\varepsilon$.
If we work with lists of statements, we assume that in this list, the statements are also separated by a semicolon, even though if-statements do not have a semicolon at the end in the syntax.
While the schedules are represented using their syntax, in the semantics, we do consider schedules to be lists separated by $<$. It follows that schedules can be empty ($\vempty$) and that $\mathit{sc} = \mathit{sc} < \vempty$. However, note that in the syntax, an empty schedule is not admitted in the type system, so every program starts with a non-empty schedule.

We define updates for functions as follows.
Given a function $f : A \to B$ and $a\in A$ and $b\in B$, then $f[a \mapsto b](a) = b$ and $f[a \mapsto b](x) = f(x)$ for all $x \neq a$. 
We lift this operation to sets of updates: $f[\{a_1 \mapsto b_1, a_2 \mapsto b_2, \dots \}] = f[a_1 \mapsto b_1][a_2 \mapsto b_2]\dots$.
Since the order of applying updates is relevant, this is only well-defined if the left-hand sides $a_1, a_2,\dots$ are pairwise distinct. 
If $B$ contains tuples, that is, $B = B_1 \times \dots \times B_n$, we can also update a single element of a tuple: if $f(a) = \langle b_1, \ldots, b_n\rangle$, then we define $f[a, i\mapsto b](a) = \langle b_1, \ldots, b_{i-1}, b, b_{i+1},\ldots,b_n\rangle$ and $f[a, i \mapsto b](x) = f(x)$ for all $x\neq a$.

We assume the existence of a parser for the concrete syntax.
Henceforth, we work on an \emph{abstract syntax tree} (AST) produced by running the parser on a well-formed \Lname program.
We thus do not concern ourselves with operator precedence and parentheses, and we assume that polymorphic elements such as \texttt{null} and $42$ are labelled with the right type for their context, \emph{viz.}, $\nil_\SynType$ is the expression $\nil$ of type $\SynType$. 

We have a number of sets containing AST elements: $\Id$ is the set of all identifiers, $\Literals$ is the set of all literals, $\Schedules$ is the set of all schedules, $\Stat$ is the set of all statements, $\Expr$ contains all expressions and $\SynOp$ contains all syntactic binary operators.
The set containing all syntactic types is $\SynTypes = \{\texttt{Nat}, \texttt{Int}, \texttt{Bool}, \texttt{String}\}\cup \Id$, corresponding to the non-terminal $\type{Type}$.

In our semantics, \emph{labels} reference concrete instances of structs (as opposed to struct definitions).
We assume some sufficiently large set $\Labels$ containing these labels.
We also have the semantic types $\Nat$, $\Int$, $\Bool$ and $\String$ corresponding to the natural numbers, the integers, the booleans and the set of all strings, respectively. We consider the natural numbers to be a subset of the integers containing all positive numbers and $0$.
All semantic values are collected in $\Values = \Labels\sqcup \Int \sqcup \Bool \sqcup \String$ (where $\sqcup$ denotes the disjoint union). The set of all semantic types is $\SemTypes = \{\Labels, \Nat, \Int, \Bool, \String\}$.
The semantic value of a literal $g\in\Literals$ is $\getValue(g)$.
We consider the semantic value of \texttt{this} to be the label $\ell\in \Labels$ of the struct instance executing the line of code in which \texttt{this} occurs.

In addition, we assume for every struct type $\StructType$ the existence of a \emph{null-label} $\nilL{\StructType} \in \Labels$, so that we can provide a default value for each syntactical type with the function $\defaultVal : \SynTypes \to \Values$, defined as:
\begin{align*}
	\defaultVal(\SynType) =
	\begin{cases}
		0               & \text{if $\SynType = \SynNat$ or $\SynType = \SynInt$}                   \\
		\false          & \text{if $\SynType = \SynBool$}                             \\
		\vempty         & \text{if $\SynType = \SynString$}                           \\
		\nilL{\SynType} & \text{if $\SynType \in \Id$}
	\end{cases}.
\end{align*}
\noindent We define the set of all null-labels to be $\NilLabels$, with $\NilLabels\subset\Labels$.

To facilitate conciseness in our operational semantics, we break down statements and  expressions into \emph{commands}, which can be viewed as atomic actions in the semantics.
\begin{definition}[Commands]
	A \emph{command} $c$ is constructed according to the following grammar:
	\begin{align*}
		c ::= \push{\mathit{v}} \mid \readg{x} \mid \writev{x} \mid \cons{x} \mid \Ifc{C} \mid \Notc \mid \Operator(o)
	\end{align*}
	where $\mathit{v} \in \Values\cup\{\this\}$ is a semantic value or $\this$, a special value, $x \in \Id$ is an identifier, $C$ is a list of commands, and $o \in \SynOp$ is an operator.
	The set of all commands is $\Commands$.
\end{definition}

Intuitively, $\this$ is the semantic equivalent to the syntactic \texttt{this}-expression.
The precise effect of each command is discussed later in this section when the inference rules are given.
Statements and expressions are compiled into a list of commands according to the following recursive interpretation function:

\begin{definition}[Interpretation function]\label{def: interp}
	Let $x, x_1, \ldots, x_n\in \Id$ be variables, $E, E_1, \ldots, E_m\in\Expr$ expressions, $g\in\Literals$ a literal, $\StructType\in\Id$ a struct type, $S \in \Stat$ a statement, $\mathcal{S} \in \Stat^{*}$ a list of statements, $\SynType \in \SynTypes$ a type and $\mathtt{op} \in \SynOp$ an operator from the syntax. Let the list $x_1;...;x_n$ be the list of $n$ variables from $x_1$~to~$x_n$.
	We define the \emph{interpretation function} $\interp{\cdot}: \Stat^{*}\cup\Expr\rarr\Commands^{*}$ transforming a list of statements or expressions into a list of commands:
	\begin{align*}
		\interp{g}                                       & = \push{\getValue(g)}                                & \interp{\mathtt{if} \; E \; \mathtt{then} \{\mathcal{S}\}} & = \interp{E};\Ifc{\interp{\mathcal{S}}}                          \\
		\interp{\mathtt{this}}                           & = \push{\this}                                       & \interp{T\,x := E}                      & = \interp{x_1.\cdots.x_0.x := E}                             \\
		\interp{\mathtt{null}_\SynType}                  & = \push{\defaultVal(\SynType)}                       & \interp{x_1.\cdots.x_n.x := E}          & = \interp{E};\interp{x_1.\cdots.x_n};\writev{x}        \\
		\interp{x_1.\cdots.x_n}                          & = \push{\this};\readg{x_1};\ldots;\readg{x_n}        & \interp{\StructType (E_1,\ldots,E_m)}   & = \interp{E_1};\ldots;\interp{E_m};\cons{\StructType}  \\
		\interp{!E}                             & = \interp{E};\Notc                                   & \interp{\varepsilon}                             & = \vempty                                              \\
		\interp{E_1 \mathop{\mathtt{op}} E_2}   & = \interp{E_1};\interp{E_2};\Operator(\mathtt{op})   & \interp{S;\mathcal{S}}                                     & = \interp{S};\interp{\mathcal{S}}                               
	\end{align*}
	Note that if $n = 0$, $\interp{x_1.\cdots.x_n} = \push{\this}$ as a special case. We require $\push{\this}$ to give us the label of the current executing struct instance at the start of dereferencing any pointer.
\end{definition}

During the runtime of a program, multiple instances of a struct definition may exist simultaneously. 
We refer to these as \emph{struct instances}.

\begin{definition}[Struct instance]
	A \emph{struct instance} is a tuple $\langle \StructType, \ComList, \Stack, \Env\rangle$ where:
	\begin{itemize}[noitemsep, nolistsep]
		\item $\StructType\in\Id$ is the type of the struct,
		\item $\ComList\in\Commands^{*}$ is a list of commands that are to be executed,
		\item $\Stack\in\Values^{*}$ is a stack that stores values during the evaluation of an expression,
		\item $\Env:\Id\rightarrow\Values$ is an environment that stores the values of local variables as well as parameters.
	\end{itemize}
	We define $\StructsSet$ as the set of all possible struct instances.
\end{definition}

A state of a program is the combination of a schedule that remains to be executed, a collection with all the struct instances that currently exist and a stack of Boolean values that are required to determine whether a fixpoint has been reached. Note that every label can refer to at most one distinct struct instance.

\begin{definition}[State]\label{def: state}
	A \emph{state} is a tuple $\langle \Sched, \Structs, \Stab\rangle$, where:
	\begin{itemize}[nolistsep, noitemsep]
		\item $\Sched\in\Schedules$ is a schedule expressed as a list,
		\item $\Structs:\Labels\rightarrow\StructsSet\cup\{\bot\}$ is a \emph{struct environment},
		\item $\Stab\in\Bool^{*}$ is a \emph{stability stack}.
	\end{itemize}
	The set of all states is defined as $\StateSpace = \Schedules \times (\Labels \to \StructsSet\cup\{\bot\}) \times \Bool^{*}$.
\end{definition}
Intuitively, the stability stack keeps track of whether a fixpoint should terminate. When starting execution of this fixpoint, a new variable is placed on the stability stack corresponding to this fixpoint. This variable is set to true every time the fixpoint iterates, and is set to false when a parameter is changed during the execution of the fixpoint. 
Consider for example the code for Prefix Sum (Listing~\ref{ex: PS3}). In it, the fixpoint for $\mathit{read} < \mathit{write}$ will place a single variable on the stability stack, which is reset to \textit{true} at the start of an iteration and only remains \textit{true} if no element updates its \textit{prev} or \textit{val} parameters to new values.

With a notion of states and struct instances, we define \emph{null-instances}:
\begin{definition}[Null-instances]\label{def: nilinst}
	Let $P = \langle \Sched, \Structs, \Stab\rangle \in \StateSpace$ be a state. Then the set of \emph{null-instances} in state $P$ is defined as $
	\{\Structs(\ell) \mid \Structs(\ell)\neq \bot \land \ell\in\NilLabels\}$. 
\end{definition}
Thus, each struct instance that is labelled with a $\nil$-label is a $\nil$-instance.

Henceforth, we fix an \Lname program $\Program\in \Programs$ s.t. $\Program = D\ \Sched_\Program$, with $D$ the definitions of the program and $\Sched_\Program$ the schedule of the program. We define $\StructTypes_\Program\subseteq\Id$ to be the set of all struct types defined in $\Program$.
The initial variable environment for a struct instance of type $\StructType$ is $\Env^0_\StructType$, defined as $\Env^0_\StructType(p) = \defaultVal(\SynType)$ for all $p \in \Par{\StructType, \Program}$ where $\SynType$ is the type of $p$ and $\Par{\StructType, \Program}$ refers to the parameters of $\StructType$ in $\Program$ and is equal to $\Par{\StructType, D}$ as used last section. 
For other variables $x \in \Id$, $\Env^0_\StructType(x)$ is left arbitrary.

The initial state of an \Lname program depends on which program is going to be executed (the state space does not depend on the program, \emph{cf.} Def.~\ref{def: state}):
\begin{definition}[Initial state]\label{def: initial}
	The \emph{initial state} of $\Program$ is $\StateP_\Program^0 = \langle\Sched_\Program, \Structs_\Program^0, \vempty\rangle$, where $\Structs_\Program^0(\nilL{\StructType}) = \langle \StructType, \vempty, \vempty, \Env^0_\StructType \rangle$ for all $\StructType \in \StructTypes_\Program$ and $\Structs_\Program^0(\ell) = \bot$ for all other labels.
\end{definition}
Intuitively, this definition states that the initial state of a program $\Program$ consists of the schedule as found in the program, a struct environment filled with $\nil$-instances for every struct type declared in $\Program$ and an empty stack.

We proceed by defining the transition relation $\sr$ by means of inference rules. In this relation, we consider every transition to be labelled by the inference rule that induced the transition. The dependence of $\sr$ on $\Program$ stems from the dependence on parameter information from $\Program$.
There are rules that define the execution of commands and rules for the execution of a schedule. We start with the former.
Command $\push{v}$ pushes value $v$ on the stack $\Stack$, and $\push{\this}$ pushes the label of the structure instance on $\Stack$: 
\begin{equation*}
	\inference[(\textbf{ComPush})]{
		\Structs(\ell) = \langle \StructType, \push{\mathit{v}};\CList, \Stack, \Env\rangle
	}{
		\langle \Sched, \Structs, \Stab \rangle\sr \langle \Sched, \Structs[\ell \mapsto \langle \StructType, \CList, \Stack;\mathit{v}, \Env\rangle], \Stab\rangle
	}
\end{equation*}
\begin{equation*}
	\inference[(\textbf{ComPushThis})]{
		\Structs(\ell) = \langle \StructType, \push{\this};\CList, \Stack, \Env\rangle
	}{
		\langle \Sched, \Structs, \Stab \rangle\sr\langle \Sched, \Structs[\ell \mapsto \langle \StructType, \CList, \Stack;\ell, \Env\rangle], \Stab\rangle
	}
\end{equation*}
The command $\readg{x}$ reads the value of variable $x$ from environment $\Env'$ of $\ell'$ and places it onto the stack:
\begin{equation*}
	\inference[(\textbf{ComRd})]{
		\Structs(\ell) = \langle \StructType, \readg{x};\CList, \Stack;\ell', \Env\rangle\\
		\Structs(\ell') = \langle \StructType', \ComList', \Stack', \Env'\rangle
	}{
		\langle \Sched, \Structs, \Stab \rangle\sr\langle \Sched, \Structs[\ell \mapsto \langle \StructType, \CList, \Stack;\Env'(x), \Env\rangle], \Stab\rangle
	}
\end{equation*}
For normal struct instances, $\writev{x}$ takes a label $\ell'$ and a value $v$ from the stack and writes this to $\Env'(x)$, the environment of the struct instance corresponding to $\ell'$.
If $x$ is a parameter (as opposed to a local variable) and writing $v$ changes its value, then any surrounding fixpoint in the schedule becomes unstable.
In that case, we set the auxiliary value $\mathit{su}$ (for \emph{stability update}) to false and clear the stability stack by setting it to $\Stab_1 \land \mathit{su};\dots;\Stab_{|\Stab|} \land \mathit{su}$.
Note that this leaves the stack unchanged if $\mathit{su}$ is true.
Below, in the update ``$[\ell', 4\mapsto \Env'[x\mapsto v]]$'', recall that $f[a,i \mapsto b]$ denotes the update of a function that returns a tuple.
\begin{equation*}
	\inference[(\textbf{ComWr})]{
		\Structs(\ell) = \langle \StructType, \writev{x};\CList, \Stack;v;\ell',
		\Env\rangle \\
		\Structs(\ell') = \langle \StructType', \ComList', \Stack', \Env'\rangle \\
		\ell' \notin \Labels^{0} \lor x\notin \Par{}(\StructType', \Program) \\
		\mathit{su} = (x\notin\Par{}(\StructType', \Program)\lor\Env'(x) = v) \\
	}{
		\begin{aligned}
			\langle \Sched, \Structs, \Stab \rangle\sr \langle \Sched, \Structs&[\ell \mapsto \langle \StructType, \CList, \Stack, \Env\rangle][\ell', 4\mapsto \Env'[x\mapsto v]], \Stab_1 \land \mathit{su};\dots;\Stab_{|\Stab|} \land \mathit{su}\rangle
		\end{aligned}
	}
\end{equation*}
The next rule skips the write if the target is a parameter of a $\nil$-instance, which ensures that the parameters of a $\nil$-instance cannot be changed. See Section~\ref{section:types} for the reasoning behind our use of $\nil$-instances. Note that we allow $\nil$-instances to use local variables, which is useful for the initialization of a system, for example.
\begin{equation*}
	\inference[(\textbf{ComWrNSkip})]{
		\Structs(\ell) = \langle \StructType, \writev{x};\CList, \Stack;v;\ell',
		\Env\rangle \\
		\Structs(\ell') = \langle \StructType', \ComList', \Stack', \Env'\rangle \\
		\ell' \in \Labels^{0} \land x\in \Par{}(\StructType', \Program) \\
	}{
		\begin{aligned}
			&\langle \Sched, \Structs, \Stab \rangle\sr \langle \Sched, \Structs[\ell \mapsto \langle \StructType, \CList, \Stack, \Env\rangle], \Stab\rangle
		\end{aligned}
	}
\end{equation*}
Let $b$ be a boolean value on the top of the stack $\chi$. A $\Notc$ command negates $b$:
\begin{equation*}
	\inference[(\textbf{ComNot})]{
		\Structs(\ell) = \langle \StructType, \Notc;\CList, \Stack;b, \Env\rangle
	}{
		\langle \Sched, \Structs, \Stab \rangle\sr\langle \Sched, \Structs[\ell \mapsto \langle \StructType, \CList, \Stack;\neg b, \Env\rangle], \Stab\rangle
	}
\end{equation*}
An $\Operator(\circ)$ command applies the semantic equivalent $o\in \{=, \neq, \leq, \geq, <, >, *, /,\linebreak[4] \%, +, -, \verb|^|, \wedge, \vee\}$ of the syntactic operator $\circ\in\SynOp$ to the two values at the top of $\chi$ (here taken to be the values $a$ and $b$), of which the result is put on top of the stack:
\begin{equation*}
	\inference[(\textbf{ComOp})]{
		\Structs(\ell) = \langle \StructType, \Operator(\circ);\CList, \Stack;a;b, \Env\rangle
	}{
		\langle \Sched, \Structs, \Stab \rangle\sr\langle \Sched, \Structs[\ell \mapsto \langle \StructType, \CList, \Stack;(a\mathop{o} b), \Env\rangle],\Stab\rangle
	}
\end{equation*}
Let $\StructType'\in\StructTypes_\Program$ be the type of a struct with $n$ parameters. 
The command $\cons{\StructType'}$ creates a new struct instance of type $\StructType'$ in the struct environment $\Structs$ with a fresh label $\ell'$, and initializes the parameters to the top $n$ values of the stack:
\begin{equation*}
	\inference[(\textbf{ComCons})]{
		\Structs(\ell) = \langle \StructType, \cons{\StructType'};\CList, \Stack;v_1;\ldots;v_n, \Env\rangle \\ \Par{}(\StructType', \Program) = \mathit{p}_1: T_1;...;\mathit{p}_n: T_n \\
		\Structs(\ell')=\bot
	}{
		\begin{aligned}
			\langle \Sched, \Structs, \Stab \rangle\sr \langle& \Sched, \Structs[\{\ell \mapsto \langle sL, \CList, \Stack;\ell', \Env\rangle,
			\\[-4pt]
			&\ell'\mapsto\langle sL', \vempty ,\vempty,\Env^{0}_{\StructType'}[\{\mathit{p}_1\mapsto v_1, \ldots, \mathit{p}_n\mapsto v_n\}]\rangle\}], \false^{|\Stab|}\rangle
		\end{aligned}
	}
\end{equation*}
The command $\Ifc{C}$ with $C\in\Commands^{*}$ adds commands $C$ to the start of $\gamma$ if the top value of the stack is $\true$. If the top value is $\false$, the command does nothing:
\begin{equation*}
	\inference[(\textbf{ComIfT})]{
		\Structs(\ell) = \langle \StructType, \Ifc{C};\CList, \Stack;\true, \Env\rangle
	}{
		\langle \Sched, \Structs, \Stab \rangle\sr\langle \Sched, \Structs[\ell \mapsto \langle \StructType, C;\CList, \Stack, \Env\rangle], \Stab\rangle
	}
\end{equation*}
\begin{equation*}
	\inference[(\textbf{ComIfF})]{
		\Structs(\ell) = \langle \StructType, \Ifc{C};\CList, \Stack;\false, \Env\rangle
	}{
		\langle \Sched, \Structs, \Stab \rangle\sr\langle \Sched, \Structs[\ell \mapsto \langle \StructType, \CList, \Stack, \Env\rangle], \Stab\rangle
	}
\end{equation*}
In the remaining rules, let $\mathit{Done}(\sigma) = \forall\ell.(\Structs(\ell) = \bot \vee \exists \StructType, \Stack, \Env. \Structs(\ell) = \langle \StructType, \vempty, \Stack, \Env\rangle)$, let $\mathit{sc}$ be a (possibly empty) schedule.
The predicate $Done(\sigma)$ holds when all commands have been executed in all struct instances in $\sigma$.

We can initiate steps globally and locally.
The global step initiation converts all statements in a step to commands for any structure instance that has that step and adds the commands to $\gamma$. Recall that $\Program = D\ \Sched$ and recall from last section that $\mathit{StepDef}(D)$ returns all pairs $(F_\StructType, \mathcal{S})$ s.t. $F$ is a step defined for $\StructType\in\StructTypes_\Program$ as the list of statements $\mathcal{S}$. Let $\StatList_{\StructType}^{F} = \vempty$ if there is no list of statements $\mathcal{S}$ s.t. $(F_\StructType, \mathcal{S})\in\mathit{StepDef}(D)$, and let $\StatList_{\StructType}^{F}$ be a list of statements s.t. $(F_\StructType, \StatList_{\StructType}^{F})\in\mathit{StepDef}(D)$ otherwise.
\begin{equation*}
	\inference[(\textbf{InitG})]{\mathit{Done}(\sigma)}{
		\begin{aligned}
			&\langle F<\mathit{sc}, \Structs, \Stab \rangle\sr \langle \mathit{sc}, \Structs[\{\ell \mapsto \langle \StructType, \interp{\StatList_{\StructType}^{F}}, \vempty, \Env\rangle \mid \Structs(\ell) = \langle \StructType, \varepsilon, \Stack, \Env\rangle\}],\Stab\rangle
		\end{aligned}
	}
\end{equation*}
The local step initiation converts the step to commands and adds those commands to $\gamma$ only for struct instances of a specified struct $\StructType$:
\begin{equation*}
	\inference[(\textbf{InitL})]{\mathit{Done}(\sigma)}{
		\begin{aligned}
			&\langle \StructType.F<\mathit{sc}, \Structs, \Stab \rangle \sr \langle \mathit{sc}, \Structs[\{\ell \mapsto \langle \StructType, \interp{\StatList_{\StructType}^{F}}, \vempty, \Env\rangle\mid \Structs(\ell) = \langle \StructType, \varepsilon, \Stack, \Env\rangle\}],\Stab\rangle
		\end{aligned}
	}
\end{equation*}
Note that neither of the step initializations flush the previous local variables from the struct environments. Instead, the type system of \Lname as found in Section~\ref{section:types} limits the scope of local variables to the steps they are declared in. If required, it can also be enforced in the semantics by defining a function which takes the value of the struct instance environment if the given variable is a parameter, and $\bot$ otherwise.

Let $\mathit{sc}_1$ be a schedule. Fixpoints are initiated when first encountered:
\begin{equation*}
	\inference[(\textbf{FixInit})]{\mathit{Done}(\sigma)}{
		\langle \mathit{Fix}(\mathit{sc})<\mathit{sc}_1, \Structs, \Stab \rangle\sr \langle \mathit{sc}<\mathit{aFix}(\mathit{sc})<\mathit{sc}_1, \Structs,\Stab;\true\rangle
	}
\end{equation*}
The symbol $\mathit{aFix}$ is a semantic symbol used to denote a fixpoint which has been initiated.
When an initiated fixpoint is encountered again, the stability stack is used to determine whether the body should be executed again:
\begin{gather*}
	\inference[(\textbf{FixIter})]{\mathit{Done}(\sigma)}{
		\langle \mathit{aFix}(\mathit{sc})<\mathit{sc}_1, \Structs,  \Stab;\false \rangle\sr\langle \mathit{sc}<\mathit{aFix}(\mathit{sc})<\mathit{sc}_1, \Structs, \Stab;\true\rangle
	}\\
	\inference[(\textbf{FixTerm})]{\mathit{Done}(\sigma)}{
		\langle \mathit{aFix}(\mathit{sc})<\mathit{sc}_1, \Structs, \Stab;\true \rangle\sr\langle \mathit{sc}_1, \Structs, \Stab\rangle
	}
\end{gather*}
With these rules, we give an operational semantics for \Lname:
\begin{definition}[Operational semantics]\label{def: os}
	The semantics of $\Program$ is the directed graph $\langle \StateSpace, \sr, P_\Program^0\rangle$, where $\StateSpace$ is the set of all states (Def.~\ref{def: state}), $\sr$ is the labelled transition relation for $\Program$ as given above and $P_\Program^0$ is the initial state of $\Program$ (Def.~\ref{def: initial}).
\end{definition}

The proofs of Section~\ref{section:props} and Section~\ref{section:proving} are based upon the semantics as defined above. Generally, when we refer to the semantics of a program $\Program$, we are talking about the semantics of $\Program$ as defined in the above definition.
	\section{Example Algorithms}\label{section:standardalgos}
In this section, we provide more intuition on how \Lname works in practice by means of five example \Lname programs.
The first creates a spanning tree, the second and third are a sorting programs and the fourth and fifth implement solutions for 3SUM.  
In \Lname, there is a struct instance for every data element. The amount of processors used is therefore assumed equal to the amount of data elements of relevance in the code. We give work (the total amount of operations), span (the length of the longest required chain of sequential operations) and running time estimates for the given programs. Other examples and implementations can be found in other \Lname papers~\cite{franken-audala-2024, leemrijse-2023}.

\subsection{Creating a spanning tree}
Given a connected directed graph $G = (V,E)$ and a root node $u \in V$, we can create a spanning tree of $G$ rooted in $u$ using \emph{breadth-first search}.
In this tree, for every node $v$, the path from $u$ to $v$ is a shortest path in $G$.
We do this by incrementally adding nodes from $G$ with a higher distance to $u$ to the spanning tree. 
\begin{lstlisting}[float=t,caption={\Lname code for creating a spanning tree}, label={ex: BFS}]
	struct (*\textit{Node}*)((*\textit{dist}*): Int, (*\textit{in}*): (*\textit{Edge}*)){} (*\label{line-bfsnode}*)
	
	struct (*\textit{Edge}*)($s$: (*\textit{Node}*), $t$: (*\textit{Node}*)){ (*\label{line-bfsedge}*)
		(*\textit{linkEdge}*){ 
			if $s$.(*\textit{dist}*) != -1 && $t$.(*\textit{dist}*) = -1 then { (*\label{line-bfs-relevant}*)
				$t$.(*\textit{in}*) := this; (*\label{line-bfs-nom}*)
			}
		}
		(*\textit{handleEdge}*) {
			if $t$.(*\textit{in}*) != null then { (*\label{line-bfs-inexist}*)
				if $t$.(*\textit{in}*) = this then { (*\label{line-bfs-nomwon}*)
					$t$.(*\textit{dist}*) := $s$.(*\textit{dist}*) + 1;
				}
				if $t$.(*\textit{in}*) != this then { (*\label{line-bfs-nomlost}*)
					$s$ := null;
					$t$ := null;
				}
	}}}(*\label{line-bfs-nomlostend}*)		
	
	Fix((*\textit{linkEdge}*) < (*\textit{handleEdge}*)) (*\label{line-bfs-sched}*)
\end{lstlisting}

We first sketch our approach.
In the $i$th BFS iteration, the algorithm adds all edges $(s,t)$ to the tree such that the distance from $u$ to $s$ is $i-1$ and the distance from $u$ to $t$ is still unknown.
If multiple such edges lead to the same $t$, the algorithm uses a \emph{race condition} to determine which edge is chosen.
As any edge will suffice, this race condition is benign.
The distance from $u$ to $t$ is then set to $i$ and we continue with the next iteration.
The program runs with $O(|V|+|E|)$ data elements. As every edge is considered only once, the amount of work performed by this algorithm is $O(|E|)$. The span of the algorithm is $O(d)$, with $d$ the diameter of the graph, as in the worst case all edges in the path which determines the diameter of the graph need to be considered sequentially. It follows that the worst case running time for the algorithm is $O(d)$.

Contained in Listing~\ref{ex: BFS} is an \Lname program that implements this approach.
The program defines the struct \textit{Node} (line~\ref{line-bfsnode}) with parameters \textit{dist}, to store the distance from root node $u$, and \textit{in}, a reference to its incoming spanning tree edge.
The struct \textit{Edge} (line~\ref{line-bfsedge}) has a source $s$ and a target $t$.

During initialization, the input should be a directed graph, with a root \textit{Node} $u$ with \textit{dist} $0$ and with the \textit{dist} parameter of the other \textit{Nodes} set to $-1$. 
For every \textit{Node}, the parameter \textit{in} should be set to null.

Both steps in the program belong to \textit{Edge}.
The first step, \textit{linkEdge}, first determines whether an \textit{Edge} $e$ from \textit{Node} $s$ to \textit{Node} $t$ is at the frontier of the tree in line~\ref{line-bfs-relevant}. 
This is the case when $s$ is in the tree, but $t$ is not.
If so, $e$ nominates itself as the \textit{Edge} connecting $t$ to the tree, $t.in$ (line~\ref{line-bfs-nom}).
This is a race condition won by only one edge for $t$, the edge which applies the semantic rule \textbf{ComWr} last.
In the second step, \textit{handleEdge}, if the nomination for $t$ has finished (line~\ref{line-bfs-inexist}) and $e$ has won the nomination (line~\ref{line-bfs-nomwon}), $e$ will update $t$'s distance to the root.
If $e$ has lost, it will remove itself from the graph, here coded as setting the source and target parameters to $\nil$ in lines~\ref{line-bfs-nomlost} to \ref{line-bfs-nomlostend}.

\begin{figure}[t]
	\centering
	\includegraphics[width=0.7\textwidth]{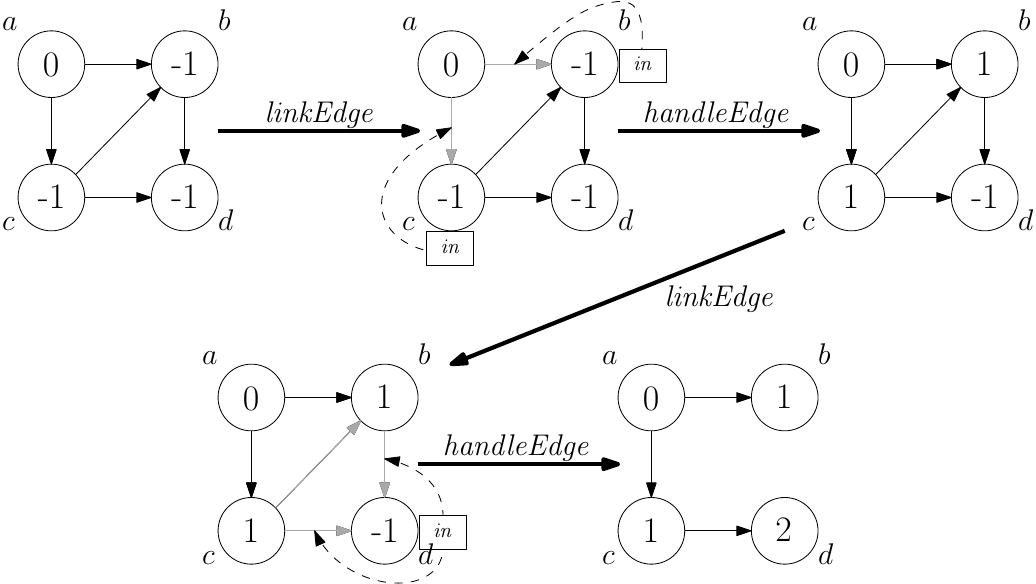}
	\caption{Execution of Listing~\ref{ex: BFS} on a small graph. Every \textit{Edge} newly considered in the current step is grey. Considered \textit{Edges} stay considered, but stable. The dotted arrows denote the possible new values for \emph{t.in} of a target node $t$. Note that the \textit{Edge} from $c$ to $d$ wins the race condition to the reference $d.in$.}
	\label{gr: BFSTree}
\end{figure}

To create a full spanning tree, this must be executed until all \textit{Nodes} have a positive \textit{dist} and all \textit{Edges} are either $t.in$ for their target \textit{Node} $t$ or have $\nil$ as their source and target. To this end, the schedule (line~\ref{line-bfs-sched}) contains a fixpoint, in which \textit{Edges} first nominate themselves and then update the distances of new \textit{Nodes}.
This fixpoint terminates, as \textit{Edges} in the spanning tree will continuously update their targets with the same information and \textit{Edges} which lost their nomination will not get past the first conditions of the two steps, causing the data elements to stabilize after all \textit{Nodes} have received a distance from $u$.
Initialization steps should be placed at the start of the  shown schedule.
An execution of the program on a small graph is shown in Figure~\ref{gr: BFSTree}, where the edges and nodes of the graph are modelled by their respective structs.

\subsection{Sorting}\label{section:sortex}
\begin{lstlisting}[float = t, caption={\Lname code for sorting}, label={ex: Sort}]
	struct (*\textit{ListElem}*)((*\textit{val}*): Int, (*\textit{next}*): (*\textit{ListElem}*), (*\textit{newNext}*): (*\textit{ListElem}*), (*\textit{comp}*): (*\textit{ListElem}*)){
		(*\textit{compareElement}*) {
			if (*\textit{comp}*) != null then {
				if ((*\textit{comp.val}*) > (*\textit{val}*) && (((*\textit{newNext}*) = null) || (*\textit{comp.val}*) < (*\textit{newNext.val}*))) then { (*\label{line-sort-check}*)
					(*\textit{newNext}*) := (*\textit{comp}*);
				}
				(*\textit{comp}*) := (*\textit{comp.next}*); (*\label{line-sort-next}*)
			}
		}
		(*\textit{reorder}*) {
			(*\textit{next}*) := (*\textit{newNext}*); (*\label{line-sort-reorder}*)
		}
	}
	
	Fix((*\textit{compareElement}*)) < (*\textit{reorder}*) (*\label{line-sort-sched}*)
\end{lstlisting}
A concise example of a \Lname program for sorting a linked list of $n$ elements can be found in Listing~\ref{ex: Sort}.
In it, the elements of the list traverse the list together, during which each element $e$ is looking for its successor in the sorted list. 
After the traversal, the successor element is saved and the link is updated to the saved element.
This reorders the list to the sorted list.
The program requires $n$ data elements, one for every list element. It performs $O(n^2)$ work, as every element is compared to every other element, and its span is $O(n)$, as it walks through a list of size $n$ sequentially. Therefore, the program runs in $O(n)$ time.
We can achieve a time complexity of $O(\log{n})$ by implementing Cole's algorithm~\cite{cole-parallel-1988} in \Lname, but that is outside the scope for this paper.

The program defines the struct \textit{ListElem}, modeling the nodes of the list, with parameters \textit{val}, \textit{next}, a reference to the next \textit{ListElem}, \textit{newNext}, a reference to the \textit{ListElem} that should come next in the sorted list, and \textit{comp}, a reference to the current \textit{ListElem} \textit{newNext} is compared to.
The initialization needs to make sure that every element has a distinct value, and that in every element, \textit{comp} is set to the first element of the list and \textit{newNext} is set to $\nil$.

To facilitate our strategy we give our \textit{ListElem} two steps, one to check an element in the list called \textit{compareElement} and one to reorder the list at the end called \textit{reorder}. 
With the step \textit{compareElement}, an element checks whether the element to which the \textit{comp} reference leads is a better next element than the current element saved in \textit{newNext} (line~\ref{line-sort-check}) and updates \textit{newNext} if that is the case.
Afterwards, the \textit{comp} reference is updated to the next element in the list (line~\ref{line-sort-next}).
With the \textit{reorder} step, an element replaces their old \textit{next} reference with \textit{newNext} (line~\ref{line-sort-reorder}).

\begin{figure}[t]
	\centering
	\includegraphics[width=0.7\textwidth]{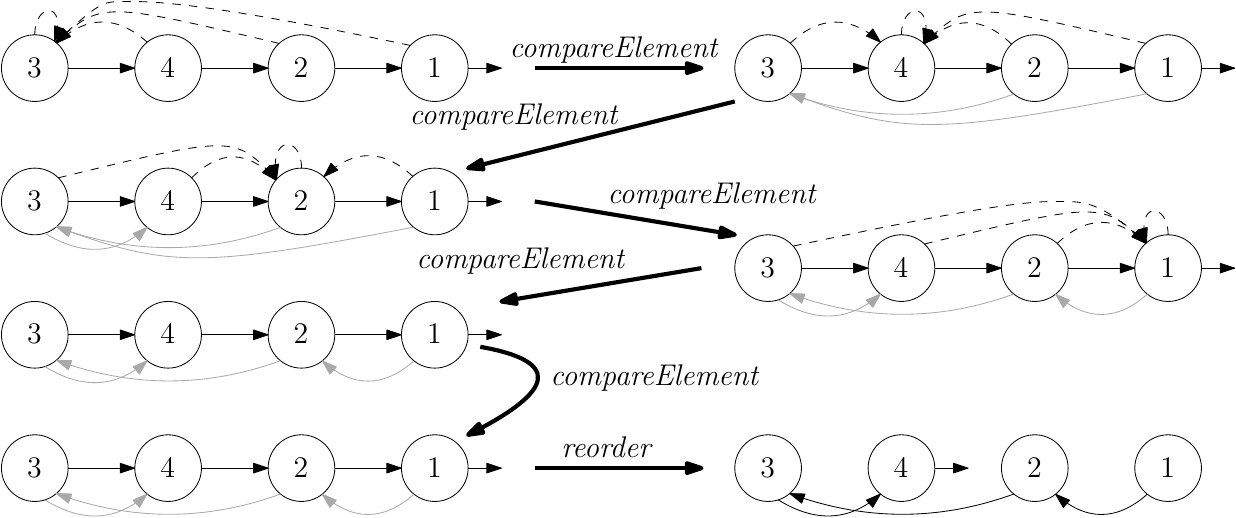}
	\caption{Execution of Listing~\ref{ex: Sort} on a small list. The parameters \textit{next}, \textit{newNext} and \textit{comp} are shown as black, grey and dashed unmarked arrows respectively, and the null-references for \textit{newNext} and \textit{comp} are not shown. The nodes corresponding to \emph{ListElems} contain the value of parameter \textit{val}.}
	\label{gr: Sort}
\end{figure}

To have the program execute our strategy, we call a fixpoint on \textit{compareElement}, such that every element checks all elements in the list. After that is done, the schedule tells the elements to \textit{reorder} (line~\ref{line-sort-sched}). An execution of the program on a small list is shown in Figure~\ref{gr: Sort}.

\subsection{Linked List Copy Sort}\label{section:sortex2}
A more complex example for sorting a linked list of $n$ elements in \Lname can be found in Listing~\ref{ex: Sort2}, which makes a sorted copy of the given list. At the start of execution, this copy consists of a single element, which encompasses the full range of the old list. In every iteration, all elements of the new list split and divide their range (when necessary), and elements from the old list linked to an element of a new list redivide themselves over that element and its new split. This continues until every new element has exactly one old element linked to it. At that point, the new list is a sorted copy of the old list. 
The idea of the algorithm is to do tree-insert for every data element at the same time, using a tree of size $O(\log{m})$, which is dynamically created and repurposed to use as few new data elements as possible.
For a list with $n$ data elements with maximum value $m$, the algorithm performs $O(n\log{m})$ work, as all data elements compare themselves to at most $\log{m}$ other elements. The span of the algorithm is $O(\log{m})$, for in the worst case, an element must go through $\log{m}$ comparisons before it is placed correctly in the new list. It follows that the program runs in $O(\log{m})$ time with $n$ data elements with maximum value $m$. As implemented, the program uses exactly $2n$ data elements.
\begin{lstlisting}[float=p,caption={\Lname code for Linked List Copy Sort}, label={ex: Sort2}]
	struct (*\textit{OldElem}*)((*\textit{val}*): Int, (*\textit{place}*): (*\textit{NewElem}*), (*\textit{move}*): Bool){
		(*\textit{checkStable}*) {
			if ((*\textit{place.p2}*) != null || ((*\textit{val}*) <= (*\textit{place.spl}*) && (*\textit{place.p1}*) != this)) then {			 
				(*\textit{place.done}*) := false;
			}
		}
		(*\textit{migrate}*) {
			if ((*\textit{move}*)) then {
				if ((*\textit{place.hasSplit}*)) then {
					(*\textit{place}*) := (*\textit{place.next}*);
				}
				(*\textit{move}*) := false;
			}
			if ((*\textit{val}*) <= (*\textit{place.spl}*)) then {
				(*\textit{place.p1}*) := this;
			}
			if ((*\textit{val}*) > (*\textit{place.spl}*)) then {
				(*\textit{place.p2}*) := this;
				(*\textit{move}*) := true;
			}
		}
	}
	struct (*\textit{NewElem}*)((*\textit{min}*): Int, (*\textit{spl}*): Int, (*\textit{max}*): Int, (*\textit{next}*): (*\textit{NewElem}*), (*\textit{p1}*): (*\textit{OldElem}*), (*\textit{p2}*): (*\textit{OldElem}*), (*\textit{hasSplit}*): Bool, (*\textit{done}*): Bool){
		(*\textit{split}*) {
			if (!(*\textit{done}*)) then {
				if ((*\textit{p1}*) != null && (*\textit{p2}*) != null) then {
					(*\textit{next}*) := (*\textit{NewElem}*)((*\textit{spl}*)+1,(*\textit{spl}*)+1+((*\textit{max}*)-((*\textit{spl}*)+1))/2,(*\textit{max}*),(*\textit{next}*),null,null,false,true);
					(*\textit{max}*) := (*\textit{spl}*);
					(*\textit{hasSplit}*) := true;				
				}
				if ((*\textit{p1}*) != null && (*\textit{p2}*) = null) then {
					(*\textit{max}*) := (*\textit{spl}*);
					(*\textit{hasSplit}*) := false;
				}
				if ((*\textit{p2}*) != null && (*\textit{p1}*) = null) then {
					(*\textit{min}*) := (*\textit{spl}*) + 1;
					(*\textit{hasSplit}*) := false;
				}
				(*\textit{spl}*) := (*\textit{min}*) + ((*\textit{max}*) - (*\textit{min}*))/2;
				(*\textit{p1}*) := null;
				(*\textit{p2}*) := null;
			}
			(*\textit{done}*) := true;
		}
	}
	
	(*\textit{migrate}*) < Fix((*\textit{checkStable}*) < (*\textit{split}*) < (*\textit{migrate}*))
\end{lstlisting}
\begin{figure}[p]
	\centering
	\includegraphics[width=0.8\textwidth]{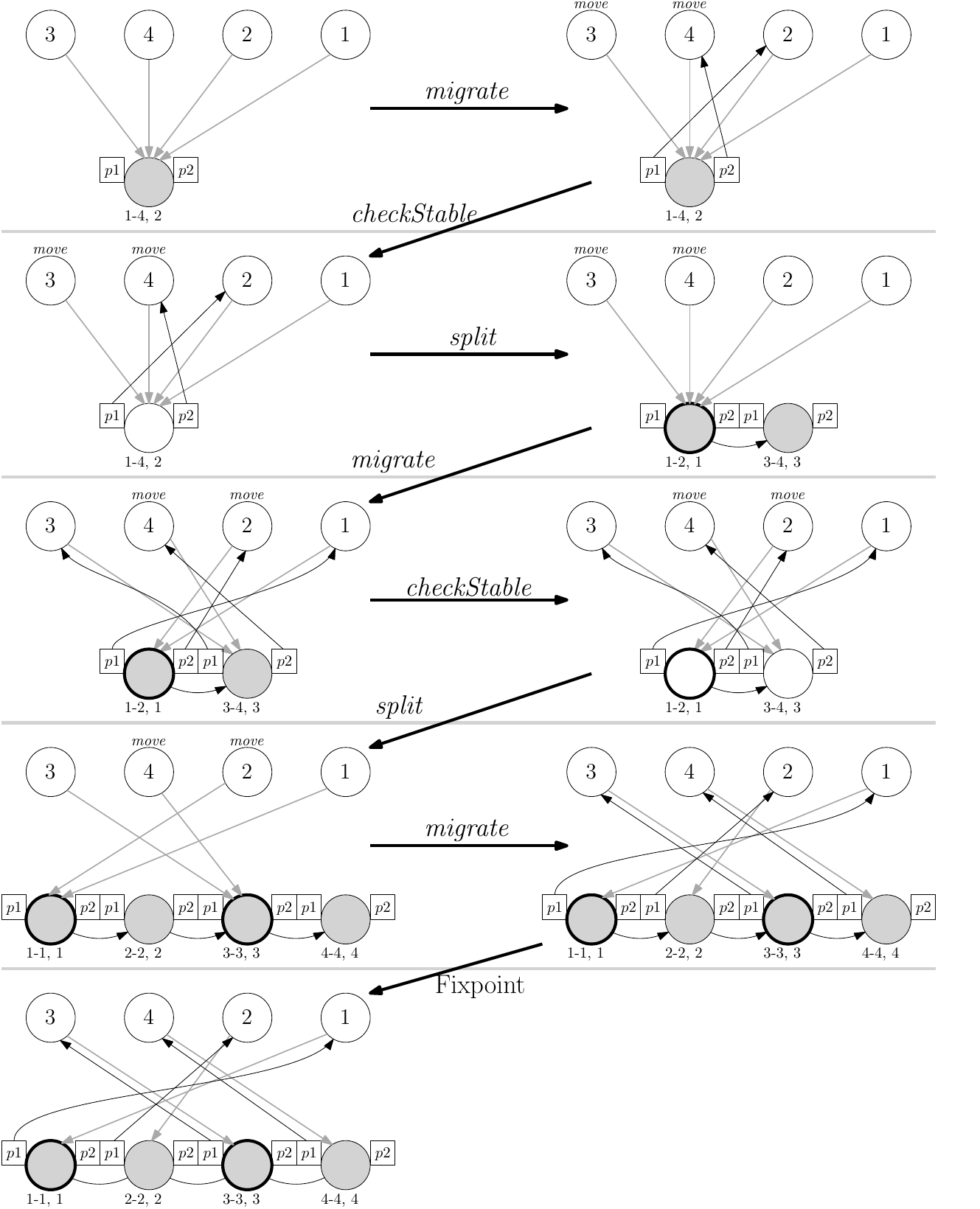}
	\caption{Execution of Listing~\ref{ex: Sort2} on a small list. The upper rows are the old elements, the lower rows the new elements. For the old elements, the parameter \textit{place} is depicted by the grey arrows. For the new elements, the numbers beneath every element are formatted ``[\textit{min}]-[\textit{max}], [\textit{spl}]''. New elements for which \textit{hasSplit} is true have a bold border, and a new element for which \textit{done} is true is grey.}
	\label{gr: Sort2}
\end{figure}
The program defines the struct \textit{OldElem}, which represents an element of the old list, and \textit{NewElem}, which represents an element of the new list. Struct \textit{OldElem} has parameters \textit{val}, which holds its value, \textit{place}, which holds the \textit{NewElem} it is currently linked to, and \textit{move}, which saves whether the old element should move to the right neighbour of its current \textit{place} during a split. Struct \textit{NewElem} has the parameters \textit{min}, \textit{spl} (split) and \textit{max}, which hold the current range of the new element and the value it will split the range on during the next split. Additionally, it has parameters \textit{next}, which holds the next new element in the list, \textit{p1}, which holds a representative old element which is in the first half of its range, \textit{p2}, which holds a representative for the second half of its range, \textit{hasSplit}, which signals to the old elements that the new element has split and \textit{done}, which saves whether the element will have to split again.

During initialization, there should be a single new element $b$ to which all old elements of the linked list are linked. For all old elements, \textit{move} should be set to $\false$. For $b$, \textit{min} should be set to the minimum value of the linked list, \textit{max} should be set to the maximum value of the linked list, \textit{spl} should be set to $(\mathit{max} - \mathit{min} + 1)/2 + \mathit{min}$. The parameter \textit{hasSplit} should be $\false$, \textit{done} should be $\true$ and \textit{next}, \textit{p1} and \textit{p2} should be set to $\nil$. Finding \textit{max} and \textit{min} can be done with a method akin to Prefix Sum.

To facilitate the strategy of splitting the new elements and dividing the old elements after the split, we use three steps. The first step, \textit{checkStable} is used to check whether a new element should split its range further. This is the case if there is still an element on the right side of the range of a new element or if there are multiple elements still on the left side of the range of a new element, which the step encodes. 
For the first part of the second step, \textit{migrate}, let $e$ be a new element which just split into $e$ and $e_2$. In the first part of \textit{migrate}, the old elements distribute themselves over $e_1$ and $e_2$ by either moving to $e_2$ or staying at $e$. In the second part of \textit{migrate} the old elements prepare themselves for the next split and distribute themselves over the left and right sides of the range of their current element.
The third step, \textit{split}, regulates the splitting of new elements. Every new element halves the range, but based on the contents of the ranges, they either change their minimum or maximum if only one half is populated, or they make a new element if both halves are populated. They then reset their representatives of both ranges and the \textit{done} parameter.

During execution, the method will execute through a fixpoint $\mathit{migrate}< \mathit{Fix}(\mathit{checkStable}<\mathit{split}<\mathit{migrate})$. The first call to \textit{migrate} divides the old elements over the range of the new elements. The fixpoint then keeps splitting until for every new element, there is only one old element (which is stored in \textit{p1}). This is helped by the parameter \textit{done}, which only allows for splits if the new element is unstable.
The execution of the first iteration of the program on a small list is shown in Figure~\ref{gr: Sort2}.

\subsection{The 3-SUM Problem}
\begin{lstlisting}[float=t,caption={\Lname implementation for naive 3SUM solution}, label={ex: 3SUM1}]
	struct (*\textit{Elem}*)((*\textit{val}*): Int, (*\textit{next}*): (*\textit{Elem}*), (*\textit{first}*): (*\textit{Elem}*), (*\textit{p1}*): (*\textit{Elem}*), (*\textit{p2}*): (*\textit{Elem}*), (*\textit{ans}*): Bool){
		(*\textit{next}*){
			if !(*\textit{ans}*) then {
				(*\textit{p1}*) := (*\textit{p1.next}*);
				(*\textit{p2}*) := (*\textit{first}*);
			}
		}
		(*\textit{walk}*) {
			if (*\textit{p1}*) != null && (*\textit{p2}*) != null  then {
				if (*\textit{val}*) + (*\textit{p1.val}*) + (*\textit{p2.val}*) = 0) then {			 
					(*\textit{ans}*) := true;
				}
			}
			if !(*\textit{ans}*) then {
				(*\textit{p2}*) := (*\textit{p2.next}*)
			} 
		}
	}
	Fix(Fix((*\textit{walk}*)) < (*\textit{next}*))
\end{lstlisting}
As our last example, we give two algorithms for the 3SUM problem~\cite{HoffmanLectureNotes, gajentaan-class-1995}: Given a set $S$ of $n$ integers, are there three elements $a, b, c\in S$ s.t. $a+b+c = 0$? Note that these elements do not need to be distinct.
The first program, in Listing~\ref{ex: 3SUM1}, contains a naive solution, and uses a nested fixpoint. This serves as an example of the use of nested fixpoints. The second program, in Listing~\ref{ex: 3SUM1} solves this in linear time (but with $O(n^2)$ work). This program is based upon the program given by Hoffman~\cite{HoffmanLectureNotes}. We assume that the set has been given to us as a list. If not, we can convert the set to a list in linear time using a fixpoint in which we nominate an element still in the set and then put it at the back of the list. What's more, our second program assumes that this list is sorted, which can happen through applying the example code from Section~\ref{section:sortex}~or~\ref{section:sortex2}.

Both programs define the struct \textit{Elem}, which represents an element of the list. In the first program, it has a value \textit{val}, a reference to the next element \textit{next}, a reference to the first element of the list \textit{first} and a reference to two other mutable elements, \textit{p1} and \textit{p2}. They also have a parameter \textit{ans}, which holds the answer to the question. After initialization, \textit{val}, \textit{next} and \textit{first} are set as specified above. The parameter \textit{ans} is set to $\false$ and \textit{p1} and \textit{p2} are set to \textit{first}.

To solve the 3SUM problem, the first program will make every list element $a$ walk through the list in a nested fashion to find elements $b$ and $c$ s.t $a$, $b$ and $c$ are not the null-\textit{Element}. For every element $b$ chosen by $a$, $a$ will \textit{walk} through the list to find $c$, if it exists. If not, it will replace $b$ with its next element in the step \textit{next}. 
This program performs $O(n^3)$ work ($n$ elements check at most $n^2$ combinations), with a span of $O(n^2)$: in the worst case, where no triple exists, $O(n)$ elements in the list need to be traversed $O(n)$ times. The code therefore executes in $O(n^2)$ time.

In the second program (Listing~\ref{ex: 3SUM2}), \textit{Elem} has the parameters \textit{next} and \textit{prev} which refer to those respective elements in the list. It also has candidate references \textit{p1} and \textit{p2}, initialized as the first and last element in the list respectively, a value \textit{val} and an integer \textit{ans} to give the answer to the problem (initialized to $0$).

\begin{lstlisting}[float=t,caption={\Lname implementation for a more efficient 3SUM solution}, label={ex: 3SUM2}]
	struct (*\textit{Elem}*)((*\textit{val}*): Int, (*\textit{next}*): (*\textit{Elem}*), (*\textit{prev}*): (*\textit{Elem}*), (*\textit{p1}*): (*\textit{Elem}*), (*\textit{p2}*): (*\textit{Elem}*), (*\textit{ans}*): Int){
		iterate {
			if (*\textit{ans}*) = 0 then {
				Int cur := (*\textit{val}*) + (*\textit{p1.val}*) + (*\textit{p2.val}*);
				if ((*\textit{p2}*) = (*\textit{prev}*) || (*\textit{p1}*) = (*\textit{next}*)) then {
					(*\textit{ans}*) := -1;
				}
				if cur = 0 then {
					(*\textit{ans}*) = 1;
				}
				if (*\textit{ans}*) = 0 && cur > 0 then {
					(*\textit{p2}*) := (*\textit{p2.prev}*);
				}
				if (*\textit{ans}*) = 0 && cur < 0 then {
					(*\textit{p1}*) := (*\textit{p1.next}*);
				}
			}
		}
	}
	
	Fix(iterate)
\end{lstlisting}

\begin{figure}[p]
	\centering
	\includegraphics[width=0.7\textwidth]{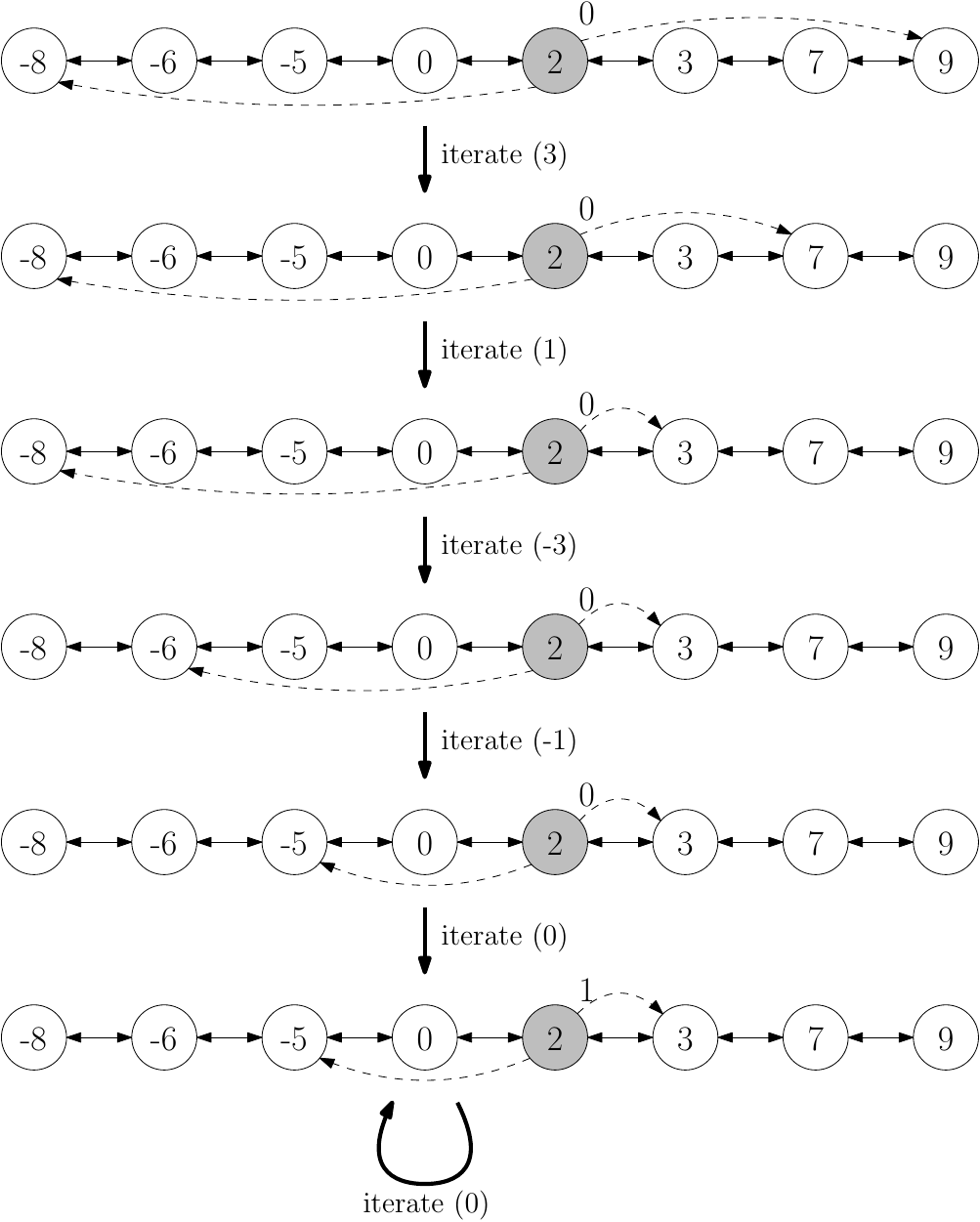}
	\caption{Execution of Listing~\ref{ex: 3SUM2} for a single element on a small list. The element considered is grey. The upper arrow points at \textit{p2}, and the lower arrow points at \textit{p1}. The value \textit{ans} has been displayed above and right of the element, and the computed value cur for an iterate step has been displayed next to the iterate steps.}
	\label{fig: 3SUM}
\end{figure}

The second program first computes the current value of the three saved elements. If the current value is $0$, a solution is found and $\mathit{ans}$ can be set to $1$. Otherwise, it will either move \textit{p1} to a node with a larger value or \textit{p2} to a node with a smaller value, depending on the current value. If \textit{p1} or \textit{p2} passes the current node, then either there is a solution with the current node as one of the side nodes or there is no solution with the current node. Either way, the node may stop checking for a solution. This is reflected in \textit{ans} being set to -$1$. At the end of the program, if there is a solution, then there will be at least one node per solution with \textit{ans} set to $1$, which is the middle node of the solution. We have depicted the fixpoint process for a single element in the list in the second program in Figure~\ref{fig: 3SUM}.

This program performs $O(n^2)$ work; every \textit{Elem} does a single (possibly incomplete) walk through the list. The span is $O(n)$, constrained by a walk through the list of an \textit{Elem}. The total running time is then $O(n)$.
\section{Properties of \Lname Programs}\label{section:props}
This section explores the behaviour of \Lname and \Lname programs by giving definitions and proving properties for \Lname.
The goal of this is to get an insight in what generic facts we can establish for this behaviour. We also use the properties established in this section in Section~\ref{section:proving} to prove two of our example algorithms correct.

We start out with some standard properties and definitions (Section~\ref{section:standardprop}). Then, we give a theorem on the properties of well-formed programs, which includes properties on the precise results of the execution of commands (Section~\ref{section:typeSafety}). From these, we can derive that the type system is \emph{safe} and we can state some corollaries that simplify reasoning about well-formed \Lname programs in proofs. 
\subsection{Standard Properties and Definitions}\label{section:standardprop}
In this section, we give some standard definitions and properties of \Lname. 
First, to simplify reasoning about semantic values and variables, we define \emph{reference notation} for semantic values, based on the dot notation as used in the syntax:
\begin{definition}[Reference Notation]
	Let $\Structs$ be a structure environment and let $x$ be a struct instance with label $\ell_x$ such that $\Structs(\ell_x) = \langle \StructType, \ComList, \Stack, \Env\rangle$. We define the notation $\ell_x.a^i$ inductively on $i$, with a variable $a$:
	\begin{enumerate}[nolistsep]
		\item $\ell_x.a^1 = \xi(a)$,
		\item $\ell_x.a^i = \Env'(a)$ for $i>1$, with $\ell_x.a^{i-1}\in\Labels$ and $\Structs(\ell_x.a^{i-1}) = \langle \StructType', \ComList', \Stack', \Env'\rangle$.
	\end{enumerate}
	We write $\ell_x.a$ for $\ell_x.a^1$ and $x.a$ for $\ell_x.a$, where $\Structs(\ell_x) = x$ and $a$ is a parameter of $x$.
\end{definition}
We then define the notions of \emph{executions} and \emph{accesses}:
\begin{definition}[\Lname Execution]\label{def: exe}
	For \Lname, an \emph{execution} from $P$ refers to a possibly infinite chain of semantic rule transitions that start in a state $P$. A \emph{finite execution} refers to a finite chain of semantic rule transitions that start in a state $P$ and ends in a state $P_1$. 
\end{definition}	
We extend this definition to fit our use of it:
As steps are finite by the definition of steps (see Sections~\ref{section:motivation} to \ref{section:semantics}), an execution of a step $F$ from $P$ refers to a chain of rule transitions from $P$ starting with an \textbf{Init}-transition for $F$ and ending in a state that contains a struct environment $\sigma$ for which $\mathit{Done}(\sigma)$ holds. A step $F$ can be executed from $P$ iff there exists an execution of $F$ from $P$.

As expressions and statements are finite, an execution of an expression $E$, a statement $S$ or a command list $C$ by some struct instance $s$ from $P$ refers to a chain of rule transitions from $P$ starting with the transition enabled by the first command of $\interp{E}$, $\interp{S}$ or $C$ for $s$ and ending with the transition enabled by the last command of, respectively, $\interp{E}$, $\interp{S}$ or $C$ for $s$. Additionally, for every command in $\interp{E}$, $\interp{S}$ or $C$, there needs to be a corresponding transition induced by that command taken by $s$. Note that this allows for interleaving with other statements or expressions; not all transitions need to correspond to a command of $\interp{E}$, $\interp{S}$ or $C$ in the command list of $s$. An expression $E$, a statement $S$ or a command list $C$ can be executed from $P$ by a struct instance $s$ iff there exists an execution of, respectively, $E$, $S$ or $C$ from $P$ for $s$.

An execution of a program $\Program$ refers to a possibly infinite chain of semantic rule transitions that start in the state $P_\Program^0$. A finite execution of a program $\Program$ starts in $P_\Program^0$ and ends when no transition is possible.
Lastly, we consider \emph{bounded} executions, where an execution of size $i$ from $P$ starts at $P$ and ends at some state $P_1$ reached after exactly $i$ transitions.

Recall that parameters and local variables are both considered variables. We then define the following:
\begin{definition}[\Lname Semantic Access]
	Let $x$ be a variable. We define every application of the semantic rules \textbf{ComRd} and \textbf{ComWr} with respect to $x$ to be an access of $x$. An application of \textbf{ComWrNSkip} for $x$ does not constitute an access. 
\end{definition}
We use this to introduce some other definitions and lemmas, starting with a notion of determinism. For this, note that we non-deterministically assign new labels to newly created struct instances. As the exact labels we assign does not matter as long as all labels are unique, which is enforced by the semantics, we can ignore the values of newly assigned labels for a notion of determinism. We use the following notions:

\begin{definition}[\Lname Determinism]
	Let $\Program$ be an \Lname program containing step $F$. Let $P$ be some state from which $F$ can be executed. Then $F$ is \emph{deterministic} for $P$ iff there exists exactly one state modulo newly assigned labels which is reached by executing $F$ from $P$.
	
	Additionally, a variable $x$ is deterministic when executing $F$ from $P$ iff in all states reached by executing $F$ from $P$ the value for $x$ is the same modulo newly assigned labels.
	Furthermore, $F$ is (fully) deterministic iff for all states $P$ that can execute $F$, $F$ is deterministic for $P$, and $x$ is (fully) deterministic if $x$ is deterministic when executing $F$ from all states that can execute $F$.
\end{definition}
From here on, we will leave out the `modulo newly assigned labels' for brevity; any mention made of something being deterministic from here on out means that it is deterministic modulo newly assigned labels.

\begin{definition}[Race Condition]
	Let $\Program$ be an \Lname program, let $F$ be a step in $\Program$ and let $P$ be a state in $\Program$ from which $F$ can be executed. Then $F$ executed from $P$ contains a \emph{race condition} for parameter $x$ iff there exist distinct struct instances $a$ and $b$ that both access a parameter $v$ in the same struct instance during an execution of $F$ from $P$ and at least one of the two instances writes to $v$. If both $a$ and $b$ write to $x$, this is a \emph{write-write race condition}, otherwise this is a \emph{read-write race condition}.
	If any step $F$ in the execution of $\Program$ has a race condition for some parameter, then $\Program$ has a race condition.
\end{definition}
Note that in the semantics, writes are atomic, which means that the result of a race condition still depends on the context and cannot be fully random (as when writes executing at the very same time on the same registers).

With the above definitions of race conditions and determinism, we are able to prove the following lemma:
\begin{lemma}[\Lname Determinism]\label{lem: det}
	A step $F$ from an \Lname program $\Program$ is deterministic for some state $P$ if $F$ does not contain a race condition from $P$.
\end{lemma}
\begin{proof}
	If $F$ is not deterministic for $P$, then there are multiple states which can result from executing $F$ from $P$ modulo newly assigned labels.
	W.l.o.g. we consider two executions $e_1$ and $e_2$ of $F$ from $P$ leading to states $P_1$ and $P_2$, which are different modulo newly assigned labels.
	If $F$ can be executed from $P$, according to the semantics, this means that either \textbf{InitG} or \textbf{InitL} is available for step $F$ from $P$. Let $P'$ be the state after calling $F$. As from $P'$, the schedule is not changed until after $F$ has executed (due to the definition of the \textbf{Com} rules in Section~\ref{section:semantics}), $P_1$ and $P_2$ must have the same schedule if they both executed $F$. Additionally, due to the definition of the rule \textbf{ComWr} and \textbf{ComCons} in Section~\ref{section:semantics}, the stability stack of $P_1$ and $P_2$ differ iff the struct environment of $P_1$ and $P_2$ differ. We therefore conclude that the difference of $P_1$ and $P_2$ is either that (w.l.o.g.) $P_1$ has a struct instance that $P_2$ does not have modulo newly assigned labels, or there exists at least one variable that has a different value in $P_1$ (w.l.o.g.) compared to its value in $P_2$.
	
	As $F$ has only one definition in $\Program$, it follows that any statement reachable in both executions is executed in the same manner by $e_1$ and $e_2$, and that any difference between the executions must stem from the values of variables read in the statements. Additionally, for any statement $S$ reachable in $e_1$ but not in $e_2$, $S$ must be in one or multiple embedded if-clauses which when executed lead to different results in $e_1$ and $e_2$, as there are no other statements possible in $s$ which allow for the skipping of statements. From this, we conclude that any difference between $P_1$ and $P_2$ must be caused by values of variables during the execution of statements. 
	
	Let $x$ be a variable which induces a difference between $P_1$ and $P_2$, and let that difference originate in some statement $S$, executed by struct instance $p$. If $S$ is executed in $e_1$ but not in $e_2$, then this recursively depends on some other variable/statement pair $(x', S')$, with $S'$ an if-statement, so we fix $S$ to be a statement executed by both $e_1$ and $e_2$. During the execution of $S$ by $p$, the only part of the statement $S$ that can have multiple possible results after execution are the references to other variables. If we follow the chain of dependency back, it follows that for any local variable on which $x$ is dependent, eventually we reach an initialization statement, which is dependent on earlier initialized local variables or parameters. It follows that the difference between $P_1$ and $P_2$ is dependent on parameter values. 
	
	As all parameters have only a single value in $P$, the difference must be dependent on a parameter $x$ changed during $F$. Then at some point this parameter value must be updated with an update statement with multiple possible results. As the sequentially consistent semantics do not allow $p$ to reorder its writes or reads in any way (as the \textbf{Com} rules of Section~\ref{section:semantics} exclusively operate on the first command of the command list), these multiple possible results cannot originate from the actions of only a single structure instance. It follows that $x$ must have been accessed by multiple struct instances during execution. As there is no method other than race conditions in the semantics which can make the interaction between these struct instances have multiple possible results, these struct instances must then have a race condition, which means $F$ contains a race condition for $x$ from $P$.
\end{proof}

We can adjust the focus from a whole step to a single parameter, leading to the following corollary:
\begin{corollary}[\Lname Parameter Determinism]\label{cor: det}
	A parameter $x$ is deterministic during the execution of a step $F$ (of an \Lname program $\Program$) from a state $P$ iff $F$ does not contain a race condition for $x$ from $P$. 
\end{corollary}
\begin{proof}
	This is true along the same lines as the proof for Lemma~\ref{lem: det}, except that we only consider differences in the value for $x$ in $P_1$ and $P_2$ and consider $x$ to be the variable used in the last two alineas of the proof.
\end{proof}

Finding race conditions can be difficult. To this end, we define \emph{potential race conditions}, which can be found on a syntax level. To do so, we first define the following:
\begin{definition}[Direct and indirect syntax access]
	Let $A$ be a parameter reference in the syntax, following $\type{Var}$. If $A$ is of the form $``a_1.\ldots.a_n.x"$, $A$ is an \emph{indirect} syntax access of $x$. If $A$ is of the form $``x"$, $A$ is an \emph{direct} syntax access of $x$. We consider a syntax access to be a \emph{write} access if it occurs in an update or variable assignment statement to the left of the assignment operator.
\end{definition}
Note that indirect syntax accesses can only be write syntax accesses in update statements. We use this to define \emph{potential data races}:
\begin{definition}[Potential Race Condition]\label{def: potdat}
	Let $\Program$ be an \Lname program and let $F$ be a step in $\Program$. Let $a$ and $b$ be a pair of struct types for which $F$ is defined, with $a = b$ being allowed. Let there be a pair of statements $S_1$ in $a$ and $S_2$ in $b$ which include syntax accesses $A$ and $B$ of some parameter $x$. Let either $A$ or $B$ be indirect and let either $A$ or $B$ be a syntax write access to $x$. Then the pair $(S_1, S_2)$ is a \emph{potential race condition} on $x$.
\end{definition}
These can be used to consider which parameters can possibly be in a data race:
\begin{lemma}\label{lem: nopot}
	Let $\Program$ be an \Lname program and let $F$ be a step in $\Program$. If $F$ does not have any potential race conditions on parameter $x$, $F$ does not have any race conditions on parameter $x$.
\end{lemma}
\begin{proof}
	If $F$ does not have any potential race conditions, there are no pairs of statements $(S_1, S_2)$ that can be taken from any executions of $F$ in any structs s.t. $S_1$ and $S_2$ are executed by different struct instances but access $x$ in the same struct instance of which at least one access is a write access. Then it follows that $F$ cannot have any race conditions on $x$.
\end{proof}
We also define a notion of \emph{instability}:

\begin{definition}[Instability]\label{def:ins}
	Let $\Program$ be an \Lname program a step $F$. Let $P$ be a state of $\Program$ from which $F$ can be executed, and let $p$ be a struct instance in $P$. A parameter $p.x$ with value $v$ before the execution of $F$ is \emph{unstable} after the execution of $F$ from $P$, denoted as $p.x^{*}$, iff a write command is executed during t that sets $p.x$ to a value $a \neq v$.  
\end{definition}

Lastly, it is important to note that in this paper, we are interested in proving programs that do not contain read-write race conditions. However, the programs given in this paper are partial programs, which do not contain an initialization. It follows that we need a better definition for what we consider a program without read-write race conditions:
\begin{definition}[RW Race Condition Free from Premise State]
	We consider an \Lname program $\Program$ to be without read-write race conditions from some premise state $P_1$ iff any execution of $\Program$ starting in $P_1$ does not contain a read-write race condition.
\end{definition}

\subsection{Properties of Well-Formed Programs}\label{section:typeSafety}
In this section, we analyse the properties of well-formed programs. We first introduce the notion of a well-formed \Lname state:
\begin{definition}[Well-Formed \Lname State]
	Let $\Program$ be any well-formed \Lname program and let $P$ be an \Lname state reachable by some execution of $\Program$. Then $P$ is a well-formed \Lname state of $\Program$.
\end{definition}

The goal of this section is to establish that the properties in the following theorem hold for well-formed \Lname state. Note that a schedule $\Sched$ is well-formed modulo $\mathit{aFix}$ if replacing all occurrences of $\mathit{aFix}$ in $\Sched$ with $\mathit{Fix}$ yields a well-formed schedule $\Sched$. We denote well-formedness modulo $\mathit{aFix}$ as $\Gamma, \Omega \underset{\mathit{aF}}{\tr} \Sched$ for some environments $\Gamma$ and $\Omega$ and schedule $\Sched$. Additionally, let $\mathit{sem}:\SynTypes \times\Program \rarr \SemTypes \cup\{\bot\}$ be defined as:
\begin{align*}
	&\mathit{sem}(\texttt{Nat}, \Program) = \Nat,\\
	&\mathit{sem}(\texttt{Int}, \Program) = \Int,\\
	&\mathit{sem}(\texttt{Bool}, \Program) = \Bool,\\
	&\mathit{sem}(\texttt{String}, \Program) = \String,\\
	&\mathit{sem}(\StructType, \Program) = \StructType \text{ if } \StructType\in\StructTypes_\Program\\
	&\mathit{sem}(\StructType, \Program) = \bot \text{ if } \StructType\notin\StructTypes_\Program
\end{align*}
\begin{theorem}[Properties of Well-Formed \Lname States]\label{thm: wellformed}
	Let $P = \langle\Sched, \Structs, \Stab\rangle$ be a well-formed \Lname state reached by some program $\Program = D\ \Sched_\Program$ s.t. $\Gamma, \Omega\tr D\ Sc_\Program$ for some $\Gamma, \Omega$. Then all of the following hold:
	\begin{enumerate}
		\item \label{prop1} Either $\Sched = \vempty$ or $\Gamma;\mathit{Steps}(D), \Omega\underset{\mathit{aF}}{\tr} \Sched$.
		\item \label{prop2} If $\mathit{Done}(\Structs)$ holds and $\Sched\neq \vempty$, there exists at least one schedule transition that can be taken from $P$ with definitions $D$. 
		\item \label{prop3} If $\mathit{Done}(\Structs)$ holds and $\Sched\neq \vempty$, then after a finite amount of transitions from $P$ either the schedule is empty or there must be an \textbf{InitG} or \textbf{InitL} transition to some state $P' = \langle\Sched', \Structs', \Stab'\rangle$ for which $\mathit{Done}(\Structs')$ does not hold.
		\item \label{prop4} If $\neg \mathit{Done}(\Structs)$, let $\ell\in\Labels$ s.t. $\Structs(\ell) = \langle\StructType, \ComList, \Stack, \Env\rangle$ and $\ComList\neq \vempty$. Let $F$ be the step currently executed by $\Structs(\ell)$. Let $S_i;\ldots;S_n$ be the statements s.t. $\interp{S_i} = c_1;\cdots;c_m$ and $\ComList = c_j;\cdots;c_m;\interp{S_{i+1};\cdots;S_n}$. Let $S_1;\ldots;S_{i-1}$ be the statements that have been executed by $\Structs(\ell)$ during the current execution of $F$. Then there exists a $c_k$ with $k\leq j$ s.t. either $c_k;\ldots;c_j = \interp{E}$ for some expression $E$ or $c_k;\ldots;c_j = \interp{S_i}$. 
		\item \label{prop5} If $\neg \mathit{Done}(\Structs)$, let $\ell\in\Labels$ s.t. $\Structs(\ell) = \langle\StructType, \ComList, \Stack, \Env\rangle$ and $\ComList\neq \vempty$. Let $F$ be the step currently executed by $\Structs(\ell)$. Let $S_i;\ldots;S_n$ be the statements s.t. $\interp{S_i} = c_1;\cdots;c_m$ and $\ComList = c_j;\cdots;c_m;\interp{S_{i+1};\cdots;S_n}$. Let $S_1;\ldots;S_{i-1}$ be the statements that have been executed by $\Structs(\ell)$ during the current execution of $F$.
		Then, with $c_k$ with $k\leq j$ s.t. $c_k;\ldots;c_j = \interp{E}$ or $c_k;\ldots;c_j = \interp{S_i}$, let $P_k = \langle \Sched_k, \Structs_k, \Stab_k\rangle$ be the state right before the transition induced by $c_k$ and let $\Structs_k(\ell) = \langle \StructType, \ComList_k, \Stack_k;\Env_k\rangle$. Then one of the following holds:
		\begin{enumerate}
			\item $c_j = \push{v}$ for some $v\in\Values$ or $c_j = \push{\this}$,
			\item $c_j = \readg{x}$ and all of the following hold:
			\begin{enumerate}
				\item $\Stack = \Stack_k;\ell'$ and $\Structs(\ell') =\langle \StructType_{\ell'}, \ComList_{\ell'}, \Stack_{\ell'}, \Env_{\ell'}\rangle$ for some $\StructType_{\ell'}, \ComList_{\ell'}, \Stack_{\ell'}, \Env_{\ell'}$ and $(x\notin\Par{\StructType_{\ell'}, \Program} \Rarr \ell' = \ell)$,
				\item either $S_q = (T\ x := E)$ for some $1\leq q < i$, $T\in\SynTypes$ and $E\in\Expr$ or $(x: T) \in \Par{\StructType_{\ell'}, \Program}$ for some $T\in \SynTypes$,
				\item either $T\in\StructTypes_\Program$ and $\Env(x)\in\Labels \land \Structs(\Env_{\ell'}(x)) = \langle T, \ComList_x, \Stack_x, \Env_x\rangle$ (for some $\ComList_x, \Stack_x$ and $\Env_x$) or $T \in\SynTypes\setminus\Id$ and $\Env'(x)\in \mathit{sem}(T)$.
			\end{enumerate} 
			\item $c_j = \writev{x}$ and all of the following hold:
			\begin{enumerate}
				\item $\Stack = \Stack_k;v;\ell' \land \Structs(\ell')=\langle \StructType_{\ell'}, \ComList_{\ell'}, \Stack_{\ell'}, \Env_{\ell'}\rangle$ for some $\StructType_{\ell'}, \ComList_{\ell'}, \Stack_{\ell'}, \Env_{\ell'}$ and $(x\notin\Par{\StructType, \Program} \Rarr \ell' = \ell)$,
				\item either $S_q = (T\ x := E)$ for some $1\leq q \leq i$, $T\in\SynTypes$ and $E\in\Expr$ or $(x: T) \in \Par{\StructType', \Program}$ for some $T\in \SynTypes$,
				\item either $T\in\StructTypes_\Program$ and $v\in\Labels \land \Structs(v) = \langle T, \ComList'', \Stack'', \Env''\rangle$ (for some $\ComList'', \Stack''$ and $\Env''$) or $T\in\SynTypes\setminus\Id$ and $v\in \mathit{sem}(T)$.
			\end{enumerate}
			\item $c_j = \cons{\StructType}$, with $\StructType\in \StructTypes_\Program\land \Par{\StructType, \Program} = p_1: T_1;\cdots;p_c: T_c$ for some $c$, and $\Stack = \Stack_k;v_1;\cdots;v_c$ and for all $p_h: T_h$, either $T_h\in \{\texttt{Nat}, \texttt{Int}, \texttt{Bool}, \texttt{String}\}$ and $v_h\in\mathit{sem}(T_h)$ or $T_h\in\StructTypes_\Program$ and $v_h\in\Labels \land \Structs(v_h) = \langle T_h, \ComList'', \Stack'', \Env''\rangle$.
			\item $c_j = \Ifc{C}$ and $\Stack = \Stack_k;b$ s.t. $b\in\mathbb{B}$.
			\item $c_j = \Notc$ and $\Stack = \Stack_k;b$ s.t. $b\in\mathbb{B}$.
			\item $c_j = \Operator(\circ)$ and $\Stack = \Stack_k;a;b$ and all of the following hold:
			\begin{enumerate}
				\item If $\circ\in\{=, \text{!=}\}$ then $a, b\in\{\Nat, \Int\}$ or there exists a $T\in\{\Nat, \Int, \Bool, \String\}\cup \StructTypes_\Program$ s.t. either $T\notin \StructTypes_\Program$ and $a, b \in T$ or $T\in\StructTypes_\Program$ and $\Structs(a) = \langle T, \ComList_a, \Stack_a, \Env_a\rangle$ and  $\Structs(b) = \langle T, \ComList_b, \Stack_b, \Env_b\rangle$,
				\item If $\circ\in\{< =, >=, <, >, =, \text{!=}, *, /, \%, +, \text{\textasciicircum}, -\}$ then $a, b\in\{\Nat, \Int\}$,
				\item If $\circ\in\{\&\&, ||\}$ then $a, b\in\Bool$.
			\end{enumerate}
		\end{enumerate}
		Additionally, there exists a transition induced by $c_j$ from $P$ to a state $P' = \langle \Sched', \Structs', \Stab'\rangle$ s.t. $\Structs'(\ell) = \langle\StructType, \ComList', \Stack', \Env'\rangle$ and all of the following hold:
		\begin{enumerate}
			\item If $c_j = \push{v}$ for some $v\in\Values$, then $\Stack' = \Stack_k;v$ and $\ComList' = c_{j+1};\ldots;c_m;\interp{S_{i+1};\cdots;S_n}$.
			\item If $c_j = \push{\this}$, then $\Stack' = \Stack_k;\ell$ and $\ComList' = c_{j+1};\ldots;c_m;\interp{S_{i+1};\cdots;S_n}$.
			\item If $c_j = \readg{x}$, with $\Stack = \Stack_k;\ell'$ and $\Structs(\ell') =\langle \StructType_{\ell'}, \ComList_{\ell'}, \Stack_{\ell'}, \Env_{\ell'}\rangle$ for some $\StructType_{\ell'}, \ComList_{\ell'}, \Stack_{\ell'}, \Env_{\ell'}$, then $\Stack' = \Stack_k;\Env'(x)$ and $\ComList' = c_{j+1};\ldots;c_m;\interp{S_{i+1};\cdots;S_n}$.
			\item If $c_j = \writev{x}$, then, with $\Stack = \Stack_k;v;\ell'$, $\Structs'(\ell') = \langle\StructType_{\ell'}, \ComList'_{\ell'}, \Stack'_{\ell'}, \Env'_{\ell'}\rangle$ s.t. if $\ell'\notin\Labels_0$, $\Env'_{\ell'} = \Env_{\ell'}[x\mapsto v]$ and $x\notin\Par{\StructType_{\ell'}, \Program} \lor \Env_{\ell'}(x)\neq v$ then $\Stab'= \false^{|\Stab|}$, and if $\ell'\in\Labels_0$, $\Env'_{\ell'} = \Env_{\ell'}$ and $\Stab' = \Stab$. Additionally, $\Stack' = \Stack_k$, $\ComList' = \interp{S_{i+1};\cdots;S_n}$.
			\item If $c_j = \cons{\StructType}$, then, with $\Stack = \Stack_k;v_1;\cdots;v_c$ and $\Par{\StructType, \Program}= p_1: T_n;...;p_c: T_c$, $\Stack'=\Stack_k;\ell'$ for some $\ell'$ s.t. $\Structs(\ell') = \bot$, $\Structs'(\ell') = \langle \StructType, \vempty, \vempty, \Env_{\StructType}^0[\{p_1\mapsto v_1, \ldots, p_c \mapsto v_c\}]\rangle$, $\ComList' = c_{j+1};\ldots;c_m;\interp{S_{i+1};\cdots;S_n}$ and $\Stab'= \false^{|\Stab|}$.
			\item If $c_j = \Ifc{C}$, with $\Stack = \Stack_k;b$, $\Stack' = \Stack_k$ and if $b = \true$, $\ComList' = C;\interp{S_{i+1};\cdots;S_n}$ and if $b = \false$, $\ComList' = \interp{S_{i+1};\cdots;S_n}$.
			\item If $c_j = \Notc$, with $\Stack = \Stack_k;b$, $\Stack' = \Stack_k;\neg b$ and $\ComList' = c_{j+1};\ldots;c_m;\interp{S_{i+1};\cdots;S_n}$.
			\item If $c_j = \Operator(\circ)$, with $\Stack = \Stack_k;a;b$, then  $\ComList' = c_{j+1};\ldots;c_m;\interp{S_{i+1};\cdots;S_n}$ and, with $o$ the semantic equivalent of $\circ$, $\Stack' = \Stack_k;c$ for some $c$ s.t. all of the following hold:
			\begin{enumerate}
				\item If $a\in T$ and $b\in T$ with $T\in \SemTypes$ and $\circ = \{=,\text{!=}\}$, then $c = a \mathop{o} b$ and $c\in\mathbb{B}$,
				\item If $a\in T_1$ and $b\in T_2$, with $T_1, T_2\in\{\mathbb{N}, \mathbb{Z}\}$ and $\circ = \{< =, >=, <, >, =, \text{!=}\}$, then $c = a\mathop{o} b$ and $c\in\mathbb{B}$,
				\item If $a\in T_1$ and $b\in T_2$, with $T_1, T_2\in\{\mathbb{N}, \mathbb{Z}\}$ and $\circ = \{*, /, \%, +, \text{\textasciicircum}, -\}$, then $c = a\mathop{o} b$ and $c\in\mathbb{Z}$,
				\item If $a, b\in \mathbb{N}$ and $\circ = \{*, /, \%, +, \text{\textasciicircum}\}$, then $c = a\mathop{o} b$ and $c\in\mathbb{N}$.
			\end{enumerate}
		\end{enumerate}
		\item\label{prop6} If $\mathit{Done}(\Structs)$ does not hold in $P$, then after a finite amount of transitions there must be a transition to some state $P' = \langle\Sched', \Structs', \Stab'\rangle$ for which $\mathit{Done}(\Structs')$ holds.
	\end{enumerate}
\end{theorem}
We prove Properties~\ref{prop1}, \ref{prop2}, \ref{prop3}, \ref{prop4}, \ref{prop5} and \ref{prop6} separately. Every proof is dependent on different parts of the semantics of $\Program$ as defined in Section~\ref{section:semantics}. All parts of the semantics of $\Program$ are used in at least one of the proofs.
\begin{proof}[Proof of Property~\ref{prop1}]
	We prove Property~\ref{prop1} by induction on the states traversed in any execution of $\Program$. Note that the schedule of $P_\Program^0$ is well-formed, as it is the schedule of $\Program$, and $\Program$ is well-formed. This proves the base case.
	
	Then, let $P_i = \langle \Sched, \Structs, \Stab\rangle$ be the $i$th state traversed in some execution of $\Program$ with definitions $D$, from which some transition is taken to a state $P_{i+1} = \langle \Sched', \Structs', \Stab'\rangle$. By induction hypothesis, we assume that $\Gamma;\mathit{Steps}(D), \Omega\underset{\mathit{aF}}{\tr}\Sched$ or that $\Sched = \vempty$. Then, we do a case distinction:
	\begin{itemize}
		\item If the transition taken is a command transition, $\Sched' = \Sched$ so $\Gamma;\mathit{Steps}(D), \Omega\underset{\mathit{aF}}{\tr}\Sched'$ or $\Sched'= \vempty$.
		\item If the transition is taken is a schedule transition, we know that $\Sched \neq \vempty$, as every schedule transition requires something in the schedule. 
		
		As $\Sched$ is well-formed, we know that it has a well-defined first element, s.t. $\Sched = \mathit{sc}$ or $\Sched = \mathit{sc} < \mathit{Sched}_1$.
		
		Then, if $\Sched = \mathit{sc}$, we know that $\Gamma,\Omega\underset{\mathit{aF}}{\tr} \mathit{sc}$, as $\Sched$ is well-formed modulo $\mathit{aFix}$. Then, we have a case distinction on the form of $\mathit{sc}$:
		\begin{enumerate}
			\item If $\mathit{sc} = F$ for some step $F$ or $\mathit{sc} = \StructType.F$ for some struct type $\StructType$ and some step $F$, then the schedule transition that is taken is \textit{InitG} or \textit{InitL} respectively. These will remove $\mathit{sc}$ from the schedule.
			\item If $\mathit{sc} = \mathit{Fix}(\mathit{sc}_1)$ for some schedule $\mathit{sc}_1$, the transition must be \textit{FixInit}, resulting in $\mathit{sc}' = \mathit{sc}_1 < \mathit{aFix}(\mathit{sc}_1)$ which replaces $\mathit{sc}$. As $\Gamma;\mathit{Steps}(D), \Omega\tr \mathit{Fix}(\mathit{sc}_1)$, it holds that $\Gamma;\mathit{Steps}(D), \Omega\underset{\mathit{aF}}{\tr} \mathit{sc}_1$. It follows that $\Gamma;\mathit{Steps}(D), \Omega\underset{\mathit{aF}}{\tr} \mathit{sc}_1<\mathit{aFix}(\mathit{sc}_1)$.
			\item If $\mathit{sc} = \mathit{aFix}(\mathit{sc}_1)$ for some schedule $\mathit{sc}_1$, the transition must be either \textit{FixIter} or \textit{FixTerm}. If the transition is \textit{FixIter}, then $\mathit{sc}' = \mathit{sc}_1 < \mathit{aFix}(\mathit{sc})$ replaces $\mathit{sc}$. As $\Gamma;\mathit{Steps}(D), \Omega\tr \mathit{aFix}(\mathit{sc}_1)$, $\Gamma;\mathit{Steps}(D), \Omega\underset{\mathit{aF}}{\tr} \mathit{sc}_1$. It follows that $\Gamma;\mathit{Steps}(D), \Omega\underset{\mathit{aF}}{\tr} \mathit{sc}_1<\mathit{aFix}(\mathit{sc}_1)$.
			If the transition is \textbf{FixTerm}, $\mathit{sc}$ is removed from the schedule.
		\end{enumerate}
		From this, we can conclude that either $\Sched' = \vempty$ or $\Gamma;\mathit{Steps}(D), \Omega\underset{\mathit{aF}}{\tr} \Sched'$, so the induction step holds.
		
		Then if $\Sched = \mathit{sc} < \Sched_1$, with $\mathit{sc}$ being a single schedule element, note that $\Gamma;\mathit{Steps}(D), \Omega\underset{\mathit{aF}}{\tr} \Sched$ iff $\Gamma;\mathit{Steps}(D), \Omega\tr \Sched_1$ and $\Gamma;\mathit{Steps}(D), \Omega\underset{\mathit{aF}}{\tr} \mathit{sc}$. Any schedule transition only affects the first schedule element, so after a single transition, $\Sched_1$ will remain unchanged. This means that we can consider $\mathit{sc}$ as a separate schedule. We then apply same case distinction as above, from which we can conclude that either the first element is removed after a transition or the transition results in a schedule $\Sched''$ s.t. $\Gamma;\mathit{Steps}(D), \Omega\underset{\mathit{aF}}{\tr} \Sched''$. In the first case, $\Sched' = \Sched_1$ and $\Gamma, \Omega\underset{\mathit{aF}}{\tr} \Sched'$. In the second case, we know that $\Gamma;\mathit{Steps}(D), \Omega\underset{\mathit{aF}}{\tr}\Sched''$ and $\Gamma;\mathit{Steps}(D), \Omega\underset{\mathit{aF}}{\tr} \Sched_1$, so we know that $\Gamma;\mathit{Steps}(D), \Omega\underset{\mathit{aF}}{\tr} \Sched'$. Therefore, in this case, the induction step holds.
	\end{itemize}
	It follows that the induction step holds, so Property~\ref{prop1} also holds.
\end{proof}
\begin{proof}[Proof of Property~\ref{prop2}]
	Let $P = \langle \Sched, \Structs, \Stab\rangle$ be a state reached in an execution of $\Program$ s.t. $\mathit{Done}(\Structs) = \true$. Assume that $\Sched \neq \vempty$. It follows from Property~\ref{prop1} that $\Sched$ is well-formed modulo $\mathit{aFix}$, so we know that $\Sched$ has a well-formed and well-defined first schedule element $\mathit{sc}$. Then, we do a case distinction on what this element is:
	\begin{enumerate}
		\item If $\mathit{sc} = F$ for some step $F$, then it follows that \textbf{InitG} is enabled: $\mathit{Done}(\Structs)$ holds, and beyond that, the only requirement is that $\mathit{sc} = F$. Same holds for $\mathit{sc} = \StructType.F$ and \textbf{InitL}.
		\item If $\mathit{sc} = \mathit{Fix}(\mathit{sc}_1)$ for some schedule $\mathit{sc}_1$, the transition \textbf{FixInit} can be taken, as $\mathit{Done}(\Structs)$ holds and $\mathit{sc}$ fits the pattern in the schedule which is the other requirement.
		\item If $\mathit{sc} = \mathit{aFix}(\mathit{sc}_1)$ for some schedule $\mathit{sc}_1$, then whether a transition can be taken depends on whether there is a value on top of the stability stack. This is, however, the case:		
		The only rule which introduces a pattern which includes $\mathit{aFix}$ in the semantics is the rule \textbf{FixInit}, which also adds a value on top of the stability stack. The only rule which removes a value of from the stability stack is \textbf{FixTerm}, which also removes an $\mathit{aFix}$ pattern. It follows that the amount of values in the stability stack corresponds exactly to the amount of $\mathit{aFix}$ patterns. It follows that there is a value on the stability stack when a $\mathit{aFix}$ pattern is encountered.
		Then, either this value is $\true$ or $\false$. If the value is $\true$, note that $\mathit{Done}(\Structs)$ holds and that an $\mathit{aFix}$ pattern is the topmost schedule element, so \textbf{FixTerm} is enabled. Along the same lines, if the value is $\false$, \textbf{FixIter} is enabled.
	\end{enumerate}
	It follows that if the schedule is not empty and $\mathit{Done}(\Structs)$ holds, there must be a well-formed first schedule element and for all forms that element can take there is an enabled transition.
\end{proof}
\begin{proof}[Proof of Property~\ref{prop3}]	
	We prove Property~\ref{prop3} by first proving through induction that after any sequence of transitions from $P$ either the schedule is empty or there must be an \textbf{InitG} or \textbf{InitL} transition. We then prove that after this \textbf{InitG} or \textbf{InitL} transition to, w.l.o.g., state $P'$, $\mathit{Done}(\Structs')$ does not hold. Note then that if either \textbf{InitG} or \textbf{InitL} are enabled, by definition of the schedule and the transitions, no other transitions are enabled, so the \textbf{InitL} or \textbf{InitG} transition must also be taken. We therefore make no distinction between the an \textbf{InitG} or \textbf{InitL} transition being enabled and being taken.
	
	We first do induction on the structure of the schedule. As base case, we take that $\Sched = \vempty$, which directly satisfies our property. We then assume as induction hypothesis 1 that for all but the first element of the schedule, any sequence of schedule transitions eventually leads to either the schedule being empty or an \textbf{InitL} or \textbf{InitG} transition being taken. 
	
	Let $\mathit{sc}$ be the first element of $\Sched$, and assume that \textbf{InitG} and \textbf{InitL} are not enabled. It follows that $\mathit{sc} = \mathit{aFix}(\mathit{sc}_1)$ or $\mathit{sc} = \mathit{Fix}(\mathit{sc}_1)$. Let the next transition taken be $t$, from $P$ to $P' = \Sched', \Structs', \Stab'$. This transition exists, by Property~\ref{prop2}.
	We then do induction on the amount of occurrences $k$ of both $\mathit{Fix}$ and $\mathit{aFix}$. If $k = 1$, It follows that $\mathit{sc}_1$ is either a step call, a typed step call or a sequence of step and typed step calls, as $\Sched$ is well-formed modulo $\mathit{aFix}$. Then if $t = \text{\textbf{FixInit}}$, (so $\mathit{sc} = \mathit{Fix}(\mathit{sc}_1)$), $\Sched'$ will start with the sequence $\mathit{sc}_1 < \mathit{aFix}(\mathit{sc}_1)$, and as $\mathit{sc}_1$ must have a step call as first element, either \textbf{InitL} or \textbf{InitG} must be enabled. The same argument holds if $t = \text{\textbf{FixIter}}$, but with $\mathit{sc} = \mathit{aFix}(\mathit{sc}_1)$. If $t = \text{\textbf{FixTerm}}$, so $\mathit{sc} = \mathit{aFix}(\mathit{sc}_1)$, $\mathit{sc}$ is removed. Then either $\Sched' = \vempty$, or we use induction hypothesis 1 to conclude that any sequence of schedule transitions eventually leads to either the schedule being empty or an \textbf{InitL} or \textbf{InitG} transition being taken. 
	
	We then assume as induction hypothesis 2 that if $k = i-1$, any sequence of schedule transitions eventually leads to either the schedule being empty or an \textbf{InitL} or \textbf{InitG} transition being taken. Let $k = i$. Then if $t = \text{\textbf{FixInit}}$, (so $\mathit{sc} = \mathit{Fix}(\mathit{sc}_1)$), $\Sched'$ will start with the sequence $\mathit{sc}_1 < \mathit{aFix}(\mathit{sc}_1)$, and as the amount of occurrences of both $\mathit{Fix}$ and $\mathit{aFix}$ in $\mathit{sc}_1$ must then be $i-1$, by induction hypothesis 2, any sequence of schedule transitions eventually leads to either the schedule being empty or an \textbf{InitL} or \textbf{InitG} transition being taken. The same argument holds if $t = \text{\textbf{FixIter}}$, but with $\mathit{sc} = \mathit{aFix}(\mathit{sc}_1)$. If $t = \text{\textbf{FixTerm}}$, so $\mathit{sc} = \mathit{aFix}(\mathit{sc}_1)$, $\mathit{sc}$ is removed. Then again either $\Sched' = \vempty$, or we use induction hypothesis 1 to conclude that any sequence of schedule transitions eventually leads to either the schedule being empty or an \textbf{InitL} or \textbf{InitG} transition being taken.
	
	It follows that after a finite amount of transitions from $P$ either the schedule is empty or there must be an \textbf{InitG} or \textbf{InitL} transition to some state $P' = \langle\Sched', \Structs', \Stab'\rangle$. By the definition of the schedule transitions, when w.l.o.g. transition \textbf{InitG} is taken for w.l.o.g. some step $F$, by definition of \textbf{InitG}, the commands of $F$ are put into the command list of every struct instance of a struct type for which $F$ is defined. As $\Sched$ is well-formed modulo $\mathit{aFix}$, there is at least one struct type for which $F$ is defined, and as there always exists a null instance for every struct type, there is then at least one struct instance which will have commands in their command list. It follows that $\neg \mathit{Done}(\Structs')$.
\end{proof}
\begin{proof}[Proof of Property~\ref{prop4}]	
	To prove the lemma, we do a case distinction on which command $c_j$ is and choose an appropriate $c_k$:
	\begin{enumerate}
		\item If $c_j = \push{\val{g}}$, it has been generated by the interpretation function in response to either the expression $g$, with $g\in\Literals$, or, if $g =\defaultVal(T)$ for some $T$, to $\nil_T$. It follows that with $c_k = c_j$, the lemma holds. Analogously, the lemma holds for if $c_j = \push{\this}$, which is generated for the expression \texttt{this}.
		\item If $c_j = \readg{x_i}$ for some variable $x_i$, then we know that $c_j$ is generated by the interpretation function in response to some variable expression $x_1.\cdots.x_n$, with $x_i \in x_1.\cdots.x_n$. Let $c_k$ be the first command of $\interp{x_1.\cdots.x_n}$. Then $c_k;\ldots;c_j = \interp{x_1.\cdots.x_i}$ and the lemma holds.
		\item If $c_j$ is a \textbf{wr}, \textbf{cons}, \textbf{if}, \textbf{not} or \textbf{op} command, it is automatically the last generated command for some expression or statement as per the definition of the interpretation function. The lemma will then hold for $c_k$ being the first command generated for the same expression or statement.
	\end{enumerate}
	As in all cases, a $c_k$ can be chosen, the lemma holds.
\end{proof}
From these lemmas, it follows that for all well-formed states of a well-formed program $\Program$, the first four properties of Theorem~\ref{thm: wellformed} hold. To prove Property~\ref{prop5}, we first prove an auxiliary lemma, which proves that if we encounter a label, we know it has a defined struct instance. This lemma depends on the fact that writes are atomic in our semantics and that struct instances cannot be deleted.
\begin{lemma}\label{lem: labelsource}
	Let $\Program$ be an \Lname program. Let $P = \langle \Sched, \Structs, \Stab\rangle$ be a state encountered during some execution of $\Program$ s.t. there exists a label $\ell$ s.t. $\Structs(\ell) =\langle \StructType, \ComList, \Stack;\ell', \Env\rangle$ for some $\StructType$, $\ComList$, $\Stack$ and $\Env$. Then $\Structs(\ell')\neq \bot$.
\end{lemma}
\begin{proof}
	We use the definitions of the lemma. If $\ell'\in\Labels_0$, then the lemma holds by the definition of $\Labels_0$ and null instances, as instances cannot be deleted in our semantics. Therefore, assume $\ell'\notin\Labels_0$. Then $\ell'$ must have been introduced to the pool of labels after the initial state, so $\ell'$ must have been introduced by some semantic transition. The only transition which can introduce new labels is \textbf{ComCons}, which also makes sure that there is a struct instance for the new label. It follows that in both cases, the lemma holds.
\end{proof}
We then prove Property~\ref{prop5} and \ref{prop6}:

\begin{proof}[Proof of Property~\ref{prop5}]
	Let $P = \langle\Sched, \Structs, \Stab\rangle$ be a well-formed \Lname state reached by some program $\Program = D\ \Sched_\Program$ s.t. $\Gamma, \Omega\tr D\ Sc_\Program$ for some $\Gamma, \Omega$. Then assume that $\neg \mathit{Done}(\Structs)$. Let $\ell\in\Labels$ be a label s.t. $\Structs(\ell) = \langle\StructType, \ComList, \Stack, \Env\rangle$ and $\ComList\neq \vempty$. Let $F$ be the step currently executed by $\Structs(\ell)$, with statements $S_1;\cdots; S_n$. Let $S_i$ be the statement s.t. $\interp{S_i} = c_1;\cdots;c_m$ and $\ComList = c_j;\cdots;c_m;\interp{S_{i+1};\cdots;S_n}$.
	Then, let $c_k$ with $k\leq j$ s.t. $c_k;\ldots;c_j = \interp{E}$ or $c_k;\ldots;c_j = \interp{S_i}$. Note that according to Property~\ref{prop4}, this $c_k$ is guaranteed to exist, so we assume $c_k$ has the largest $k$ possible for $c_j$. Let $P_k = \langle \Sched_k, \Structs_k, \Stab_k\rangle$ be the state right before the transition induced by $c_k$ and let $\Structs_k(\ell) = \langle \StructType, \ComList_k, \Stack_k;\Env_k\rangle$.
	
	To prove Property~\ref{prop5}, we do strong induction over the distance $d$ between $c_k$ and $c_j$, with $d = j-k$:
	\begin{enumerate}[leftmargin=7ex]
		\item[$d = 0$] Then it follows that $c_j = c_k$. This means that $c_j$ is both the first and the last command of some expression or statement. By the definition of the interpretation function, this can only happen for the expressions $g$ with $g\in\Literals$, $\this$, $\nil_T$ and $x_1.\cdots.x_0$. In all of those cases, $c_j = \push{v}$ for some value $v\in \Values$ or $c_j = \push{this}$. This satisfies the first part of Property $\ref{prop5}$. Then, if $c_j = \push{v}$ for some $v\in\Values$, a transition \textbf{ComPush} will be enabled to a state $P' = \langle \Sched', \Structs', \Stab'\rangle$ with $\Structs'(\ell) = \langle\StructType, c_{j+1};\ldots;c_m;\interp{S_{i+1};\cdots;S_n};\ComList', \Stack;v, \Env'\rangle$. As $\Stack = \Stack_k$, this satisfies the property. If $c_j = \push{\this}$, then a transition \textbf{ComPushThis} will be enabled to a state $P' = \langle \Sched', \Structs', \Stab'\rangle$ with $\Structs'(\ell) = \langle\StructType, c_{j+1};\ldots;c_m;\interp{S_{i+1};\cdots;S_n}, \Stack;\ell, \Env'\rangle$. As $\Stack = \Stack_k$, this satisfies the property.
		\item[$d > 0$] As induction hypothesis, we assume that the property holds for any command $c_j'$ with a corresponding command $c_k'$ s.t. the distance $d'$ for $c_j'$ and $c_k'$ is smaller than $d$. As $c_k$ has the largest possible $k$ s.t. $c_k;\ldots;c_j = \interp{E}$ or $c_k;\ldots;c_j = \interp{S_i}$ and $d>0$, it follows that $c_j$ is not a \textbf{push} command; if $c_j$ was a push command, it must have been introduced as the first command of an expression $g$ with $g\in\Literals$, $\this$, $\nil_T$ or $x_1.\cdots.x_n$ (by the definition of the interpretation function), so either $d = 0$ or there exists a larger $k$ s.t. $c_k;\ldots;c_j$ is the list of commands generated from an expression. We then do a case distinction on the commands $c_j$ can be:
		\begin{itemize}
			\item[\textbf{rd}] If $c_j = \readg{x}$ for some $x$, then $c_k = \push{\this}$ and the expression $E$ is a variable expression $x_1.\cdots.x$, as seen in the proof for Property~\ref{prop4}. As $\readg{x}$ was generated by the interpretation function, it follows that there exists a variable $x_h$ with $h \geq 0$ s.t. $E = x_1.\cdots.x_h.x$. It follows that $E = E'.x$ for $E'=x_1.\cdots.x_h$, and that we can apply the induction hypothesis on command $c_{j-1}$, as $c_k;\ldots;c_{j-1} = \interp{E'}$ and the distance between $c_k$ and $c_{j-1}$ is $d-1$. From Property~\ref{prop5} for $c_{j-1}$ we then know that there was a transition from a state $P_h = \langle\Sched_h, \Structs_h, \Stab_h\rangle$ with some $\ell_h$ s.t. $\Structs_h(\ell_h) = \langle\StructType_h, \ComList_h, \Stack_h, \Env_h\rangle$ and $\Stack = \Stack_k;\Env_h(x_h)$, for some $x_h$ defined in $S_i$. We also know that $\Env_h(x_h)\in\Labels$, as otherwise we would have a variable expression $x_1.\cdots.x_h.x$ in $\Program$ where $x_h$ resolves to a value, not a reference, which is not provable in the type system of $\Program$. As $\Program$ is well-formed, this cannot happen. By Lemma~\ref{lem: labelsource}, we know that $\Structs(\Env_h(x_h)) = \langle \StructType_h, \ComList_h, \Stack_h, \Env_h\rangle$ for some $\StructType_h, \ComList_h, \Stack_h, \Env_h$ and we also know that, with $\StructType_h$ the struct type of $\Structs(\Env_h(x_h))$, $x\notin\Par{\StructType_h,\Program} \Rarr \Env_h(x_h) = \ell$, as otherwise $x_1.\cdots.x_h.x$ would not be admitted by the type rule \textbf{VarR}. 
			
			As $\Program$ is well-formed, we know that there was either a statement $S_q = (T\ x := E_x)$ with $1\leq q < i$ and $E_x\in\Expr$ or that $(x: T)\in\Par{\StructType_h, \Program}$ for some $T\in\SynTypes$, as otherwise $x_1.\cdots.x_h.x$ would not be provable in the type system. Then, if $T\in\StructTypes_\Program$, we know that $\Env(x)\in \Labels$. By Lemma~\ref{lem: labelsource} we then know that $\Structs(\Env_h(x)) = \langle \StructType_x, \ComList_x, \Stack_x, \Env_x\rangle$ for some $\StructType_x, \ComList_x, \Stack_x, \Env_x$ and we know that as $\Program$ is well-formed, $\StructType_x = T$, as otherwise $x_1.\cdots.x_h.x$ would not be provable in the type system. 
			
			Then there exists a transition \textbf{ComRd} to a state $P' = \langle \Sched', \Structs', \Stab'\rangle$ with \\$\Structs'(\ell) = \langle\StructType, c_{j+1};\ldots;c_m;\interp{S_{i+1};\cdots;S_n}, \Stack_k;\Env_h(x), \Env'\rangle$, by definition of \textbf{ComRd}.
			\item[\textbf{wr}] If $c_j = \writev{x}$ for some $x$, then $c_j = c_m$, by the definition of the interpretation function. It follows that $c_k = c_1$ and that $c_k;\ldots;c_m = \interp{S_i}$. By the interpretation function, it follows that $\interp{S_i} = \interp{E'};\interp{x_1.\cdots.x_h};\writev{x}$ for some expression $E'$ and some variable expression $x_1.\cdots.x_h$. By the induction hypothesis on the commands $\interp{E'} = c_1;\ldots;c_a$, we know that after command $c_a$, we are in a state $P_a = \langle \Sched_a, \Structs_a, \Stab_a\rangle$, with $\Structs_a(\ell) = \langle\StructType, c_{a+1};\ldots;c_m;\interp{S_{i+1};\ldots;S_n}, \Stack_k;v, \Env_a\rangle$ for some value $v\in \Values$. By the induction hypothesis on the commands $\interp{x_1.\cdots.x_h} = c_{a+1};\ldots;c_{m-1}$ and analogous reasoning as in the case above, we know that after command $c_{m-1}$, we are in a state $P_b = \langle \Sched_b, \Structs_b, \Stab_b\rangle$, with $\Structs_b(\ell) = \langle\StructType,c_m;\interp{S_{i+1};\ldots;S_n}, \Stack_k;v;\ell_h, \Env_a\rangle$ for some $\ell_h\in\Labels$, with $\Structs(\ell_h) = \langle\StructType_h, \ComList_h, \Stack_h, \Env_h\rangle$ for some $\StructType_h, \ComList_h, \Stack_h, \Env_h$ and with $x\notin\Par{\StructType_h, \Program} \Rarr \ell' = \ell$.
			
			Again, as $\Program$ is well-formed, we know that there was either a statement $S_q = (T\ x := E_x)$ with $1\leq q < i$ and $E_x\in\Expr$ or that $(x: T)\in\Par{\StructType_h, \Program}$ for some $T\in\SynTypes$, as otherwise $x_1.\cdots.x_h.x$ would not be provable in the type system. Then, if $T\in\StructTypes_\Program$, we know that $v\in \Labels$. By Lemma~\ref{lem: labelsource} we then know that $\Structs(v) = \langle \StructType_v, \ComList_v, \Stack_v, \Env_v\rangle$ for some $\StructType_v, \ComList_v, \Stack_v, \Env_v$ and we know that as $\Program$ is well-formed, $\StructType_v = T$, as otherwise $x_1.\cdots.x_h.x := E'$ would not be provable in the type system. 
			
			Then if $\ell_h\notin\Labels_0$, there exists a transition \textbf{ComWr} to a state $P' = \langle \Sched', \Structs', \Stab'\rangle$ with\\$\Structs'(\ell) = \langle\StructType,\interp{S_{i+1};\cdots;S_n}, \Stack_k, \Env'\rangle$ and $\Structs'(\ell') = \langle\StructType_{\ell'}, \ComList_{\ell'}, \Stack_{\ell'}, \Env_{\ell'}'\rangle$ s.t. $\Env_{\ell'}' = \Env_{\ell'}[x\mapsto v]$ and if $x\notin\Par{\StructType_{\ell'}, \Program} \lor \Env_{\ell'}(x)\neq v$ then $\Stab'= \false^{|\Stab|}$ by definition of \textbf{ComWr}.
			
		    If $\ell_h\in\Labels_0$, there exists instead a transition \textbf{ComWrNSkip} to a state $P' = \langle \Sched', \Structs', \Stab\rangle$ with\\$\Structs'(\ell) = \langle\StructType,\interp{S_{i+1};\cdots;S_n}, \Stack_k, \Env'\rangle$ and $\Structs'(\ell') = \langle\StructType_{\ell'}, \ComList_{\ell'}, \Stack_{\ell'}, \Env_{\ell'}\rangle$ by definition of \textbf{ComWrNSkip}.
			\item[\textbf{cons}] If $c_j = \cons{\StructType}$ for some struct type $\StructType$, then it must have been generated as the last command of an expression or statement $\StructType (E_1, \ldots, E_h)$. It then follows that the largest possible $c_k$ is the first command of $\StructType (E_1, \ldots, E_h)$. Then let $\Par{\StructType, \Program} = (x_1: T_1); \ldots; (x_h: T_h)$. It follows from repeated application of the induction hypothesis that every expression $E_i\in E_1, \ldots, E_h$ leaves a value $v\in \Values$ on the stack, s.t. $\Stack = \Stack_k;v_1;\ldots;v_h$. We also know for every expression $E_i\in E_1, \ldots, E_h$ that either $T_i\in\{\texttt{Nat}, \texttt{Int}, \texttt{Bool}, \texttt{String}\}$ and $v_i\in\mathit{sem}(T_i)$ or $T_i\in\StructTypes_\Program$ and $v_i\in\Labels$, as otherwise $\Program$ would not be well-formed as the constructor statement or expression would not be provable in the type system. We then know that if $v_i\in\Labels$, by Lemma~\ref{lem: labelsource} $\Structs(v_i) = \langle \StructType_i, \ComList_i, \Stack_i, \Env_i\rangle$, and we know by the well-formedness of $\Program$ that $\StructType_i = T$.
			
			Then by the definition of the transition \textbf{ComCons} with a fresh label $\ell'$, there exists a transition to a state $P' = \langle \Sched', \Structs', \Stab'\rangle$ with $\Structs'(\ell) = \langle\StructType, c_{j+1};\ldots;c_m;\interp{S_{i+1};\cdots;S_n}, \Stack_k;\ell', \Env'\rangle$ and $\Structs'(\ell') = \langle \StructType, \vempty, \vempty, \Env_{\StructType}^0[\{p_1\mapsto v_1, \ldots, p_h \mapsto v_h\}]\rangle$ and  $\Stab'= \false^{|\Stab|}$.
			\item[\textbf{if}] If $c_j = \Ifc{C}$ for some command list $C$, then $c_j$ must have been generated as part of an if-statement, s.t. $c_j = c_m$ and $\interp{S_i} = \interp{E'};\Ifc{\interp{\mathcal{S}}}$ for some expression $E'$ and some list of statements $\mathcal{S}$. It follows that $c_k = c_1$. Then by application of the induction hypothesis on $\interp{E'} = c_1;\ldots;c_{m-1}$, we know that $\Stack = \Stack_k;b$, where $b\in\Bool$ because $\Program$ would not be well-formed otherwise.
			
			Then by the definition of the transitions \textbf{ComIfT} and \textbf{ComIfF}, there exists a transition to a state $P' = \langle \Sched', \Structs', \Stab'\rangle$ with $\Structs'(\ell) = \langle\StructType, \ComList', \Stack_k, \Env'\rangle$ s.t. if $b = \true$, $\ComList' = C; \interp{S_{i+1};\cdots;S_n}$. and if $b = \false$,  $\ComList' = \interp{S_{i+1};\cdots;S_n}$.
			\item[\textbf{not}] If $c_j = \Notc$, it must have been generated as the last command of an expression $!E'$ for some expression $E'$. It follows that the largest possible $c_k$ is the first command of $\interp{!E} = \interp{E};\Notc$. By application of the induction hypothesis on $\interp{E'}$, we then know that $\Stack = \Stack_k;b$, and we know that $b\in\Bool$ because otherwise $\Program$ would not be well-formed. 
			
			Then by the definition of the transition \textbf{ComNot} there exists a transition to a state $P' = \langle \Sched', \Structs', \Stab'\rangle$ with $\Structs'(\ell) = \langle\StructType, \interp{S_{i+1};\cdots;S_n}, \Stack_k;\neg b, \Env'\rangle$.
			\item[\textbf{Op}] If $c_j = \Operator(\circ)$ for some syntactic operator $\circ$, it must have been generated as the last command of an expression $E_1 \circ E_2$. It follows that the largest possible $c_k$ is the first command of $\interp{E_1\circ E_2} = \interp{E_1};\interp{E_2};\Operator(\circ)$. Then, by induction hypothesis on $\interp{E_1}$ and $\interp{E_2}$, we know that $\Stack = \Stack_k;a;b$ for $a, b\in \Values$. We then know that if $\circ\in\{=, \text{!=}\}$ then $a, b\in\{\Nat, \Int\}$ or there exists a $T\in\{\Nat, \Int, \Bool, \String\}\cup \StructTypes_\Program$ s.t. either $T\notin \StructTypes_\Program$ and $a, b \in T$ or $T\in\StructTypes_\Program$, by the type rules \textbf{Eq} and \textbf{Comp}. We also know that and $\Structs(a) = \langle T, \ComList_a, \Stack_a, \Env_a\rangle$ and  $\Structs(b) = \langle T, \ComList_b, \Stack_b, \Env_b\rangle$ by Lemma~\ref{lem: labelsource} and the type rule \textbf{Eq}. By the type rules \textbf{Comp}, \textbf{NArith} and \textbf{IArith}, we know that if $\circ\in\{< =, >=, <, >, =, \text{!=}, *, /, \%, +, \text{\textasciicircum}, -\}$ then $a, b\in\{\Nat, \Int\}$, and by the type rule \textbf{BinLog} we know that if $\circ\in\{\&\&, ||\}$ then $a, b\in\Bool$.
			
			Then by the definition of \textbf{ComOp}, there exists a transition to a state $P' = \langle \Sched', \Structs', \Stab'\rangle$ with $\Structs'(\ell) = \langle\StructType, c_{j+1}; \ldots; c_m\interp{S_{i+1};\cdots;S_n}, \Stack_k;c \Env'\rangle$ s.t. if $a\in T$ and $b\in T$ with $T\in \SemTypes$ and $\circ = \{=,\text{!=}\}$, then $c = a \mathop{o} b$ and $c\in\mathbb{B}$, if $a\in T_1$ and $b\in T_2$, with $T_1, T_2\in\{\mathbb{N}, \mathbb{Z}\}$ and $\circ = \{< =, >=, <, >, =, \text{!=}\}$, then $c = a\mathop{o} b$ and $c\in\mathbb{B}$, if $a\in T_1$ and $b\in T_2$, with $T_1, T_2\in\{\mathbb{N}, \mathbb{Z}\}$ and $\circ = \{*, /, \%, +, \text{\textasciicircum}, -\}$, then $c = a\mathop{o} b$ and $c\in\mathbb{Z}$ and if $a, b\in \mathbb{N}$ and $\circ = \{*, /, \%, +, \text{\textasciicircum}\}$, then $c = a\mathop{o} b$ and $c\in\mathbb{N}$.
		\end{itemize}
		It follows that the step holds for all possible cases, so the induction holds.
	\end{enumerate}
	As the induction holds, Property~\ref{prop5} holds.
\end{proof}
\begin{proof}[Proof of Property~\ref{prop6}]
	Let $P = \langle\Sched, \Structs, \Stab\rangle$ be a well-formed \Lname state reached by some program $\Program = D\ \Sched_\Program$ s.t. $\Gamma, \Omega\tr D\ Sc_\Program$ for some $\Gamma, \Omega$. Then assume that $\neg \mathit{Done}(\Structs)$. Let $k$ be the total amount of commands still present in some command list of some struct instance. $k>0$, as $\mathit{Done}(\Structs) = \false$. Then the amount of transitions enabled directly corresponds to the commands $c$ s.t. for some $\ell'\in\Labels$, $\Structs(\ell') = \langle \StructType', c;\ComList', \Stack', \Env\rangle$. It follows that after a transition to a state $P_1$, there are $k-1$ commands in total still present in some command list of some struct instance, and the transitions enabled are similarly bound to commands as in $P$. As $k$ must be a finite number, as $\Program$ is well-formed (with a finite syntax), it follows that after a finite amount of transitions, we arrive in a state $P' = \langle \Sched',\Structs', \Stab'\rangle$ s.t. $\mathit{Done}(\Structs'') = \true$.
\end{proof}

As all properties hold, we conclude that Theorem~\ref{thm: wellformed} holds. With this theorem, we can prove type safety. According to Pierce~\cite[Section 8.3]{pierce2002types}, a type system can be called safe if it satisfies two properties, \textit{progress} and \textit{preservation}. Progress requires a well-formed program to not get stuck. Preservation requires that when a well-formed syntactical term with a type gets evaluated semantically, the result has the expected type. Both properties follows from Theorem~\ref{thm: wellformed}:
\begin{corollary}[Type Safety]\label{cor: safety}
\Lname's type system is \emph{safe}.
\end{corollary}
\begin{proof}
	First of all, for progress, let $P = \langle \Sched, \Structs, \Stab\rangle$ be a well-formed state reachable by a program $\Program$. Then, if $\mathit{Done}(\Structs)$ does not hold, there is at least one label $\ell\in\Labels$ s.t. $\Structs(\ell) = \langle \StructType, \ComList, \Stack, \Env\rangle$ and $\ComList\neq \vempty$. Then $\ComList = c;\ComList'$ and by Property~\ref{prop5} of Theorem~\ref{thm: wellformed}, we know that there exists at least one transition induced by $c$ from $P$ to another state $P'$.	
	If $\mathit{Done}(\Structs)$ holds but $\Sched \neq \vempty$, then by Property~\ref{prop2} of Theorem~\ref{thm: wellformed}, we know there is at least one schedule transition enabled for $P$.
	
	Then, for preservation, note that we have two categories of syntactical terms with types: the schedule and the expression. We know by Property~\ref{prop1} that the schedule of all well-formed states are well-formed modulo $\mathit{aFix}$, which satisfies preservation for schedules. For expressions, we know by Property~\ref{prop5} that for every expression $E$ with a last command $c_j\in\{\push{v}, \push{\this}, \readg{x}, \cons{\StructType}, \Notc, \Operator(\circ)\}$, for some $v\in\Values$, variable $v$, $\StructType\in\StructTypes_\Program$ and syntactic operator $\circ$, a value $v$ with a specific semantic type is put on the stack of the executing struct instance. This semantic type always respects the syntactic type of $E$: for \textbf{push}-commands, this type is directly derived from $E$, for the \textbf{rd}-command, $v$'s type depends on the value read, which respects the type of the variable it is read from, which respects the type of $E$, for the \textbf{cons}-command, the label put on the stack has struct type $\StructType$, for the $\Notc$-command the resulting type is $\Bool$, and for the \textbf{op}-command the resulting semantic type is defined by the operator and the types of the input values in a way that is derived from the type rules \textbf{Eq}, \textbf{IArith}, \textbf{NArith} and \textbf{BinLog}. It follows that preservation also holds for expressions.
	
	As both progress and preservation hold, the type system is safe.
\end{proof}

Additionally, we can shift the focus from commands to expressions and statements. For expressions, we define \emph{resulting values} for expressions:
\begin{definition}[Expression Results]
	Let the result $v$ of an execution of an expression $E$ in a step $F$ in a well-formed program $\Program$ by a struct instance $s$ be the value $v$ resulting from the transition induced by the last command $c_m$ of $\interp{E}$ in the command list of $s$ as per Theorem~\ref{thm: wellformed}.
\end{definition}
Note that there always is at least one resulting value for every expression, by Theorem~\ref{thm: wellformed}. However, this value does not have to be deterministic, because we allow for interleaving during the execution of an expression $E$ by struct instance $s$. After all, if $E$ contains a read to a variable which is not deterministic during $F$, then the result of $E$ cannot be deterministic either. However, the following holds:
\begin{lemma}\label{lem: detexp}
	Let $E$ be an expression in a step $F$ in a well-formed program $\Program$ executed by a struct instance $s$ from some state $P$. Let $x_1, \ldots, x_n$ be the parameters referenced during $E$. If all parameters $x_i\in x_1, \ldots, x_n$ are deterministic during the execution of $F$ from $P$, then the result of $E$ executed by $s$ is also deterministic during the execution of $F$ from some state $P$.
\end{lemma}
\begin{proof}
	If all of the parameters of $E$ are deterministic, it follows that any computations on them have a deterministic outcome, so the result is also deterministic.
\end{proof} 
In this paper, we are interested in proving programs without read-write race conditions from some premise state $P_1$. Note that in program executions without read-write race conditions from $P_1$, all resulting values from all expressions are deterministic, as the fact that parameters are read in an expression $E$ executed by a struct instance $s$ during the execution of a step $F$ from some state $P$ means they cannot be written to during $F$, which leads to the parameters being deterministic. So in a program without read-write race conditions from $P_1$, expression results are by default deterministic.

On a more general level, we know from Theorem~\ref{thm: wellformed} that the execution of a statement $S$ in a step $F$ in a well-formed program $\Program$ by a struct instance $s$ leads to the effects in Property~\ref{prop5} depending on the last command in $\interp{S}$. Intuitively, this means the following:
\begin{corollary}[Execution Effects]\label{cor: statef}
	Let $S$ be a statement executed during the execution of a step $F$ in a well-formed program $\Program$ by a struct instance $s$. Let $P'$ be the state resulting from the transition induced by the last command of $\interp{S}$ for $s$. Then:
	\begin{enumerate}
		\item If $S$ is an update statement $x_1.\ldots.x_n.x := E$, then, with $x$ the variable in the struct instance belonging to a result of an execution of $x_1.\ldots.x_n$, if $x$ is not a parameter of a null-instance, $x$ will be updated to a result of an execution of $E$ in $P'$, and all values in the stability stack will be reset to $\false$ if this means that a parameter has changed. If $x$ is a parameter of a null-instance, $E$ is executed and its effects are still present in $P'$, but $x$ has not been updated.
		\item If $S$ is a variable assignment statement $T\ x:= E$, then it has the same effect as executing the update statement $x:= E$.
		\item If $S$ is an if-statement, in $P'$, $E$ will be executed to a result $b$. If $b = \true$, the if-clause will be taken. If $b = \false$, the if-clause will be skipped.
		\item If $S$ is a constructor statement $\StructType(E_1, \ldots, E_m)$, then in $P'$, there will be a new struct instance for a fresh label $\ell$ with type $\StructType$ and its parameters set to results of executions of $E_1, \ldots, E_m$, and all values in the stability stack will be reset to $\false$. Additionally, $\ell$ will be the last value on the stack of $s$.
	\end{enumerate}
	Additionally, let $\mathcal{E}$ be the expressions executed as a part of $S$. For any expression $E\in\mathcal{E}$, if $E$ is a constructor expressions $\StructType(E_1, \ldots, E_m)$, its result will be a fresh label $\ell'$ s.t. in $P'$, there is a struct instance for $\ell'$ with type $\StructType$ and its parameters set to results of executions of $E_1, \ldots, E_m$. Furthermore, if there exists a constructor expression $E$ in $\mathcal{E}$, all values in the stability stack will be reset to $\false$. 
	Lastly, the effects of a statement will be deterministic if all results from all expressions in $S$ are deterministic.
\end{corollary}
\begin{proof}
	Follows directly from applications of Theorem~\ref{thm: wellformed} and Lemma~\ref{lem: detexp}.
\end{proof}
As in this paper we assume our programs do not have read-write race conditions from some premise state, this means that the direct effects of all statements are deterministic. However, we do not exclude write-write race conditions, so we need to weaken the above Corollary to find the permanent results of executing update statements in a step:
\begin{corollary}[Permanent Update Results]\label{cor: steppar}
	Let $\Program$ be a well-formed program without read-write race conditions from some premise state $P_1$, with a step $F$. Let $F$ be executed from some state $P$ and let $s$ be a struct instance in $P$. Let $S$ be the last update statement of some parameter $s'.x$ of some struct instance $s'$ by $s$, which updates $s'.x$ to a value $v$. Let $\mathcal{S}$ be the last update statements of $s'.x$ across all definitions of $F$ for any struct type $\StructType$. Then we can guarantee that after the execution of $F$ from $P$:
	\begin{itemize}
		\item[a.] If $s'$ is a $\nil$-instance, $s'.x = a = \nil$.
		\item[b.] If $s'$ is not a $\nil$-instance:
		\begin{itemize}
			\item[i.] If $s'.x$ is not involved in a write-write race condition, $s'.x = b$.
			\item[ii.] If $s'.x$ is involved in a write-write race condition in $F$, let $N$ be the set of all possible values written to $s'.x$ during the execution of any statement from $\mathcal{S}$ by any struct instance during the execution of $F$ from $P$. Then $s'.x\in N$.
		\end{itemize}
	\end{itemize} 
	Additionally, we know that if $s'.x$ has had its value changed during the execution of $F$, all values in the stability stack have been reset to $\false$ during the execution of $F$.
\end{corollary}

\paragraph{Implications of Theorem~\ref{thm: wellformed} and its corollaries.} With the theorems, lemmas and corollaries proven in this section, one can prove the effects of an \Lname step $F$ by hand on the level of the syntactic code of $F$ in the program, without transforming that behaviour into commands, using the above corollaries to formally establish the behaviour of statements. For variables $x$ deterministic during the execution of $F$, one only has to consider the final value of $x$ relative to a single struct instance, which can then be generalised to all struct instances. This can be done by sequentially walking through the code of $F$. For variables $x$ involved in write-write race conditions, one will need to construct the set $N$ as defined in Corollary~\ref{cor: steppar}.
	\section{Proving \Lname Code Correct}\label{section:proving}
In this section, we prove the \Lname code of Listing~\ref{ex: PS3} (Section~\ref{section:prefixproof}) and Listing~\ref{ex: Sort2} (Section~\ref{section:sortproof}) correct, using the definitions, properties and lemmas of Section~\ref{section:props} and concepts from Sections~\ref{section:syntax}, \ref{section:types} and \ref{section:semantics}. We then conclude this section with some more general insights in the structure of proving \Lname programs correct (Section~\ref{section:proofconcs}).
During the proofs, we consider the program in a vacuum; we assume that if something is not changed by the program as a whole, it is not changed at all. With this, we rule out outside interference by hardware. As this paper is concerned with the theoretical foundations of \Lname and we intend to prove the correctness of \Lname programs, not \Lname program implementations we consider this to be a fair assumption. We also omit the type system proof derivation to prove both programs well-formed.

\subsection{Proving Prefix Sum Correct.}\label{section:prefixproof}
In this subsection, we prove that the code as given in Listing~\ref{ex: PS3} correctly executes the Prefix Sum function. We start by stating the problem formally:

\begin{problem}[Prefix Sum]
	Given a sequence of elements $p_1, \ldots, p_n$, with values $x_1, \ldots, x_n$ for these elements, compute for each position $1 \leq i \leq n$ the sum $\Sigma_{k = 1}^i x_k$. The sequence is given as a single, finite list without loops.
\end{problem}

As the code does not describe initialization, we need to specify the requirements of a premise state from which this code can be executed with the intended effect. These requirements have already been stated in Section~\ref{section:standardalgos}, but are formalized here:

\begin{definition}[Premise State]\label{def: pps}
	We define a \emph{premise state} of the Prefix Sum code as a state $P_1 = \langle \Sched, \Structs, \Stab\rangle$ s.t. there are labels $\ell_1, \ldots, \ell_n$ for positions $p_1, \ldots, p_n$ where $\Structs(\ell_1) = \langle \textit{Position}, \vempty, \vempty, \Env_1\rangle$ is the first position $p_1$, with $p_1.\mathit{val} = x_1$ and $p_1.\mathit{prev} = \nil$, and for every $p_i$ with $1<i\leq n$, $p_i = \Structs(\ell_i) = \langle \textit{Position}, \vempty, \vempty, \Env_i\rangle$, with $p_i.\mathit{val} = x_i$ and $p_i.\mathit{prev} = \ell_{i-1}$.
\end{definition}

To prove the code in Listing~\ref{ex: PS3} correct, we first consider the steps of this code, for which we prove the effects. The strategy is to first consider which parameters are in a race condition, if any, using information on the states from which the step is executed. Then, we can establish the effects of the step on the parameters using Corollary~\ref{cor: steppar}. For parameters not involved in race conditions, this consists of walking through the step sequentially and establishing the last value written to the parameter. 
After we have considered the effects of all the steps, we combine them into fixpoint inner contracts. These we then use to prove the effects of fixpoints, from which we establish the final result of executing the code. If this code corresponds to the requirements as stated in the problem, it is correct provided the state from which it is executed is a valid premise state.

For Listing~\ref{ex: PS3}, instead of considering variables separately we prove that all steps are deterministic as a whole:
\begin{lemma}
	All steps in Listing~\ref{ex: PS3} are deterministic.
\end{lemma}
\begin{proof}
	This follows from the fact that no step in Listing~\ref{ex: PS3} has potential race conditions:
	\begin{itemize}
		\item In \textit{read}, a potential race condition would need to include either \textit{auxval} or \textit{auxprev}, as those are the parameters written to in \textit{read}. However, these parameters are not accessed indirectly, so there can be no potential race conditions on them, following from Definition~\ref{def: potdat}. Therefore, \textit{read} has no potential race conditions.
		\item In \textit{write}, no parameters used are accessed indirectly, which is required for a potential race condition to exist, according to Definition~\ref{def: potdat}. Therefore, \textit{write} has no potential race conditions.
	\end{itemize}
	Therefore, no step in Listing~\ref{ex: PS3} has potential race conditions. It follows from Lemma~\ref{lem: nopot} that no step in Listing~\ref{ex: PS3} has race conditions. It follows by Lemma~\ref{lem: det} that all steps in Listing~\ref{ex: PS3} are deterministic.
\end{proof}

As all steps are deterministic, we can walk through a step sequentially to determine its effects. That way, we establish the following contracts for \textit{read} and \textit{write} in Listing~\ref{ex: PS3}:
	\begin{lemma}[\textit{read} Contract]\label{lem: psrc}
	Let $P = \langle \Sched, \Structs, \Stab\rangle$ be an \Lname state and let $\Labels_\mathit{Position}$ be the labels of \textit{Position} structs in $P$. Executing \emph{read} starting at state $P$ results in a state $P' = \langle \Sched, \Structs', \Stab\rangle$. In $P'$, for every $\ell_p\in\Labels_\mathit{Position}$ s.t. $p = \Structs(\ell_p) = \langle \mathit{Position}, \ComList, \Stack, \Env_p\rangle$, let $p'$ be $\Structs'(\ell_p)$. Then $p'.\mathit{auxval} = p.\mathit{prev}.\mathit{val}$ and $p'.\mathit{auxprev} = p.\mathit{prev}.\mathit{prev}$.
\end{lemma}
\begin{proof}
	Let $P$, $P'$, $p$ and $p'$ be as defined in the lemma. As \textit{read} is deterministic, we can sequentially walk through the statements of the \textit{read} step. 
	
	By applying Corollary~\ref{cor: statef} to the first statement of \textit{read}, it follows that $p'.\mathit{auxval} = p.\mathit{prev.val}$. By applying it to the second statement of \textit{read}, we know that $p'.\mathit{auxprev}$ is equal to $p.\mathit{prev.prev}$. 
\end{proof}
\begin{lemma}\label{lem: pswc}
	Let $P = \langle \Sched, \Structs, \Stab\rangle$ be an \Lname state and let $\Labels_\mathit{Position}$ be the labels of \textit{Position} structs in $P$. Executing \emph{write} starting at state $P$ results in a state $P' = \langle \Sched, \Structs', \Stab\rangle$. In $P'$, for every $\ell_p$ s.t. $p = \Structs(\ell_p) = \langle \mathit{Position}, \ComList, \Stack, \Env_p\rangle$, let $p'$ be $\Structs'(\ell_p)$. Then $p'.\mathit{val} = p.\mathit{val} + p.\mathit{auxval}$ and $p'.\mathit{prev} = p.\mathit{auxprev}$.
\end{lemma}
\begin{proof}
	Let $P$, $P'$, $p$ and $p'$ be defined in the lemma. As \textit{write} is deterministic, we can sequentially walk through the statements of the \textit{write} step. By applying Corollary~\ref{cor: statef} to all statements, we get that $p'.\mathit{val} = p.\mathit{val} + p.\mathit{auxval}$ and $p'.\mathit{prev} = p.\mathit{auxprev}$. 
\end{proof}

We then consider the schedule. As the fixpoint in the schedule first executes \textit{read} and then \textit{write}, we define the following combined contract for $\mathit{read} < \mathit{write}$:
\begin{lemma}[Fixpoint execution]\label{lem: contract}
	Let $P = \langle \Sched, \Structs, \Stab\rangle$ be an \Lname state and let $\Labels_\mathit{Position}$ be the labels of \textit{Position} structs in $P$. Executing \emph{read$<$write} starting at state $P$ results in a state $P' = \langle \Sched, \Structs', \Stab\rangle$. In $P'$, for every $\ell_p$ s.t. $p = \Structs(\ell_p) = \langle \mathit{Position}, \ComList, \Stack, \Env_p\rangle$, let $p'$ be $\Structs'(\ell_p)$. Then $p'.\mathit{val} = p.\mathit{val} + p.\mathit{prev.val}$, $p'.\mathit{prev} = p.\mathit{prev.prev}$, $p'.\mathit{auxval} = p.\mathit{prev.val}$ and $p'.\mathit{auxprev} = p.\mathit{prev.prev}$.
\end{lemma}
\begin{proof}
	Follows from Lemmas~\ref{lem: psrc} and \ref{lem: pswc}.
\end{proof}

We now prove the following two lemmas. As the proof for these lemmas can be efficiently combined, we prove them together:
\begin{lemma}[Termination]\label{lem: pst}
	Executing \emph{Fix(read $<$ write)} terminates.
\end{lemma}
\begin{lemma}[Result]\label{lem: psr}
	Executing \emph{Fix(read $<$ write)} starting at premise state $P_1 = \langle \Sched_1, \Structs_1, \Stab_1\rangle$ results in a state $P' = \langle \Sched', \Structs', \Stab'\rangle$. In $P'$, for every $\ell_i$ s.t. position $p_i = \Structs(\ell_{i})$ with $1\leq i\leq n$, let $p'_i$ be $\Structs'(\ell_{p_i})$. Then $p'_i.\mathit{val} = \sum_{k=1}^i p_k.val$.
\end{lemma}
\begin{proof}[Proof of Termination and Result]
	For the first lemma, the fixpoint terminates when an iteration is reached which does not change any of the parameters. We need to prove that this iteration is eventually reached.
	The second lemma is more straightforward.
	
	To prove both, we will first prove that after every iteration $j$ of the fixpoint, $p_i'.\mathit{val} = \sum_{k=h}^{i} p_k.\mathit{val}$, with $h = \max{(1, i-2^j+1)}$ and $p_i'.\mathit{prev} = \ell_{i - 2^{j}}$ or $\nil$ if $i-2^{j} \leq 0$ for any $1\leq i\leq n$, by induction on $j$.
	
	\begin{description}
		\item[$j = 1$.] For the first iteration, we start at premise state $P_1$ and end at a state $P_2$. In $P_2$, according to Lemma~\ref{lem: contract}, $p_i'.\mathit{val} = p_i.\mathit{val} + p_i.\mathit{prev.val} = p_i.\mathit{val} + p_{i-1}.\mathit{val}$, with $p_{i-1} = \Structs_1(\ell_{i-1})$. If $i-1 < 1$, then $p_{i-1} = \nil$, as $p_1.\mathit{prev} = \nil$, so $p_{i-1}.\mathit{val} = 0$ and then $p_i.\mathit{val} + p_{i-1}.\mathit{val} = p_1.\mathit{val}$. Therefore $p_i'.\mathit{val} = p_i.\mathit{val} + p_{i-1}.\mathit{val} = \sum_{k = h}^i p_k.\mathit{val}$, with $h = \max{(1, i-1)} = \max{(1, i-2^j+1)}$, as $j = 1$.
		
		We also know from Lemma~\ref{lem: contract} that $p_i'.\mathit{prev} = p_i.\mathit{prev.prev}$. If $i-2 > 0$, then $i - 1 > 0$, so with $p_i.\mathit{prev} = \ell_{i-1}$ and $p_{i-1} = \Structs(\ell_{i-1})$, $p_{i-1}.\mathit{prev} = \ell_{i-2}$, so $p_i'.\mathit{prev} = \ell_{i-2} = \ell_{i - 2^j}$. If $i-2 \leq 0$, then either $i-1 \leq 0$, in which case $p_i.\mathit{prev.prev} = \nil.\mathit{prev} = \nil$, or $i - 1 > 0$, in which case $p_{i-1}.\mathit{prev} = \nil$, so $p_i.\mathit{prev.prev} = \nil$ and $p_i'.\mathit{prev} = \nil$.
		
		\item[$j > 1$.] By the induction hypothesis, when we start at state $P_{j-1} = \langle \Sched, \Structs, \Stab\rangle$ for iteration $j$, for any $p_i = \Structs(\ell_i)$, $p_i.\mathit{val} = \sum_{k=h}^{i} p_k.\mathit{val}$ with $h = \max{(1, i-2^{j-1}+1)}$ and $p_i.\mathit{prev} = \ell_{i - 2^{j-1}}$ or $\nil$ if $i - 2^{j-1}\leq 0$. 
		Due to Lemma~\ref{lem: contract}, it follows that we end up in a state $P_j = \langle \Sched', \Structs', \Stab'\rangle$ s.t. for $p'_i = \Structs(\ell_i)$, $p_i'.\mathit{val} = p_i.\mathit{val} + p_i.\mathit{prev.val}$ and $p_i'.\mathit{prev} = p_i.\mathit{prev.prev}$.
		
		From the IH we know that $p_i.\mathit{val} = \sum_{k=h}^{i} p_k.\mathit{val}$ with $h = \max{(1, i-2^{j-1}+1)}$. 
		Then if $i - 2^{j-1} \leq  0$ we know that $i-2^j\leq 0$ and that $p_i.\mathit{prev} = \nil$, so $p_i'.\mathit{prev} = \nil.\mathit{prev} = \nil$. Additionally, $h = 1$, so $p_i.\mathit{val} = \sum_{1}^{i} p_k.\mathit{val}$, and $p_i'.\mathit{val} = p_i.\mathit{val} + p_i.\mathit{prev.val} = \sum_{1}^{i} p_k.\mathit{val} + 0 = \sum_{1}^{i} p_k.\mathit{val}$. 
		
		If $i - 2^{j-1} > 0$, then we know that $p_i.\mathit{prev} = \ell_{i-2^{j-1}}$ with $p_{i-2^{j-1}} = \Structs(\ell_{i-2^{j-1}})$. Then either $i - 2^j \leq 0$ or not. 
		
		If $i - 2^j \leq 0$, then $(i - 2^{j-1}) - 2^{j-1} \leq 0$, and then by the IH we know that $p_{i-2^{j-1}}.\mathit{prev} = \nil$, so $p_i'.\mathit{prev} = q_i.\mathit{prev.prev} = \nil$. Otherwise, $p_{i-2^{j-1}}.\mathit{prev} = \ell_{i-2^{j}}$, so $p_i'.\mathit{prev} = \ell_{i-2^j}$.
		
		As $i - 2^{j-1} > 0$, we know that $p_i.\mathit{val} = \sum_{k={i-2^{j-1} + 1}}^{i} p_k.\mathit{val}$. We also know that $p_{i-2^{j-1}}.\mathit{val} = \sum_{k = g}^{i-2^{j-1}}p_k.\mathit{val}$ with $g = \max{(1, i-2^j + 1)}$ by the IH and the facts established in the last paragraph. Then $p_i'.\mathit{val} = p_i.\mathit{val} + p_{i-2^{j-1}}.\mathit{val} = \sum_{k={i-2^{j-1} + 1}}^{i} p_k.\mathit{val} + \sum_{k = g}^{i-2^{j-1}}p_k.\mathit{val} = \sum_{k = g}^{i} p_k.\mathit{val}$.
		
		As this proves all parts of the IH for iteration $j$, the induction step holds.
	\end{description}
	Note that $\nil.\mathit{prev} = \nil$ and $\nil.\mathit{val} = 0$. As per the induction, for any $p_i$ and any iteration $j$ such that $2^{j}\geq i$, $p_i.\mathit{prev} = \nil$. Therefore the $\mathit{prev}$ parameter for $p_i$ will be stable in iteration $\lceil\log_2{(i)}\rceil + 1$. 
	
	Also note that in iteration $\log_2{(i)}+1$, $p_i.\mathit{val}$ will be stable, as per the induction. Therefore, Lemma~\ref{lem: pst} holds; the fixpoint will terminate after $\lceil\log_2{n}\rceil + 1$ iterations, as the last of the $n$ elements is the last to become stable.
	
	After $\log_2{(n)}+1$ iterations, $p_i'.\mathit{val} = \sum_{k=1}^{i}p_k.\mathit{val}$, as per the induction, so Lemma~\ref{lem: psr} holds. 
\end{proof}
This results in our conclusion:
\begin{theorem}[Correctness of Listing~\ref{ex: PS3}]
	\label{thm:correctness-prefix-sum}
	When executing the \Lname code as given in Listing~\ref{ex: PS3} from a premise state according to Definition~\ref{def: pps} for a set of elements $p_1,\ldots, p_n$ with values $x_1, \ldots, x_n$, in the resulting state, every \textit{Position} $q_i$ corresponding to element $p_i$ with $1\leq i \leq n$ has its \textit{val} parameter set to $\sum_{k=1}^{i} x_k$.
\end{theorem}
\begin{proof}
	Follows from Lemma~\ref{lem: psr} and the premise, as $\sum_{k=1}^i p_k.val = \sum_{k=1}^i x_k$.
\end{proof}
\subsection{Proving Linked List Copy Sort Correct}\label{section:sortproof}
In this subsection, we prove that the code as given in Listing~\ref{ex: Sort2} sorts correctly. We state the problem formally as:

\begin{problem}[Sorting]
	Given a sequence $S_1$ of elements $p_1,\ldots, p_n$ with distinct positive values $x_1,\ldots, x_n$, rearrange these elements to a sequence $S_2$ $p_1',\ldots, p_n'$ with values $x_1',\ldots, x_n'$ s.t. for every element $p_i\in S_1$, a copy of that element $p_j'$ exists s.t. $p_j'\in S_2$ and $x_i'< x_{i+1}'$ for every $1\leq i < n$. 
\end{problem}

The premise state of Listing~\ref{ex: Sort2} is given as:
\begin{definition}[Premise of Listing~\ref{ex: Sort2}]\label{def: Sort2Prem}
		We define a \emph{premise state} of the Linked List Copy Sort code as a state $P_1 = \langle \Sched, \Structs, \Stab\rangle$ s.t:
	\begin{itemize}
		\item For every element $p_i$ in $p_1,\ldots, p_n$ there exists a struct instance $a_i$ (in $a_1,\ldots a_n$) with label $\ell_i$ (in $\ell_1, \ldots, \ell_n$) s.t. $a_i = \langle \mathit{OldElem}, \vempty, \vempty, \Env_i\rangle$ and $\Structs(\ell_i) = a_i$.
		\item There exists a non-$\nil$ struct instance $b$ with label $\ell_b$ s.t. $b = \langle \mathit{NewElem}, \vempty, \vempty, \Env_b\rangle$.
		\item For all $a_i$, $a_i.\mathit{val} = x_i$ with $x_i\in x_1,\ldots x_n$, $a_i.\mathit{place} = b$ and $a_i.\mathit{move} = \false$.
		\item $b.\mathit{min} = \min(x_1, \ldots, x_n)$, $b.\mathit{max} = \max(x_1, \ldots, x_n)$ (which can be done with a method akin to prefix-sum), $b.\mathit{spl} = \lfloor (b.\mathit{max} - b.\mathit{min})/2\rfloor + b.\mathit{min}$, $b.\mathit{next} = \ell_\mathit{NewElem}^0$, $b.\mathit{p1} = \ell_\mathit{OldElem}^0$, $b.\mathit{p2} = \ell_\mathit{OldElem}^0$, $b.\mathit{hasSplit} = \false$ and $b.\mathit{done} = \true$.
	\end{itemize}
\end{definition}

Considering the steps, note that to both \textit{checkStable} and \textit{migrate}, write-write race conditions are integral to its function. Therefore, the determinism of steps in Listing~\ref{ex: Sort2} can be encapsulated by the following lemma:
\begin{lemma}\label{lem: Sort2det}
	All of the following hold:
	\begin{enumerate}
		\item Only the parameter \textit{place.done} can be involved in a race condition in the step \textit{checkStable}, and if it is, it is involved in a write-write race condition.
		\item Only the parameters \textit{place.p1} and \textit{place.p2} can be involved in a race condition in the step \textit{migrate}, and if they are, they are involved in a write-write race condition.
		\item The step \textit{split} is deterministic.
	\end{enumerate}
\end{lemma}
\begin{proof}
	We prove all of the points separately:
	\begin{enumerate}
		\item From Listing~\ref{ex: Sort2}, the only struct executing \textit{checkStable} is the struct \textit{OldElem}, and through the implementation of \textit{checkStable} in \textit{OldElem}, we know that \textit{place.done} is the only parameter that is written to, so only \textit{place.done} has potential race conditions. As \textit{place.done} is not read in the code of \textit{checkStable}, any race condition \textit{place.done} is involved in must be a write-write race condition.
		\item This point holds following the same reasoning as the first point, except with the parameters \textit{place.p1} and \textit{place.p2} and the step \textit{migrate}.
		\item In \textit{split}, no parameters used are accessed indirectly, which is required for a potential race condition to exist, according to Definition~\ref{def: potdat}. Therefore, \textit{split} has no potential race conditions. It follows that \textit{split} has no race conditions (Lemma~\ref{lem: nopot}) and therefore, \textit{split} is deterministic (Lemma~\ref{lem: det}).
		\qedhere
	\end{enumerate}
\end{proof}
With this lemma, we can establish contracts for the steps in Listing~\ref{ex: Sort2}.
\begin{lemma}[\textit{migrate} Contract]\label{contract: migrate}
	Let $P = \langle \Sched, \Structs, \Stab\rangle$ be an \Lname state and let $\Labels_\mathit{OldElem} \subseteq \Labels$ be the labels of \textit{OldElem} structs in $P$. Executing \textit{migrate} starting at state $P$ results in a state $P' = \langle \Sched, \Structs', \Stab\rangle$. In $P'$, for every $\ell_p$ s.t. $p = \Structs(\ell_p) = \langle \mathit{OldElem}, \ComList, \Stack, \Env_p\rangle$, $p' = \Structs'(\ell_p)$, all of the following hold:
	\begin{enumerate}
		\item $p.\mathit{move} \land p.\mathit{place.hasSplit}\iff p'.\mathit{place} = p.\mathit{place.next}$,
		\item $(p.\mathit{val} > p.\mathit{place.spl})\iff  p'.\mathit{move} = \true$
		\item $(p.\mathit{val} \leq p.\mathit{place.spl})\implies \exists\ell\in\Labels_\mathit{OldElem}.(\Structs'(\ell) = q \land q.\mathit{place} = p'.\mathit{place} \land p'.\mathit{place.p1} = \ell)$,
		\item $(p.\mathit{val} > p.\mathit{place.spl})\implies \exists\ell\in\Labels_\mathit{OldElem}.(\Structs'(\ell) = q \land q.\mathit{place} = p'.\mathit{place} \land p'.\mathit{place.p2} = \ell)$.
	\end{enumerate}
\end{lemma}
\begin{proof}
	Let $P$, $P'$, $p$ and $p'$ be as defined in the lemma. As the only parameters accessed by $p$ involved in a race condition are $p'.\mathit{place.p1}$ and $p'.\mathit{place.p2}$, which do not have any bearing on point $1$ of the lemma, we can walk deterministically through the lines relevant for point $1$, lines 7 to 13. Through application Corollary~\ref{cor: statef}, point $1$ follows, as \textit{place} can only be updated when the if-statement resolves to true.
	
	For point 2 and 3, note that the parameter $\mathit{move}$ is not involved in any race conditions (as per Lemma~\ref{lem: Sort2det}), we then know through application Corollary~\ref{cor: statef} that $\mathit{move}$ is set to $\false$ for the struct instance of $\ell_p$ after line 13 and that $p.\mathit{val} > p.\mathit{place.spl} \iff p'.\mathit{move} = \true$.
	
	Then to prove that $(p.\mathit{val} \leq p.\mathit{place.spl})\implies \exists\ell\in\Labels_\mathit{OldElem}.(\Structs'(\ell) = q \land q.\mathit{place} = p'.\mathit{place} \land p'.\mathit{place.p1} = \ell)$, let $p_1$ be the struct instance of the label $\ell_p$ at the moment of executing line 15 of the code. Then the update of $p_1.\mathit{place.p1}$ may be in a race condition with another struct instance executing line 15, $q_1$ with label $\ell_q$, as long as $q_1.\mathit{place} = p_1.\mathit{place}$. Therefore, by Corollary~\ref{cor: statef} and Corollary~\ref{cor: steppar}, we know that after executing line 15, there exists a struct instance $q_1$ s.t. $q_1.\mathit{place} = p_1.\mathit{place}$ and $p_1.\mathit{place.p1} = \ell_q$. 	
	As after line 13, the \textit{place} parameter is not further updated in the step code, $p_1.\mathit{place} = p'.\mathit{place}$ and with $q = \Structs'(\ell_q)$, $q.\mathit{place} = q_1.\mathit{place}$, it follows by Corollary~\ref{cor: statef} and Corollary~\ref{cor: steppar} that after executing line 15, $\exists\ell\in\Labels_\mathit{OldElem}.(\Structs'(\ell') = q \land q.\mathit{place} = p'.\mathit{place} \land p'.\mathit{place.p1} = \ell)$.
	
	When $p.\mathit{val} > p.\mathit{place.split}$, we know that $\ell_p\notin \Labels_0$, as then $p.\mathit{val} = 0$ and $p.\mathit{place.split} = 0$. Therefore, we know through Corollary~\ref{cor: statef} with line 19 that if $(p.\mathit{val} > p.\mathit{place.split})$, $p'.\mathit{move} = \true$. Proving that $(p.\mathit{val} > p.\mathit{place.spl})\implies \exists\ell\in\Labels_\mathit{OldElem}.(\Structs'(\ell) = q \land q.\mathit{place} = p'.\mathit{place} \land p'.\mathit{place.p2} = \ell)$ uses the same reasoning as proving that $(p.\mathit{val} \leq p.\mathit{place.spl})\implies \exists\ell\in\Labels_\mathit{OldElem}.(\Structs'(\ell) = q \land q.\mathit{place} = p'.\mathit{place} \land p'.\mathit{place.p1} = \ell)$, so point 4 holds.
\end{proof}
\begin{lemma}[\textit{checkStable} Contract]\label{lem: Sort2cSc}
	Let $P = \langle \Sched, \Structs, \Stab\rangle$ be an \Lname state and let $\Labels_\mathit{OldElem} \subseteq \Labels$ be the labels of \textit{OldElem} structs in $P$. Executing \textit{checkStable} starting at state $P$ results in a state $P' = \langle \Sched, \Structs', \Stab\rangle$. In $P'$, for every $\ell_p$ s.t. $p = \Structs(\ell_p) = \langle \mathit{OldElem}, \ComList, \Stack, \Env_p\rangle$ and $p' = \Structs'(\ell_p)$, it holds that 
	\begin{equation*}
		\begin{split}
			(p.\mathit{place.p2} \neq \ell^0_\mathit{OldElem}) &\lor (p.\mathit{val} \leq p.\mathit{place.split} \land p.\mathit{place.p1} \neq \ell_p) \implies\\ &p'.\mathit{place.done} = \false
		\end{split}
	\end{equation*}
\end{lemma}
\begin{proof}
	Holds by application of Corollaries~\ref{cor: statef} and \ref{cor: steppar} Note that applying Corollary~\ref{cor: steppar} is this simple because only $\false$ can be written to $\mathit{place.done}$.
\end{proof}
\begin{corollary}[\textit{checkStable} Reverse Contract]\label{lem: Sort2cScq}
	Let $P = \langle \Sched, \Structs, \Stab\rangle$ be an \Lname state and let $\Labels_\mathit{NewElem} \subseteq \Labels$ be the labels of \textit{NewElem} structs in $P$ and let $\Labels_\mathit{OldElem} \subseteq \Labels$ be the labels of \textit{OldElem} structs in $P$. Executing \textit{checkStable} starting at state $P$ results in a state $P' = \langle \Sched, \Structs', \Stab\rangle$. In $P'$, for every $\ell_q$ s.t. $q = \Structs(\ell_q) = \langle \mathit{NewElem}, \ComList, \Stack, \Env_p\rangle$ and $q' = \Structs'(\ell_q)$, it holds that 
	\begin{equation*}
		\begin{split}
			q'.\mathit{done} = \false \iff &\\
			(q.\mathit{p2} \neq \ell^0_\mathit{OldElem}) &\lor \exists \ell_r\in\Labels_\mathit{OldElem}(\Structs(\ell_r).\mathit{val} \leq q.\mathit{split} \land q.\mathit{p1} \neq \ell_r) 
		\end{split}
	\end{equation*}
\end{corollary}
\begin{proof}
	Follows from applying Lemma~\ref{lem: Sort2cSc} on all old elements from the perspective of the result, which allows us to make the implication a bi-implication.
\end{proof}
\begin{lemma}[\textit{split} Contract]\label{lem: Sort2splitc}
	Let $P = \langle \Sched, \Structs, \Stab\rangle$ be an \Lname state and let $\Labels_\mathit{NewElem} \subseteq \Labels\setminus$ be the labels of \textit{NewElem} structs in $P$. Let $\Labels_P$ be all labels used in $P$. Executing \textit{split} starting at state $P$ results in a state $P' = \langle \Sched, \Structs', \Stab\rangle$. In $P'$, for every $\ell_p$ s.t. $p = \Structs(\ell_p) = \langle \mathit{OldElem}, \ComList, \Stack, \Env_p\rangle$ and $p' = \Structs'(\ell_p)$, all of the following hold:
	\begin{enumerate}
		\item If $\ell_p\in\Labels_0$, then $p = p'$ and no struct instance is created during the execution of \textit{split} by the struct instance of $\ell_p$.
		\item Iff $\ell_p\notin\Labels_0$, all of the following hold:
		\begin{enumerate}
			\item $p'.\mathit{done} = \true$.
			\item Iff $p.\mathit{done} = \false$:
			\begin{enumerate}
				\item $p.\mathit{p1} \neq \ell^0_\mathit{OldElem} \land p.\mathit{p2} \neq \ell^0_\mathit{OldElem} \iff \exists \ell\notin\Labels_P.(\Structs'(\ell) = \langle \mathit{NewElem}, \vempty, \vempty, \Env_\ell\rangle \land \Env_\ell = \Env^0_\mathit{NewElem}[\{\mathit{min} \mapsto \mathit{p.spl}+1, \mathit{spl} \mapsto \mathit{p.spl} + (\mathit{p.max}-\mathit{p.spl})/2, \mathit{max}\mapsto \mathit{p.max}, \mathit{next}\mapsto\mathit{p.next}, \mathit{done}\mapsto\true \}] \land p'.\mathit{next} = \ell) \land p'.\mathit{max} = \mathit{p.spl} \land p'.\mathit{hasSplit} = \true$,
				\item $p.\mathit{p1} \neq \ell^0_\mathit{OldElem} \land p.\mathit{p2} = \ell^0_\mathit{OldElem} \iff p'.\mathit{max} = \mathit{p.spl}\land p'.\mathit{hasSplit} = \false$,
				\item $p.\mathit{p1} = \ell^0_\mathit{OldElem} \land p.\mathit{p2} \neq \ell^0_\mathit{OldElem} \iff p'.\mathit{min} = \mathit{p.spl}+1\land p'.\mathit{hasSplit} = \false$,
				\item $p'.\mathit{spl} = \mathit{p'.min} + (\mathit{p'.max} - \mathit{p'.min})/2 \land p'.\mathit{p1} = \ell^0_\mathit{OldElem} \land p'.\mathit{p2} = \ell^0_\mathit{OldElem}$.
			\end{enumerate}
		\end{enumerate}
	\end{enumerate}
\end{lemma}
\begin{proof}
	As, by Lemma~\ref{lem: Sort2det}, \textit{split} does not have any race conditions, point $2$ follows from multiple applications of Corollary~\ref{cor: statef}. For the first point, note that if $\ell_p\in\Labels_0$, $p.\mathit{p1} = \ell_\mathit{OldElem}^0$ and $p.\mathit{p2} = \ell^0_\mathit{OldElem}$. Therefore, the struct instance of $\ell_p$ will not create a new struct instance as per line 27. That $p = p'$ again follows from multiple applications of Corollary~\ref{cor: statef}. Therefore, point $1$ holds.
\end{proof}
We then consider the schedule. We first fix the state after the first execution of \textit{migrate}:
\begin{lemma}[Fixpoint premise]\label{lem: Sort2FixPremise}
Let $P_1$ be the premise state and let $P_2$ be the state after the first execution of \textit{migrate}. Then in $P_2$, all of the following hold:
\begin{enumerate}
	\item For every element $p_i$ in $p_1,\ldots, p_n$ there exists a struct instance $a_i$ (in $A = a_1,\ldots a_n$) with label $\ell_i$ (in $L = \ell_1, \ldots, \ell_n$) s.t. $a_i = \langle \mathit{OldElem}, \vempty, \vempty, \Env_i\rangle$ and $\Structs(\ell_i) = a_i$.
	\item There exists a non-$\nil$ struct instance $b$ with label $\ell_b$ s.t. $b = \langle \mathit{newElem}, \vempty, \vempty, \Env_b\rangle$.
	\item For all $a_i$, $a_i.\mathit{val} = x_i$ with $x_i\in x_1,\ldots x_n$ and $a_i.\mathit{place} = b$.
	\item $b.\mathit{min} = \min(x_1, \ldots, x_n)$, $b.\mathit{max} = \max(x_1, \ldots, x_n)$, $\b.\mathit{spl} = (b.\mathit{max} - b.\mathit{min})/2 + b.\mathit{min}$, $b.\mathit{next} = \ell_\mathit{NewElem}^0$, $b.\mathit{hasSplit} = \false$ and $b.\mathit{done} = \true$.
	\item Let $A_1$ be the set of $a_i\in A$ s.t. $a_i.\mathit{val} \leq b.\mathit{spl}$, with labels $L_1$, and let $A_2 = A\setminus A_1$, with labels $L_2$. Then for all $a_i\in A_1$, $a_i.\mathit{move} = \false$, for all $a_i\in A_2$, $a_i.\mathit{move} = \true$, if $|A_1| > 0$, $\exists \ell_i\in L_1.(b.\mathit{p1} = \ell_i)$, and if $|A_2| > 0$, $\exists \ell_i\in L_2.(b.\mathit{p2} = \ell_i)$.
\end{enumerate}
\end{lemma}
\begin{proof}
	The first three points are directly derived from Definition~\ref{def: Sort2Prem}. The last point follows from Lemma~\ref{contract: migrate}, as when $|A_1| > 0$, there is at least one old element for which point $2$ of Lemma~\ref{contract: migrate} applies, and if $|A_2| > 0$, there is at least one old element for which point $3$ of Lemma~\ref{contract: migrate} applies. Note that point $1$ applies to none of the elements from the premise. 
\end{proof}

We then prove the following lemma specifying the effect of a single fixpoint execution:
\begin{lemma}[Fixpoint execution contract]\label{lem: Sort2fixcontract}
	Let $P = \langle \Sched, \Structs, \Stab\rangle$ be an \Lname state, let $\Labels_\mathit{OldElem} \subseteq \Labels$ be the labels of \textit{OldElem} structs in $P$ and let $\Labels_\mathit{NewElem} \subseteq \Labels$ be the labels of \textit{NewElem} structs in $P$. Let $\Labels_P$ be all labels used in $P$. Executing an iteration of the fixpoint starting at state $P$ results in a state $P' = \langle \Sched, \Structs', \Stab\rangle$. In $P'$, for every $\ell_p$ s.t. $p = \Structs(\ell_p) = \langle \mathit{OldElem}, \ComList, \Stack, \Env_p\rangle$ and $p' = \Structs'(\ell_p)$ and $\ell_q$ s.t. $q = \Structs(\ell_q) = \langle \mathit{NewElem}, \ComList, \Stack, \Env_q\rangle$, $q' = \Structs'(\ell_q)$ and $p.\mathit{place} = q$, all of the following hold:
	\begin{enumerate}
		\item If $\ell_p\in\Labels_0$, $p = p'$, and if $\ell_q\in\Labels_0$, $q = q'$ and no structure instance is created during the execution of split by the struct instance of $\ell_q$.
		\item Iff $\ell_p\notin\Labels_0$ and $\ell_q\notin\Labels_0$, all of the following hold:
		\begin{enumerate}
			\item $q'.\mathit{done} = \true$
			\item Iff $(q.\mathit{p2} \neq \ell^0_\mathit{OldElem}) \lor \exists \ell_r\in\Labels_\mathit{OldElem}(\Structs(\ell_r).\mathit{val} \leq q.\mathit{split} \land q.\mathit{p1} \neq \ell_r)$ all of the following hold:
			\begin{enumerate}
				\item $q.\mathit{p1} \neq \ell^0_\mathit{OldElem} \land q.\mathit{p2} \neq \ell^0_\mathit{OldElem} \iff \exists \ell\notin\Labels_P.(\Structs'(\ell) = \langle \mathit{NewElem}, \vempty, \vempty, \Env_\ell\rangle \land \Env_\ell = \Env^0_\mathit{NewElem}[\{\mathit{min} \mapsto \mathit{q.spl}+1, \mathit{spl} \mapsto \mathit{q.spl}+1 + (\mathit{q.max}-(\mathit{q.spl}+1))/2, \mathit{max}\mapsto \mathit{q.max}, \mathit{next}\mapsto\mathit{q.next}, \mathit{done}\mapsto\true \}]\land q'.\mathit{next} = \ell) \land q'.\mathit{max} = \mathit{q.spl} \land q'.\mathit{hasSplit} = \true$
				\item $q.\mathit{p1} \neq \ell^0_\mathit{OldElem} \land q.\mathit{p2} = \ell^0_\mathit{OldElem} \iff q'.\mathit{max} = \mathit{q.spl}\land q'.\mathit{hasSplit} = \false$,
				\item $q.\mathit{p1} = \ell^0_\mathit{OldElem} \land q.\mathit{p2} \neq \ell^0_\mathit{OldElem} \iff q'.\mathit{min} = \mathit{q.spl}+1\land q'.\mathit{hasSplit} = \false$,
				\item $q'.\mathit{spl} = \mathit{q'.min} + (\mathit{q'.max} - \mathit{q'.min})/2 \land q'.\mathit{p1} = \ell^0_\mathit{OldElem} \land q'.\mathit{p2} = \ell^0_\mathit{OldElem}$.
				\item $q'.\mathit{done}^{*}$
			\end{enumerate}
			\item $p.\mathit{move} \land q'.\mathit{hasSplit} \iff p'.\mathit{place} = q'.\mathit{next}$,
			\item $(p.\mathit{val} > p'.\mathit{place.spl})\iff  p'.\mathit{move} = \true$,
			\item $(p.\mathit{val} \leq p'.\mathit{place.spl})\implies  \exists\ell\in\Labels_\mathit{OldElem}.(\Structs'(\ell) = r \land r.\mathit{place} = p'.\mathit{place} \land p'.\mathit{place.p1} = \ell)$,
			\item $(p.\mathit{val} > p'.\mathit{place.spl})\implies \exists\ell\in\Labels_\mathit{OldElem}.(\Structs'(\ell) = r \land r.\mathit{place} = p'.\mathit{place} \land p'.\mathit{place.p2} = \ell)$.
		\end{enumerate}
	\end{enumerate}
\end{lemma}
\begin{proof}
	This is a straightforward combination of the three contracts that make up the contracts for the fixpoint. If $\ell_p\in\Labels_0$, then all of its parameters are $\nil$, so none of the updates of $\ell_p$ will lead to changed parameters. If $\ell_q\in\Labels_0$, then point $2$ holds by Lemma~\ref{lem: Sort2splitc}. Point $2a$ holds by Lemma~\ref{lem: Sort2splitc}. The condition of $2b$ is from Corollary~\ref{lem: Sort2cScq}; as \textit{checkStable} determines whether \textit{done} is $\false$ and the execution of \textit{split} directly depends on that, the clauses $i$-$iv$ are lifted from \textit{split} and therefore hold due to Lemma~\ref{lem: Sort2splitc}. Clause $v$ is a result of the interaction between \textit{checkStable} and \textit{split}; if the condition in $2b$ is true, then \textit{checkStable} will change $q.\mathit{done}$ to $\false$, which is later overwritten in \textit{split} with $\true$. If the condition is $\false$, then $q.\mathit{done}$ will not be changed, but will stay constantly $\true$. Clauses $2c$, $2d$, $2e$, and $2f$ hold due to Lemma~\ref{contract: migrate}. 
\end{proof}

We then use this lemma to prove the following lemma for executions of the fixpoint from the fixpoint premise:
\begin{lemma}[Fixpoint execution properties]\label{lem: Sort2props}
	Let $F_{i-1} = \langle \Sched, \Structs, \Stab\rangle$ be the state before the $i$th execution of the fixpoint from $P_2$ as defined in Lemma~\ref{lem: Sort2FixPremise}, with $F_0 = P_2$. Let $\Labels_\mathit{OldElem} \subseteq \Labels$ be the labels of \textit{OldElem} structs in $F_{i-1}$ and let $\Labels_\mathit{NewElem} \subseteq \Labels$ be the labels of \textit{NewElem} structs in $F_{i-1}$. Let $\Labels_P$ be all labels used in $P$. Let $\ell_p$ be a label of an \textit{OldElem} struct instance, where $\Structs(\ell_p) = p$. Then the following all hold:
	\begin{enumerate}
		\item $p.\mathit{val} > p.\mathit{place.spl} \iff p.\mathit{move} = \true$.
		\item $(p.\mathit{val} \leq p.\mathit{place.spl})\implies  \exists\ell\in\Labels_\mathit{OldElem}.(\Structs(\ell) = r \land r.\mathit{place} = p.\mathit{place} \land p.\mathit{place.p1} = \ell)$,
		\item $(p.\mathit{val} > p.\mathit{place.spl})\implies \exists\ell\in\Labels_\mathit{OldElem}.(\Structs(\ell) = r \land r.\mathit{place} = p.\mathit{place} \land p.\mathit{place.p2} = \ell)$.
		\item $p.\mathit{place.spl} = p.\mathit{place.min} + \lfloor(p.\mathit{place.max} - p.\mathit{place.min})/2\rfloor$.
		\item If $p.\mathit{place.next} = \ell_\mathit{NewElem}^0$, then $\nexists \ell_t\in\Labels_\mathit{NewElem}.(\Structs(\ell_t) = t \land p.\mathit{place} \neq \ell_t\land t.\mathit{next} = \ell_\mathit{NewElem}^0)$.
		\item If $p.\mathit{place.next} \neq \ell_\mathit{NewElem}^0$, then $p.\mathit{place.max} < p.\mathit{place.next.min}$.
		\item $p.\mathit{place.min} \leq p.\mathit{val} \leq p.\mathit{place.max}$.
	\end{enumerate}
	Additionally, let let $F_{i} = \langle \Sched', \Structs', \Stab'\rangle$ be the state after the $i$th execution of the fixpoint from $P_2$, with $\Structs'(\ell_p) = p'$. Then all of the the following properties, describing the relation between $F_{i-1}$ and $F_i$, also hold:
	\begin{itemize}
		\item[\textbf{C1}] $p'.\mathit{val} = p.\mathit{val}$ and $p'.\mathit{val}$ is stable.
		\item[\textbf{R1}] If $\exists \ell_t\in\Labels_\mathit{OldElem}\setminus\{\ell_p\}.(\Structs(\ell_t) = t \land t.\mathit{place} = p.\mathit{place}) \lor p.\mathit{val} > p.\mathit{place.spl}$, then $p'.\mathit{place.max} - p'.\mathit{place.min} \leq \lceil p.\mathit{place.max} - p.\mathit{place.min}/2\rceil$.
		\item[\textbf{R2}] If $\nexists \ell_t\in\Labels_\mathit{OldElem}\setminus\{\ell_p\}.(\Structs(\ell_t) = t \land t.\mathit{place} = p.\mathit{place}) \land p.\mathit{val} \leq p.\mathit{place.spl}$, with $p.\mathit{place} = \ell_q$, then the struct instances of $\ell_p$ and $\ell_q$ do not make new struct instances in execution $i$ and all parameters of $\ell_p$ and $\ell_q$ are stable after execution $i$, $p'.\mathit{place} = \ell_q$ and $q'.\mathit{p1} = \ell_p$.
	\end{itemize}
\end{lemma}
\begin{proof}
	First, \textbf{C1} directly follows from the fact that $p'.\mathit{val}$ is not assigned a value in the contract in Lemma~\ref{lem: Sort2fixcontract}. We use this to prove points $1$-$7$ hold for all possible $F_i$ by induction on $i$:
	\begin{description}
		\item[$i=0$.] Then $F_i = P_2$. Point $1$, $2$ and $3$ hold by point $5$ of the properties of $P_2$. Point $4$ holds by point $4$ of the properties of $P_2$ as given in Lemma~\ref{lem: Sort2FixPremise}. As there is only one \textit{NewElem} $b$ s.t. for all possible $p$, $p.\mathit{place} = b$ in $P_2$ (point $5$), and $b.\mathit{next} = \ell_\mathit{NewElem}^0$ in $P_2$ (point $6$), points $5$ and $6$ of the lemma hold. As $b.\mathit{min}$ is the minimum value of all old elements and $b.\mathit{max}$ is the maximum value of all old elements, point $7$ of the lemma also holds.
		\item[$i > 0$.] Then as induction hypothesis, we assume points $1$-$7$ hold for $F_{i-1}$. Then we prove the points separately:
		\begin{enumerate}
			\item Assume that $p'.\mathit{val} > p'.\mathit{place.spl}$. Then as the $\mathit{val}$ parameter never gets written to in Listing~\ref{ex: Sort2}, $p'.\mathit{val} > p'.\mathit{place.spl} \iff p.\mathit{val} > p'.\mathit{place.spl}$. Then by point $2d$ of Lemma~\ref{lem: Sort2fixcontract}, $p'.\mathit{val} > p'.\mathit{place.spl} \iff p'.\mathit{move} = \true$. 
			\item We assume that $(p'.\mathit{val} \leq p'.\mathit{place.spl})$, and prove that $\exists\ell\in\Labels_\mathit{OldElem}.(\Structs'(\ell) = r \land r.\mathit{place} = p'.\mathit{place} \land p'.\mathit{place.p1} = \ell)$. From Lemma~\ref{lem: Sort2fixcontract}, due to the fact that $p'.\mathit{val} = p.\mathit{val}$, we can use point $2e$ directly to establish this point.
			\item Proven along the same lines as the above point, only using point $2f$ from Lemma~\ref{lem: Sort2fixcontract} instead.
			\item To prove that $p'.\mathit{place.spl} = p'.\mathit{min} + \lfloor(p'.\mathit{max} - p'.\mathit{min})/2\rfloor$, note that $p'.\mathit{place}$ is either equal to $p.\mathit{place}$ or not. We do a case distinction on these cases. If $p'.\mathit{place} = p.\mathit{place}$, according to point $2b$-$iv$ in Lemma~\ref{lem: Sort2fixcontract} and the fact that \Lname only does integer calculations (and therefore the result is automatically rounded down), this point holds. If $p'.\mathit{place} \neq p.\mathit{place}$, let $p.\mathit{place} = \ell_q$ and let $q = \Structs(\ell_q)$ and $q' = \Structs'(\ell_q)$. Then as $p'.\mathit{place} \neq p.\mathit{place}$ and as the only possible value is then $p'.\mathit{place} = q'.\mathit{next}$ by point $2c$ of Lemma~\ref{lem: Sort2fixcontract}, we know from $2c$ that $p.\mathit{move} = \true$ and $q'.\mathit{hasSplit} = \true$. It then follows by point $2b$-$i$ of Lemma~\ref{lem: Sort2fixcontract} that $q'.\mathit{next.spl} = q.\mathit{spl} + 1 + (q.\mathit{max} - (q.\mathit{spl}+1))/2 = q'.\mathit{next.min} + (q'.\mathit{next.max} - q'.\mathit{next.min})/2$. Then, as again \Lname only does integer calculations, the point also holds in this case.
			\item By the induction hypothesis, if $p.\mathit{place.next} = \ell_\mathit{NewElem}^0$, then $\nexists \ell_t\in\Labels_\mathit{NewElem}.(\Structs(\ell_t) = t \land p.\mathit{place} \neq \ell_t\land t.\mathit{next} = \ell_\mathit{NewElem}^0)$. Then as no $\mathit{next}$ parameter is set to $\nil$ in Listing~\ref{ex: Sort2}, if there are two \textit{NewElem}s in $F_i$ with an empty $\mathit{next}$ parameter, then both must be related to $p.\mathit{place}$. Let $p.\mathit{place} = \ell_q$, with $\Structs'(\ell_q) = q'$. Then as the struct instance of $\ell_q$ can create only one new struct instance in execution $i$, the \textit{NewElem} struct instances with an empty $\mathit{next}$ parameter must be $\ell_q$ and a struct instance with a label $\ell$ made by $\ell_q$. But then by Lemma~\ref{lem: Sort2fixcontract}, this can only be $q'.\mathit{next}$, which means that $\ell_q$ does have a non-$\nil$ value for $\mathit{next}$. Therefore, this point holds.
			\item By the induction hypothesis, if $p.\mathit{place.next} \neq \ell_\mathit{NewElem}^0$, then $p.\mathit{place.max} < p.\mathit{place.next.min}$. Let $\ell_q = p.\mathit{place.next}$, with $\Structs(\ell_q) = q$ and $\Structs'(\ell_q) = q'$. Then $q.\mathit{min} \leq q'.\mathit{min}$ and $p'.\mathit{max} \leq p.\mathit{max}$, as seen in cases $2b$-($i$-$iv$) of Lemma~\ref{lem: Sort2fixcontract}. It follows that $p'.\mathit{max} < q'.\mathit{min}$ so this point holds.
			\item 
			By the induction hypothesis, $p.\mathit{place.min} \leq p.\mathit{val} \leq p.\mathit{place.max}$. 
			Let $p.\mathit{place} = \ell_q$. Let $q = \Structs(\ell_q)$ and $q' = \Structs'(\ell_q)$. Then either $p'.\mathit{place} = \ell_q$ or not. We use a case distinction: If $p'.\mathit{place} =\ell_q$, then if the condition of $2b$ of Lemma~\ref{lem: Sort2fixcontract} does not hold for $q$, $q' = q$, and as $p'.\mathit{val} = p.\mathit{val}$, this point holds. We therefore assume that the condition of $2b$ does hold for $q$. 
			
			Then either $p.\mathit{move} = \true$ or not. In the case where $p.\mathit{move} = \true$, we know that $p'.\mathit{val} = p.\mathit{val} \geq q.\mathit{spl} + 1 \geq q'.\mathit{min}$. To prove that $p'.\mathit{val} \leq q'.\mathit{max}$, note that by the induction hypothesis, it follows from $p.\mathit{val} \geq q.\mathit{spl}+1$ that $q.\mathit{p2} \neq \ell^0_\mathit{OldElem}$ (point $3$ of this lemma). As we know that $p.\mathit{move} = \true$ but that $p'.\mathit{place} = \ell_q$, it follows that $q'.\mathit{hasSplit} = \false$. From this it follows that $2b$-$iii$ holds and $2b$-$i$ and $2b$-$ii$ do not hold, from which we conclude that $q'.\mathit{max} = q.\mathit{max}$. It follows that $p'.\mathit{place.min} \leq p'.\mathit{val} \leq p'.\mathit{place.max}$.
			If $p.\mathit{move} = \false$, it follows that $p'.\mathit{val} = p.\mathit{val} \leq q.\mathit{spl} \leq q'.\mathit{max}$. We then also know that $q.\mathit{p1}\neq \ell^0_\mathit{OldElem}$ (point $2$ in this lemma), from which it follows that either $2b$-$i$ or $2b$-$iii$ of Lemma~\ref{lem: Sort2fixcontract} can hold for $\ell_p$ and $\ell_q$, but not $2b$-$ii$. In both cases, $q'.\mathit{min} = q.\mathit{min}$, from whcih it follows that $p'.\mathit{place.min} \leq p'.\mathit{val} \leq p'.\mathit{place.max}$.
			
			It rests to prove the point if $p'.\mathit{place} \neq \ell_q$. In this case, we know that $p'.\mathit{place} = q'.\mathit{next}$ (the only value other than $\ell_q$ it can have according to Lemma~\ref{lem: Sort2fixcontract}), from which it follows that $p.\mathit{move} = \true$ and $q'.\mathit{hasSplit} = \true$. From $p.\mathit{move} = \true$, we get that $p.\mathit{val} \geq q.\mathit{spl} + 1$ (point $1$ of this lemma). Then we know that $q.\mathit{p2}\neq \ell^0_\mathit{OldElem}$ (point $3$ of this lemma), from which it follows that the condition for $2b$ holds for $\ell_q$. As $q'.\mathit{hasSplit} = \true$, it follows that $2b$-$i$ holds for $q$, from which it follows that $p'.\mathit{place.min} = q.\mathit{spl} + 1$ and $p'.\mathit{place.max} = q.\mathit{max}$. As $p'.\mathit{val} = p.\mathit{val}$, it follows that $p'.\mathit{place.min} \leq p'.\mathit{val} \leq p'.\mathit{place.max}$.			
			As then in all cases $p'.\mathit{place.min} \leq p'.\mathit{val} \leq p'.\mathit{place.max}$, this point holds.
		\end{enumerate}
		As all points hold, the step case holds.
	\end{description}
	As the induction holds, points $1$-$7$ hold. We use these points to prove points \textbf{R1} and \textbf{R2}, for $F_{i-1}$ and $F_i$:
	\begin{itemize}
		\item[\textbf{R1}.] If $\exists \ell_t\in\Labels_\mathit{OldElem}.(\Structs(\ell_t) = t \land t.\mathit{place} = p.\mathit{place}) \lor p.\mathit{val} > p.\mathit{place.spl}$, let $p.\mathit{place} = \ell_q$, with $\Structs(\ell_q) = q$ and $\Structs'(\ell_q) = q'$. Then if $p.\mathit{val} > p.\mathit{place.spl}$, we know from point $3$ of this lemma that $q.\mathit{p2} \neq \ell^0_\mathit{OldElem}$ and from point $1$ of this lemma that $p.\mathit{move} = \true$. Then either $2b$-$i$ or $2b$-$iii$ of Lemma~\ref{lem: Sort2fixcontract} holds for $p$, $p'$, $q$ and $q'$. Regardless of the exact case, it follows that $p'.\mathit{place.min} = q.\mathit{spl} + 1$ and $p'.\mathit{place.max} = q.\mathit{max}$. From point $4$ of this lemma, we know that $q.\mathit{spl} = q.\mathit{min} + \lfloor (q.\mathit{max} + q.\mathit{min})/2\rfloor$, so 
		\begin{equation*}
			\begin{split}
				&p'.\mathit{place.max} - p'.\mathit{place.min} \\
				=\;&q.\mathit{max} - (q.\mathit{min} + \lfloor (q.\mathit{max} - q.\mathit{min})/2\rfloor + 1)\\
				\leq\;&(q.\mathit{max} - q.\mathit{min}) - (\lfloor (q.\mathit{max} - q.\mathit{min})/2\rfloor)\\
				=\;&\lceil (q.\mathit{max} - q.\mathit{min})/2\rceil\\
				=\;&\lceil (\mathit{p.place.max} - \mathit{p.place.min})/2\rceil.
			\end{split}
		\end{equation*}
		Then if $p.\mathit{val} \leq p.\mathit{place.spl}$, it holds that $\exists \ell_t\in\Labels_\mathit{OldElem}.(\Structs(\ell_t) = t \land t.\mathit{place} = p.\mathit{place})$, so we know that there are multiple \textit{OldElems} which qualify for $q.\mathit{p1}$. Then the condition of $2b$ holds, and either $2b$-$i$ holds or $2b$-$ii$ holds. In both cases it follows that $p'.\mathit{place.max} = q.\mathit{spl}$ and $p'.\mathit{place.min} = q.\mathit{min}$. From point $4$ of this lemma, we know that $q.\mathit{spl} = q.\mathit{min} + \lfloor (q.\mathit{max} + q.\mathit{min})/2\rfloor$, so 
		\begin{equation*}
			\begin{split}
				&p'.\mathit{place.max} - p'.\mathit{place.min}\\
				=\;&q.\mathit{spl} - q.\mathit{min}\\
				=\;&q.\mathit{min} - q.\mathit{min} + \lfloor (q.\mathit{max} + q.\mathit{min})/2\rfloor\\
				\leq\;&\lceil (q.\mathit{max} + q.\mathit{min})/2\rceil \\
				=\;&\lceil p.\mathit{place.max} - p.\mathit{place.min}/2\rceil.
			\end{split}
		\end{equation*}
		It follows that \textbf{R1} holds.
		\item[\textbf{R2}.] Assume $\nexists \ell_t\in\Labels_\mathit{OldElem}\setminus\{\ell_p\}.(\Structs(\ell_t) = t \land t.\mathit{place} = p.\mathit{place}) \land p.\mathit{val} \leq p.\mathit{place.spl}$. The condition holds straightforwardly when $\ell_p \in\Labels_0$ or $\ell_q\in\Labels_0$, by point $1$ of Lemma~\ref{lem: Sort2fixcontract} and the fact that no struct instances are ever created by $p$ during the execution of the fixpoint as per the same lemma. Then if both $\ell_p$ and $\ell_q$ are not $\nil$-labels, as $p$ does not create struct instances and the condition for $2b$ does not hold for $q$ due to the assumption, no struct instances are created. 
		
		To prove that the parameters are stable, we also consider Lemma~\ref{lem: Sort2fixcontract}. As the condition for $2b$ does not hold, we know that only $q.\mathit{p1}$ and $q.\mathit{p2}$ can have changed, through $2e$ and $2f$. $q'.\mathit{done}$ cannot have changed, as only when the condition of $2b$ holds $q'.\mathit{done}$ can be unstable.
		
		Due to our assumption, we know that $2f$ cannot apply and that $q'.\mathit{p1}$ can only be $\ell_q$. Additionally, from point $2$ of this lemma and our assumption, we also know that $q.\mathit{p1} = \ell_p$. Therefore, all parameters of $q$ are stable and $q'.\mathit{p1} = q.\mathit{p1} = \ell_p$.
		
		For the parameters of $\ell_p$, note that $p.\mathit{place} = \ell_q$. Due to point $1$ of this lemma, we know due to the assumption that $p.\mathit{move} = \false$, from which it follows by $2c$ of Lemma~\ref{lem: Sort2fixcontract} that $p'.\mathit{place} = p.\mathit{place} = \ell_q$. Additionally, from the assumption and $2d$ of Lemma~\ref{lem: Sort2fixcontract} it follows that $p'.\mathit{move} = \false$ as well. From point \textbf{\textit{C}} of this lemma, $p.\mathit{val}$ has not changed in the iteration and is stable. Therefore, all of the parameters of $\ell_p$ are stable in $p'$ and $p'.\mathit{place} = \ell_q$.
		
		Then if $\nexists \ell_t\in\Labels_\mathit{OldElem}\setminus\{\ell_p\}.(\Structs(\ell_t) = t \land t.\mathit{place} = p.\mathit{place}) \land p.\mathit{val} \leq p.\mathit{place.spl}$, all parameters of $\ell_p$ and $\ell_q$ are stable after the execution of the fixpoint iteration, $p'.\mathit{place} = \ell_q$ and $q.\mathit{p1} = \ell_p$.
		\end{itemize}
		With all points proven, the lemma holds.
\end{proof}
We use this lemma to prove the following lemma:
\begin{lemma}[Fixpoint termination and result]\label{lem: Sort2fix}
	The execution of the fixpoint starting at $P_2$ as specified in Lemma~\ref{lem: Sort2FixPremise} terminates. Let $P' = \langle \Sched, \Structs', \Stab\rangle$ be the state after the execution of the fixpoint. Then the following all hold:
	\begin{enumerate}
		\item For every \textit{NewElem} element $q\in\Structs'$ with label $\ell_q$, there is exactly one \textit{OldElem} element $p\in\Structs'$ with label $\ell_p$ s.t. $p.\mathit{place} = \ell_q$ and $p.\mathit{place.min}\leq p.\mathit{val} \leq p.\mathit{place.max}$. This element can be referenced by $q.\mathit{p1}$.
		\item For every \textit{NewElem} element $q\in\Structs'$ s.t. $q.\mathit{next} \neq \ell_\mathit{NewElem}^0$, $q.\mathit{max} < q.\mathit{next.min}$.
		\item There is exactly one \textit{NewElem} element $q\in\Structs'$ s.t. $q.\mathit{next} = \ell_\mathit{NewElem}^0$.
	\end{enumerate}
\end{lemma}
\begin{proof} 
	First of all, in any state $P$ with struct environment $\sigma$ during or after the execution of the fixpoint, there is always at least one \textit{OldElem} $p\in\sigma$ s.t. $p.\mathit{place} = \ell_q$ for every \textit{NewElem} $q\in\sigma$ with label $\ell_q$. This is due to new elements only being made if there are \textit{OldElem}s in both $q.\mathit{p1}$ and $q.\mathit{p2}$ as representatives of both halves of the range of $q$ and the range of $q$ being dynamically updated otherwise (Lemma~\ref{lem: Sort2fixcontract}). 
	
	Due to \textbf{R1} of Lemma~\ref{lem: Sort2props} We know that if there are multiple elements with $\ell_q$ as value of their \textit{place} parameter before an iteration of the fixpoint, the range of $\ell_q$ will halve during the iteration of the fixpoint. As we deal with distinct numbers, after enough halvings of the range, there will always eventually be exactly one \textit{OldElem} that has $\ell_q$ as its \textit{place}, and due to the dynamic range halving, this element is eventually referenced by the \textit{p1} parameter of $\ell_q$. Due to \textbf{R2} of Lemma~\ref{lem: Sort2props} it follows that that element and $\ell_q$ are then stable. As this holds for any $\ell_q$, all \textit{OldElems} and \textit{NewElems} will eventually be stable and the fixpoint will terminate. 
	
	From this and from point $7$ of Lemma~\ref{lem: Sort2props}, it also follows that point $1$ of this lemma holds.
	Point $2$ follows from point $6$ of Lemma~\ref{lem: Sort2props}, and point $3$ holds from point $5$ of Lemma~\ref{lem: Sort2props}. Therefore, the entire lemma holds.
\end{proof}

It then follows that:
\begin{theorem}[Correctness of Listing~\ref{ex: Sort2}]
	\label{thm:correctness-sorting}
	When executing the \Lname code as given in Listing~\ref{ex: Sort2} from a premise state according to Definition~\ref{def: Sort2Prem} for a set of elements $p_1,\ldots, p_n$ with distinct positive values $x_1, \ldots, x_n$, it results in a state in which the elements are arranged into a sequence $S_2 = p'_1,\ldots, p'_n$ with values $x'_1,\ldots,x'_n$ s.t. for every $p_i\in S_1$, $p_i\in S_2$ and $x'_i < x'_{i+1}$ for every $1\leq i \leq n$.
\end{theorem}
\begin{proof}
	The sequence is the sequence of \textit{NewElems}, with the corresponding \textit{OldElem} referenced through parameter \textit{p1}. In this sequence, every old element is represented due to point $1$ of Lemma~\ref{lem: Sort2fix}, the sequence is complete due to point $2$ and the sequence is sorted due to point $3$ and point $1$.
\end{proof} 
\subsection{Conclusions on Proving \Lname programs}\label{section:proofconcs}
In this section, we identify some similarities of the proofs we have presented. The first thing we establish is that we always maintain roughly the following structure, which we expect can be applied to most \Lname proofs:
\begin{enumerate}
	\item Establish the overall premise.
	\item Establish the degree of determinism and find out which parameters are not directly deterministic.
	\item Establish a contract for each step:
	\begin{itemize}
		\item Each step is considered separately. 
		\item The syntax directly corresponds to the form of the step contract.		
		\item Contracts can be combined by merging the relevant predicates; for example, a contract of a sequence of two steps can be formed by first simplifying the precondition predicate for the second step with the postcondition predicate for the first step and then adding the remaining predicate to the precondition predicate of the first step.
		\item Contracts are \emph{relative}: they require a basis element, which is standardly an abstract element executing a step, but which can also be another involved element (as in the case of Corollary~\ref{lem: Sort2cScq}). It is therefore often possible to abstract away from which specific struct instance is executing a step.
	\end{itemize}
	\item For every fixpoint sequentially:
	\begin{itemize}
		\item Establish the premise state before the fixpoint.
		\item Make a combined contract for the steps inside the fixpoint.
		\item Prove fixpoint termination (if possible).
		\item Prove the fixpoint result (which can be combined with the previous step).
	\end{itemize}
	\item Relate the result of executing the last step/fixpoint to the overall result, and relate that back to the original premise state.	
\end{enumerate}
This gives us some insight in how we can further develop verification systems for \Lname. First of all, as step contracts are relative to a basis and do not necessarily need to make assumptions about the context other than the available structs and their parameters, they can be reasoned about locally. They are also \emph{modular}, for as the step contracts do not require context, they can be made separately and combined where necessary. As steps are finite by definition, generating step contracts can be automated. 
Similarly, as determinism is dependent on the syntax, this can possibly be automated as well.
Then, for fixpoints, as they are always dependent on stability of the entire system for termination, we estimate that the termination proof often comes down to proving that there is a monotonic function over all the data which eventually stabilizes. Finding this function also gives insight in the result of the fixpoint. 
\section{Related Work}\label{section:relatedwork}
Conceptually, our work is related to the Parallel Pointer Machine (PPM)~\cite{cook-parallel-1993, goodrich-sorting-1996}, which models memory as a graph that is traversed by processors.
\Lname differs from this in that processors are implicit and data is the main focus.

The concept of cooperating data elements is present in the Chemical Abstract Machine\cite{berry-chemical-1992}, based on the $\Gamma$-language~\cite{banatre-gamma-1990, banatre-parallel-1988}.
In the data-autonomous paradigm these components are coordinated by a schedule as opposed to the Chemical Abstract Machine, where the data elements float around freely.
By extension, \Lname is related to the $\Gamma$-Calculus Parallel Programming Framework~\cite{gannouni-gamma-calculus-2015}. 

The data-autonomous paradigm shares the same focus on data as \emph{message passing} languages like Active Pebbles~\cite{willcock-active-2011}, ParCel-2~\cite{cagnard-parcel-2-2000} and AL-1~\cite{marcoux-1-1988}, but differs in using shared variables instead of synchronisation and messages.
It also does not allow the use of data as passive elements, like in the messages of MPI~\cite{clarke-mpi-1994}.

The \emph{specialist-parallel} approach~\cite{carriero-how-1989} models a problem as a network of relatively autonomous nodes which perform one specified task. In comparison, the data-autonomous paradigm defines their specialists around data instead of tasks and data elements perform multiple or no tasks depending on their steps.

In \Lname, the relations between data elements can be viewed as a graph, which is also the case for \emph{graph based} languages, such as DDG~\cite{tran-parallel-2000}, a scheduling language, and GraphGrind~\cite{sun-graphgrind-2017}, a graph partitioning language. 
The Connection Machine~\cite{hillis-connection-1989} uses a graph-based hardware architecture for parallel computation. Similarly, the way data is expressed in Legion~\cite{bauer-legion-2012} and OP2~\cite{mudalige-op2-2012} is similar to \Lname. However, these two languages work top down from a main process that calls functions on data, which is unlike the data-autonomous paradigm.

Since the data-autonomous paradigm extends data-parallelism (see Figure~\ref{fig:paradigms}), \Lname shares concepts with other data-parallel languages like CUDA~\cite{garland-parallel-2008, harris-parallel-2007} and OpenCL~\cite{chong-sound-2014}. 
It has the most in common with object-oriented approaches to data-parallelism, like the POOL family of languages~\cite{america-parallel-1990}, languages in which small elements do parallel computations based on their neighbours, like \textsc{ReLaCS}~\cite{raimbault-relacs-1993}, PPC~\cite{maresca-programming-1993}, Chestnut~\cite{stromme-chestnut-2012} and the ParCel languages~\cite{vialle-parcel-1-1996, cagnard-parcel-2-2000}, and \emph{actor languages}.

Actor languages, like Ly~\cite{ungar-harnessing-2010}, ParCel-1~\cite{vialle-parcel-1-1996}, PObC++~\cite{pinho-object-oriented-2014} and A-NETL~\cite{baba--netl-1995}, treat objects as independent, collaborating actors, in a similar way as how the data-autonomous paradigm treats data.
Often, these languages use the \emph{message passing} model to cooperate, which \Lname does not. 
Of those who do not, OpenABL~\cite{cosenza-openabl-2018} uses agents similar to data elements, but gives the agents to functions instead of functions to agents.
\emph{Active Object} languages~\cite{boer-complete-2007, boer-survey-2017} do give their objects functions, which is very closely related to data elements. 
The execution of functions in these objects however, is fully asynchronous: objects can activate other objects by calling methods in them for them to execute.
This is less structured than in \Lname, in which the functions to be executed are defined in the schedule.
As a result, \Lname does not use futures, unlike most active object languages.

The use of a schedule in the data-autonomous paradigm relates \Lname to some more functional data-parallel languages as well, like Halide~\cite{ragan-kelley-halide-2017}, which uses a schedule as well, and even Futhark~\cite{henriksen-futhark-2017}, in which the manipulation of an array has some similarity to calling a step in \Lname. The schedule can be considered as a coordination language~\cite{arbab-coordination-1998} for the paradigm and \Lname, but is fully integrated and required for both to function. It also does not need to create channels between components, like for example Reo~\cite{arbab-reo-2004}.

Similar to our motivation, ICE~\cite{ghanim-easy-2018}, which is a framework for implementing PRAM algorithms, sets the goal of bridging the gap between algorithms and implementation.
However, as ICE is based on a PRAM, it is not data autonomous.

In our approach for proving \Lname programs, we take cues from \emph{Hoare logic}~\cite{hoare-axiomatic-1969, apt-assessing-2021}, which has previously been used and expanded upon to prove programs correct or find correct programs~\cite{owicki-verifying-1976, owicki-axiomatic-1976, feijen-method-2013, de-vries-reverse-2011, myreen-hoare-2007, stirling-generalization-1988, lamport-hoare-1980} and of which some principles seem to suit \Lname particularly well.

Our proofs consist of two parts. The first part, in which we generate step contracts, is akin to \emph{symbolic execution}~\cite{king-symbolic-1976, baldoni-survey-2018}, albeit a variant which does not need to consider loops and needs to include parallelism. Our approach, like \emph{separation logic}~\cite{brookes-concurrent-2016, ohearn-separation-2019}, has a local focus: contracts are generated relative to their executing elements, consider only the parameters accessed by their executing elements and work with pointers. However, the way this locality is achieved is orthogonal; where separation logic partitions the memory to achieve locality for processes, our approach partitions the process into the steps and achieves the locality by only considering one step at a time.

The second part of the proofs aligns more with the standard practices of the Hoare logic family, where there is a precondition (defined by the initialization), an envisioned postcondition and the proof needs to bridge the gap between them. The difference is that instead of predefined rules (as used by Hoare logic family methods), the rules used here are the step contracts generated in the first part.

In our approach, we make a distinction between write-write race conditions and read-write race conditions, where we consider write-write race conditions benign enough to include and read-write race conditions as harmful. Distinctions between race conditions, whether it is good to have them and methods to find the harmful ones have been discussed before~\cite{narayanasamy-automatically-2007, boehm-how-2011}.

	\section{Conclusion}\label{section:conclusion}
In this paper, we presented the data-autonomous paradigm and introduced it by means of the Autonomous Data Language, by giving examples of standard algorithms and discussing the syntax, type system and semantics. We have given the basics for proving algorithms correct in \Lname, and demonstrated their use. Furthermore, we have proven the safety of the type system of \Lname.

This contributes to the research on the data-autonomous paradigm. With this paper, we have shown that the data-autonomous paradigm can be simple and verifiable. This combines with earlier work~\cite{franken-audala-2024}, where we show that \Lname is Turing Complete, to show that the paradigm is not less expressive than other paradigms. We have also defined a weak memory semantics for \Lname in~\cite{leemrijse-formalisation-2024}, which is used to create a compiler from \Lname to CUDA in~\cite{leemrijse-2023}, which allows for feasibly fast parallel execution of \Lname programs, demonstrating that the data-autonomous paradigm can be used practically.

One extension which we think is quite important for the practical use of \Lname is an extension to allow inheritance and packages in \Lname. Currently \Lname does not support these, which can lead to quite large struct definitions which have implementations of multiple different algorithms and functions in them. These algorithms also need to be reimplemented for every new struct type. With inheritance, it would be easier to make multiple subtypes of structs, which would ease this problem, while packages would allow structs to refer to different files for implementations. Note that these packages would need to also define a set of parameters required to be able to use the implementation in them.

For another practical extension, note that currently \Lname is focused on design of programs and not on the debugging of programs. It does not give error messages or halt when the given program does unexpected things which may not have been the intention of the programmer (such as have data races or write to $\nil$). The current view is that the programmer should understand what the program does from its design, not its effects. However, an extension of \Lname could be the inclusion of error messages and possible interruptions when a possibly unintended behaviour happens. Note that this will have an impact on Theorem~\ref{thm: wellformed}.

With the strong modularity of the steps and the structure of the proofs given in Section~\ref{section:proving}, we were strongly reminded of Hoare Logic~\cite{hoare-axiomatic-1969} and Concurrent Separation Logic~\cite{brookes-concurrent-2016, ohearn-separation-2019}. In the future, we intend to formalize a proof method for \Lname using, or based on, a suitable Hoare Logic or CSL, using the insights gained from the lemmas of Section~\ref{section:props} and the proofs of Section~\ref{section:proving}.

In other future work, we can extend on \Lname by finding and implementing algorithms to reduce the manual work it takes to generate \Lname step contracts. We can also utilize \Lname's commands to create data-race finding methods or extend \Lname. It would also be interesting to mechanize some of the proofs in a proving framework. Lastly, we can also look further into which domains of programs are particularly well expressed in \Lname, and compare \Lname to languages of that domain.
	
	\paragraph{Funding}
	Thomas Neele is supported by NWO grant VI.Veni.232.224.
	%
	%
	%
	 \bibliographystyle{elsarticle-num}
	 \bibliography{refs}
	%
	
\end{document}